\documentclass[numberedappendix,apj,twocolumn]{emulateapj} 

\usepackage{graphicx}
\usepackage{color}

\begin{document}

\setlength{\textwidth}{6.5in}
\setlength{\textheight}{9.0in}
\setlength{\oddsidemargin}{+0.2in}
\setlength{\evensidemargin}{0.0in}
\setlength{\topmargin}{-0.0in} 

\font\ksmrm=cmr10 scaled \magstephalf
\newcommand\n {\noindent}
\newcommand\m {\medskip}
\newcommand\s {\smallskip} 
\newcommand\bi{\bigskip\indent}
\newcommand\mi{\medskip\indent}
\newcommand\si{\smallskip\indent}
\newcommand\bn{\bigskip\noindent}
\newcommand\mn{\medskip\noindent}
\newcommand\sn{\smallskip\noindent}
\newcommand\cl{\centerline} 
\newcommand\ve{\vfill\eject}
\newcommand\degree   {{\ifmmode^\circ\else$^\circ$\fi}}
\newcommand\arcm     {{\ifmmode {'\ }\else$'     $\fi} } 
\newcommand\arcs     {{\ifmmode{''\ }\else$''    $\fi} } 
\newcommand\arcmpt   {{$\buildrel{\prime}\over .$}}
\newcommand\arcspt   {{$\buildrel{\prime\prime}\over .$}}
\newcommand\bII      {{$b^{II}$} }
\newcommand{\bul}    {$\bullet$\ }
\newcommand\cf       {{\it cf.} }
\newcommand\cge      {{$_ >\atop{^\sim}$}}
\newcommand\cle      {{$_ <\atop{^\sim}$}}
\newcommand\degsq    {{$deg^{2}$}}
\newcommand\eg       {{\it e.g.}, }
\newcommand\emin     {{$e^{-}$} }
\newcommand\etal     {{et\thinspace al.}}
\newcommand\Fnu      {{$F_{\nu}$} }
\newcommand\Ho       {{$H_{0}$} }
\newcommand\kmsMpc   {{\ $km\ s^{-1}\ Mpc^{-1}$} }
\newcommand\lII      {{$l^{II}$} }
\newcommand{\HST}    {\emph{HST}}
\newcommand\ie       {{\it i.e.}, }
\newcommand\kms      {{\ $km\ s^{-1}$} }
\newcommand\Lya      {{Ly$\alpha$} }
\newcommand\Lstar    {{L$^{*}$} }
\newcommand\Mstar    {{M$^{*}$} }
\newcommand\mAB      {{$m_{AB}$} }
\newcommand\MV       {{$M_{V}$} }
\newcommand\magarc   {{\ mag\ arcsec$^{-2}$} }
\newcommand\mum      {{$\mu$m}}
\newcommand\Msun     {{\ $M_{\odot}$} }
\newcommand\re       {{$r_e$}}
\newcommand\UV       {{$UV$} }
\newcommand\Teff     {{T$_{eff}$}}
\newcommand\BVizYsJH {{BVizY$_{s}$JH}}
\newcommand\Ys       {{Y$_{s}$}}
\newcommand\YsJ      {{Y$_{s}$J}}
\newcommand\YsJH     {{Y$_{s}$JH}}
\newcommand\YJH      {{YJH}}

\title{The Hubble Space Telescope Wide Field Camera 3 Early Release Science 
data: Panchromatic Faint Object Counts for 0.2--2 microns wavelength 
\footnote{{Based on observations made with the NASA/ESA Hubble Space
Telescope, which is operated by the Association of Universities for
Research in Astronomy, Inc., under NASA contract NAS 5-26555. }} 
}


\author{
Rogier A. Windhorst~\altaffilmark{1}, 
Seth H. Cohen~\altaffilmark{1},
Nimish P. Hathi~\altaffilmark{2},
Patrick J. McCarthy~\altaffilmark{3},
Russell E. Ryan, Jr.~\altaffilmark{4},
Haojing Yan~\altaffilmark{5},
Ivan K. Baldry~\altaffilmark{6}, 
Simon P. Driver~\altaffilmark{7},
Jay A. Frogel~\altaffilmark{8}, 
David T. Hill~\altaffilmark{7}, 
Lee S. Kelvin~\altaffilmark{7}, 
Anton M. Koekemoer~\altaffilmark{9},
Matt Mechtley~\altaffilmark{1},
Robert W. O'Connell~\altaffilmark{10},
Aaron S. G. Robotham~\altaffilmark{7}, 
Michael J. Rutkowski~\altaffilmark{1},
Mark Seibert~\altaffilmark{3}, 
Amber N. Straughn~\altaffilmark{11}, 
Richard J. Tuffs~\altaffilmark{12}, 
Bruce Balick~\altaffilmark{13}, 
Howard E. Bond~\altaffilmark{9}, 
Howard Bushouse~\altaffilmark{9}, 
Daniela Calzetti~\altaffilmark{14}, 
R. Mark Crockett~\altaffilmark{15}, 
Michael J. Disney~\altaffilmark{16}, 
Michael A. Dopita~\altaffilmark{17}, 
Donald N. B. Hall~\altaffilmark{18}, 
Jon A. Holtzman~\altaffilmark{19}, 
Sugata Kaviraj~\altaffilmark{15}, 
Randy A. Kimble~\altaffilmark{11}, 
John W. MacKenty~\altaffilmark{9}, 
Max Mutchler~\altaffilmark{9}, 
Francesco Paresce\altaffilmark{20}, 
Abihit Saha~\altaffilmark{21}, 
Joseph I. Silk~\altaffilmark{14}, 
John T. Trauger~\altaffilmark{22}, 
Alistair R. Walker~\altaffilmark{23}, 
Bradley C. Whitmore~\altaffilmark{9},
\& Erick T. Young~\altaffilmark{24}
}

\slugcomment{Resubmitted to the Astrophysical Journal Supplement Series,
December, 2010}

\email{Rogier.Windhorst@asu.edu}


\altaffiltext{1}{School of Earth and Space Exploration, Arizona State University,
P.O. Box 871404, Tempe, AZ 85287-1404}

\altaffiltext{2}{Department of Physics \& Astronomy, University of California,
Riverside, CA 92521}

\altaffiltext{3}{Observatories of the Carnegie Institution of Washington,
Pasadena, CA 91101-1292}

\altaffiltext{4}{Department of Physics, University of California, One Shields
Avenue, Davis, CA 95616}

\altaffiltext{5}{Center for Cosmology and AstroParticle Physics, The Ohio State
University, Columbus, OH 43210}

\altaffiltext{6}{Astrophysics Research Institute, Liverpool John Moores
University, Birkenhead CH41 1LD, United Kingdom}

\altaffiltext{7}{School of Physics and Astronomy, University of St Andrews, Fife
KY16 9SS, UK}

\altaffiltext{8}{Association of Universities for Research in Astronomy,
Washington, DC 20005}

\altaffiltext{9}{Space Telescope Science Institute, Baltimore, MD 21218}

\altaffiltext{10}{Department of Astronomy, University of Virginia,
Charlottesville, VA 22904-4325}

\altaffiltext{11}{NASA--Goddard Space Flight Center, Greenbelt, MD 20771}

\altaffiltext{12}{Max Planck Institute for Nuclear Physics (MPIK),
Saupfercheckweg 1, D-69117 Heidelberg, Germany}

\altaffiltext{13}{Department of Astronomy, University of Washington, Seattle, WA
98195-1580}

\altaffiltext{14}{Department of Astronomy, University of Massachusetts, Amherst,
MA 01003}

\altaffiltext{15}{Department of Physics, University of Oxford, Oxford OX1 3PU,
United Kingdom}

\altaffiltext{16}{School of Physics and Astronomy, Cardiff University, Cardiff
CF24 3AA, United Kingdom}

\altaffiltext{17}{Research School of Astronomy \& Astrophysics, The Australian
National University, Weston Creek, ACT 2611, Australia}

\altaffiltext{18}{Institute for Astronomy, University of Hawaii, Honolulu, HI
96822}

\altaffiltext{19}{Department of Astronomy, New Mexico State University, Las
Cruces, NM 88003}

\altaffiltext{20}{INAF--IASF Bologna, Via Gobetti 101, 40129 Bologna, Italy}

\altaffiltext{21}{National Optical Astronomy Observatories, Tucson, AZ
85726-6732}

\altaffiltext{22}{NASA--Jet Propulsion Laboratory, Pasadena, CA 91109}

\altaffiltext{23}{Cerro Tololo Inter-American Observatory, La Serena, Chile}

\altaffiltext{24}{NASA--Ames Research Center, Moffett Field, CA 94035}

\ve 

\begin{abstract}

We describe the Hubble Space Telescope (HST) Wide Field Camera 3 (WFC3) Early
Release Science (ERS) observations in the Great Observatories Origins Deep Survey
(GOODS) South field. The new WFC3 ERS data provide calibrated, drizzled mosaics
in the UV filters F225W, F275W, and F336W, as well as in the near-IR filters
F098M (\Ys), F125W (J), and F160W (H) with 1--2 HST orbits per filter. Together
with the existing HST Advanced Camera for Surveys (ACS) GOODS-South mosaics in
the BViz filters, these panchromatic 10-band ERS data cover 40--50 square arcmin
at 0.2--1.7 \mum\ in wavelength at 0\arcspt 07--0\arcspt 15 FWHM resolution and
0\arcspt 090 Multidrizzled pixels to depths of AB$\simeq$26.0--27.0 mag
(5-$\sigma$) for point sources, and AB$\simeq$25.5--26.5 mag for compact
galaxies. 

In this paper, we describe: a) the scientific rationale, and the data taking plus
reduction procedures of the panchromatic 10-band ERS mosaics; b) the procedure of
generating object catalogs across the 10 different ERS filters, and the specific
star-galaxy separation techniques used; and c) the reliability and completeness
of the object catalogs from the WFC3 ERS mosaics. The excellent 0\arcspt
07--0\arcspt 15 FWHM resolution of HST/WFC3 and ACS makes star-galaxy separation
straightforward over a factor of 10 in wavelength to AB$\simeq$25--26 mag from
the UV to the near-IR, respectively. 

Our main results are: 1) Proper motion of faint ERS stars is detected over
6-years at 3.06$\pm$0.66 m.a.s./year (4.6-$\sigma$), consistent with Galactic
structure models; 2) Both the Galactic star counts and the galaxy counts show
mild but significant trends of decreasing count slopes from the mid--UV to the
near-IR over a factor of 10 in wavelength; 3) Combining the 10-band ERS counts
with the panchromatic Galaxy and Mass Assembly (GAMA) survey counts at the bright
end (10\cle AB\cle 20 mag) and the Hubble Ultra Deep Field (HUDF) counts in the
\BVizYsJH\ filters at the faint end (24\cle AB\cle 30 mag) yields galaxy counts
that are well measured over the entire flux range 10\cle AB\cle 30 mag for 0.2--2
\mum\ in wavelength; 4) Simple luminosity+density evolution models can fit the
galaxy counts over this entire flux range. However, no single model can explain
the counts over this entire flux range in all 10 filters {\it simultaneously}.
More sophisticated models of galaxy assembly are needed to reproduce the overall
constraints provided by the current panchromatic galaxy counts for 10\cle AB\cle
30 mag over a factor of 10 in wavelength. 

\end{abstract}


\section{Introduction}


\n The study of the formation and evolution of galaxies and large scale structure
are amongst the most active interfaces between theory and observation in modern
astrophysics. Galaxies are believed to have formed gradually over cosmic time
from a combination of gas infall and mergers (Hopkins \etal\ 2006), regulated by
feedback from stellar winds, supernovae, and/or AGN (\ie Scannapieco \etal\ 2005;
di Matteo, Springel, \& Hernquist 2005). The origin of the Hubble sequence is not
yet fully understood (\eg Driver \etal\ 1998), but is likely related to the
balance between major mergers versus minor accretion events and steady infall
(\ie Conselice \etal\ 2003; Hopkins \etal\ 2010; Peng \etal\ 2010). The critical
epoch for the assembly of massive galaxies appears to be the $\sim$4 Gyr span
from redshift z$\simeq$3 to z$\simeq$1, where also the cosmic star-formation
history seems to have peaked (Madau, Pozzetti, \& Dickinson 1998; Hopkins 2004;
Hopkins \& Beacom 2006).

At redshifts \cge 2--3, deep Hubble Space Telescope (HST) imaging surveys and
ground-based spectroscopy have revealed a paucity of massive galaxies and few
classical disks or spheroids (\eg Law \etal\ 2007, 2009). In contrast, large
ground-based spectroscopic surveys targeting redshifts z\cle 1 --- coupled with
HST imaging --- have shown that by this epoch massive galaxies are largely
mature, and that the Hubble sequence has been mostly established (\eg Abraham
\etal\ 1996, 1999, 2007; Driver \etal\ 1995, 1998; Glazebrook \etal\ 1995; Lilly
\etal\ 1998). While substantial growth ($\sim$50\%) in the stellar mass of all
galaxy types --- including spheroids --- may have occurred during the last
$\sim$7 Gyr, the process of major galaxy assembly was well underway by z$\simeq$1
(\eg de Lucia \etal\ 2006; Dickinson \etal\ 2003). The interim period --- from
redshifts of z$\simeq$3 to z$\simeq$1 --- is the era in which much of the stellar
mass in galaxies is accumulated (\eg Dickinson \etal\ 2003; Abraham \etal\ 2007),
and when galaxies acquire the characteristic structural and dynamical properties
that define them today. The HST Wide Field Camera 3 (WFC3) was optimized to study
this critical period of galaxy assembly.


WFC3 was successfully installed into HST on May 14, 2009, by the astronauts
on-board Space Shuttle Atlantis during the Space Transportation System mission
125 (STS-125). This shuttle mission was the fifth Servicing Mission of HST,
however, for historical reasons, it is referred to as SM4. Many of the current
co-authors were members of the WFC3 Science Oversight Committee (SOC) from July
1998 through November 2009. Our main role as the SOC was to define the WFC3
science requirements and goals, monitor them during the pre-launch phases of the
project, and to oversee the design, implementation, integration, and testing
(both ground-based and on-orbit) of the WFC3 instrument.

The WFC3 provides a unique opportunity to compare the galaxy populations in the
local and distant universe. With its wide spectral coverage (0.2--1.7 \mum), very
high spatial resolution (0\arcspt 04 FWHM at 0.2\mum\ to 0\arcspt 16 FWHM at
1.6\mum), fine pixel sampling (0\arcspt 039/pixel in the UVIS channel and
0\arcspt 13/pixel in the IR channel), and high sensitivity (AB$\simeq$26--27 mag
in 2 orbits; 5-$\sigma$ for point sources), many new interesting questions and
outstanding problems can be addressed with the WFC3 data. By sampling the vacuum
UV with high sensitivity and the very high angular resolution afforded by the
diffraction limited 2.4 m Hubble Space Telescope, WFC3 can observe star-forming
regions in galaxies over most of the Hubble time. The near-IR channel on WFC3
allows one to do restframe visible-band photometry of distant galaxies to low
luminosities and over areas large enough to provide representative samples.
Together, the panchromatic images produced by WFC3 allow the user to decompose
distant galaxies into their constituent substructures, examine their internal
stellar populations, and help constrain their dust content. In this Early Release
Science (ERS) program, the UVIS and IR channels of WFC3 are used to provide a
small, but representative sampling of the capabilities of WFC3 to examine the
formation and evolution of galaxies in the critical galaxy assembly epoch of
z$\simeq$1--3, when the universe was only 6--2 Gyrs old, respectively. 

Details of the HST WFC3 ERS program (PID \#11359; PI R. O'Connell) can be found
on this $URL$\footnote{http://www.stsci.edu/cgi-bin/get-proposal-info}. The
current ERS program was specifically conceived to make maximum use of these WFC3
capabilities, and to make an optimal comparison between the intermediate and high
redshift galaxy samples identified in the current ERS program and nearby galaxies
imaged in other HST programs. These capabilities were important WFC3 science
drivers, while the instrument was designed and constructed from 1998 to 2008. 

In the year that the intermediate redshift WFC3 ERS data has been available, a
number of papers have appeared or submitted that use this panchromatic data set.
For instance, Ryan \etal\ (2010) discuss the evolution of passive galaxies using
the WFC3 ERS observations, and make a detailed study of their size evolution over
cosmic time. Rutkowski \etal\ (2011) present a panchromatic catalog of early-type
galaxies at intermediate redshifts (z$\simeq$0.3--1.5) from the WFC3 ERS data,
and derive their rest-frame (FUV--V) and (NUV--V) colors as a function of
redshift. Cohen \etal\ (2011) present a ten-band photometric study of distant
galaxies in the WFC3 ERS data, measure reliable photometric redshifts, and derive
their physical properties with cosmic time. Hathi \etal\ (2010) discuss
UV-dropout galaxies in the GOODS-South at redshifts z$\simeq$1.5--3 from the WFC3
ERS data, and summarize the evolution of the faint-end luminosity function (LF)
slope $\alpha$ and characteristic luminosity \Lstar from z$\simeq$8 to
z$\simeq$0. Oesch \etal\ (2010b) similarly discuss the evolution of the
ultraviolet luminosity function from z$\sim$0.75 to z$\sim$ 2.5. Straughn \etal\
(2011) study faint emission-line galaxies from the WFC3 ERS IR grism
observations. Van Dokkum \& Brammer (2010) discuss WFC3 grism spectra and images
of one growing compact galaxy at z$\simeq$1.9. Finkelstein \etal\ (2011) discuss
spatially resolved imaging of \Lya emission line objects at z$\simeq$4.4 through
parallel ACS F658N narrow-band images to the WFC3 ERS data, and their
constraints as to how Lyman continuum photons escape from such objects.

Labb\'e \etal\ (2010) discuss the star formation rates and stellar masses of
z$\simeq$7--8 galaxies from IRAC observations of the WFC3/IR ERS and Hubble Ultra
Deep Field (HUDF) fields. Robertson (2010) estimates how to best improve the LF
constraints from high redshift galaxy surveys using WFC3 ERS data and from
additional deep WFC3 survey data yet to be obtained. Bouwens \etal\ (2010a)
discuss potentially very blue UV-continuum slopes of low luminosity galaxies at
z$\simeq$7 from the WFC3 ERS IR data, and their possible implications for very
low metallicities in these objects. Bouwens \etal\ (2010b) also discuss
z$\simeq$8 galaxy candidates seen in the ultradeep WFC3/IR observations of the
HUDF. Yan \etal\ (2011) probe the bright-end of the galaxy LF at z\cge 7 using
HST pure parallel observations, and discuss these in context of the WFC IR
observations in the ERS and in the HUDF. Yan \etal\ (2010) discuss galaxy
formation in the reionization epoch from the WFC3 observations of the HUDF, and
suggest a LF at z$\simeq$8--10 that differs from that found by Bouwens \etal\
(2010) and Oesch \etal\ (2010a). Wyithe \etal\ (2011) explain that part of this
discrepancy may arise due to distortion of the very high redshift galaxy number
counts through gravitational lensing by random foreground galaxies at
z$\simeq$1--2. This boosts the number of z\cge 8--10 objects that become
observable in the WFC3 IR samples, a non-negligible fraction of which therefore
must be sought close to foreground galaxies at z$\simeq$1--2. These many examples
show the great potential of the WFC3 ERS data presented in the current paper, and
we refer the reader to these other papers for in-depth studies of the ERS data
that are beyond the scope of the current paper. It suffices to say that the
quality of and the scientific results from the ERS data exceeded the expectations
of the WFC3 SOC, even though we have been involved with the WFC3 instrument from
its conception in 1998. 

In \S 2 of this paper, we present the WFC3 ERS survey strategy, the filters used
and their achieved depths. In \S 3, we present the observations in both the WFC3
UVIS and IR channels, and the pointings and their areal coverage. In \S 4, we
present the WFC3 data reduction procedures, their reliability and completeness,
and their current limitations. In \S 5, we present the object finding procedures 
and catalog generation, and the star-galaxy separation procedure used and its
reliability in the 10 ERS filters. In \S 6, we present the panchromatic ERS star
counts and discuss the faint ERS stellar proper motion results. In \S 7, we
present the panchromatic ERS galaxy counts from 0.2--1.7 \mum\ to AB$\simeq$26-27
mag, and compare these to the 10-band ground-based Galaxy and Mass Assembly
(GAMA) survey counts for 10\cle AB\cle 20 mag at the bright end, and to the HUDF
counts in the \BVizYsJH\ filters (defined in section 2.1) for 24\cle AB\cle 30
mag at the faint end. We also present the panchromatic ERS images for interesting
individual objects. In \S 8, we summarize our main results and conclusions.
Throughout this paper, we use WMAP-year7 cosmology (Komatsu \etal\ 2010), or
\Ho=71\kmsMpc, $\Omega_o$=0.26, and $\Lambda$=0.74, and the $AB_{\nu}$ magnitude
system (Oke 1974). 

\section{WFC3 and its Capabilities} 

\subsection{The ERS Filter Set} 

\mn \mi In the current ERS program, the unique panchromatic capabilities of WFC3
are used to survey the structure and evolution of galaxies at the peak of the
galaxy assembly epoch at z$\simeq$1--3. Deep ultraviolet and near-IR imaging, and
slitless near-IR spectroscopy of existing deep multi-color GOODS-S/ACS fields are
used to gauge star-formation and the growth of stellar mass as a function of
galaxy morphology, structure and surrounding density in this critical cosmic
epoch at redshifts 1\cle z\cle 3. 

The total HST filter set provided by the WFC3 ERS imaging of the Great
Observatories Origin Deep Survey (GOODS) South field is shown in Fig. 1a, and its
properties are summarized in Tables 1 and 2. WFC3 adds the F225W, F275W and F336W
filters in the WFC3 UVIS channel, and the F098M, F125W and F160W filters
(hereafter \YsJH) in the WFC3 IR channel. Together with the existing GOODS ACS
F435W, F606W, F775W and F850LP images (Giavalisco \etal\ 2004), the new WFC3 UVIS
and IR filters provide a total of 10 HST filters that span the wavelength range
$\lambda$$\simeq$0.2--1.7 \mum\ nearly contiguously. We refer to this entire
10-band survey hereafter as the `'ERS'', to these 10 filters as the ``ERS
filters'', and to the 7 reddest ERS filters as the ``\BVizYsJH'' filters
throughout. Details of the GOODS survey can be found in Giavalisco \etal\ (2004)
and references therein. The top panel of Fig. 1a compares the ERS filters to the
spectral energy distribution of two single burst model galaxies (middle and
bottom panels of Fig. 1a) with ages of 0.1 and 1 Gyr at redshifts of z=0, 2, 4,
6, 8, respectively. 

The ERS images in the WFC3 UVIS filters F225W, F275W, and F336W are used to
identify galaxies at redshifts z\cge 1.5 from their UV continuum breaks, which
between the F225W and F275W filters is sampled at redshifts as low as 
z$\simeq$1.5--1.7 (see Fig. 1). These filters provide star-formation
indicators tied directly to both local and z\cge 3 galaxy populations, which
are the ones best observed through their Lyman breaks from the ground at
$\lambda$\cge 350 nm. The critical new data that the WFC3 UVIS channel can
provide are thus very high resolution, deep images for 0.2\cle $\lambda$\cle
0.36 \mum, as illustrated in Fig. 1a. 

The ERS images in the WFC3 near-IR filters F098M, F125W and F160W are used to
probe the Balmer and 4000 \AA\ breaks and stellar mass function well below 10$^9$
\Msun for mass-complete samples in the critical redshift range of z$\simeq$1--3.
The unique new data that the WFC3 IR channel can provide are high resolution,
very sensitive near-IR photometry over fields larger than those possible with HST
NICMOS, or over wide fields with adaptive optics from the ground (\eg Steinbring
\etal\ 2004; Melbourne \etal\ 2005). 

\subsection{ERS Grisms}

In addition to these broad-band ERS filters, we used the WFC3 near-IR grisms G102
and G141 to obtain slitless spectroscopy of hundreds of faint galaxies at a
spectral resolution of R$\simeq$210--130, respectively. The WFC3 near-IR grism
data can trace the primary indicators of star-formation --- the Lyman-$\alpha$
and H-$\alpha$ emission-lines --- in principle over the redshift range for
z$\simeq$5--13 and z$\simeq$0.2--1.7, respectively. WFC3 can also trace the Lyman
break and the rest-frame UV continuum slope, as well as the Balmer and 4000
\AA-breaks over the redshift range z$\simeq$1--9 and z$\simeq$0--2.5,
respectively. The ERS grism program thus at least covers the peak of the cosmic
star formation history at redshifts 1\cle z\cle 2, using some of the most
important star-formation and post-starburst indicators, while also providing some
metallicity-independent reddening indicators. Both IR grism dispersers provide
capabilities that cannot be reproduced from the ground: slitless spectroscopy of
very faint objects (AB$\simeq$25--26 mag) over a contiguous wide spectral range
in the near-IR, that is not affected by atmospheric night-sky lines. The ERS
grism observations are 2 orbits in depth each, covering a single WFC3 field, which
was also covered in a previous ACS G800L grism survey (Straughn \etal\ 2009). An
example of the ERS G141 and G102 grism spectra is shown in the figures of
Appendix B.2. Further details of the ERS grism data reduction and the analysis of
the faint emission line galaxies are given in Appendix B.2 and by Straughn \etal\
(2011). 

WFC3 UVIS G280 UV-prism observations were not made as part of the 104 orbit 
intermediate redshift ERS program, because of its much lower throughput and the 
significant overlap of its many spectral orders (Bond \& Kim Quijano 2007; Wong 
\etal\ 2010). Currently, one Cycle 17 GO program (11594; PI J. O'Meara) is using 
the WFC3 G280 prism to carry out a spectroscopic survey of Lyman limit absorbers 
at redshifts 1.8\cle z\cle 2.5. Readers interested in the WFC3 G280 prism 
performance should follow the results from that program. 

\subsection{ERS UVIS Filter Red-Leaks and IR Filter Blue-Leaks}

In the context of the WFC UVIS and IR channel performance for intermediate to
high redshift early- and late-type galaxies, it is useful to briefly summarize
here the possible effects of UVIS channel filter red-leaks and IR channel filter
blue-leaks on the measured fluxes of these objects. UVIS red-leaks are defined as
the fraction of flux longwards of 400 nm of an SED of given effective temperature
\Teff\ that makes it erroneously into the UV filter. The IR blue-leaks are
defined as the fraction of flux short-wards of 830 nm of an SED of given
effective temperature \Teff\ that makes it erroneously into the IR filter or
grism (for details, see Wong \etal\ 2010). 

The WFC3 UVIS filters were designed with great attention to minimize their
red-leaks, which were much larger in the earlier generation WFPC2 UV filters.
Similarly, the WFC3 IR filters and grisms were designed to minimize the
blue-leaks. For both sets of WFC3 filters, lower out-of-band transmission
usually goes at the expense of lower in-band transmission, and vice versa.
Hence, both the WFC3 and IR filters were designed and fabricated such that the
in-band transmission was optimized as much as possible, while keeping the
out-of-band transmission to acceptable or correctable levels for all SEDs
expected in the astrophysical relevant situations. 

The resulting WFC3 UVIS red-leaks are acceptably small (\cle 10\%) for all
zero-redshift SEDs with \Teff\cge 5000 K for the F225W filter, \Teff\cge 4000 K
for the F275W filter, and \Teff\cge 2000 K for the F336W filter, respectively
(Wong \etal\ 2010). For cooler (\Teff\cle 2000--5000 K) zero-redshift SEDs, some
red-leak correction thus has to be applied to the observed F336W, F275, and F225W
fluxes, respectively. However, for objects at substantial redshifts (z\cge
0.75--1), the SED will shift out of the UVIS sensitivity regime quickly enough to
significantly reduce the red-leak. Hence, the WFC3 UVIS red-leaks in general only
need to be corrected for the reddest (\Teff\cle 5000 K), lower-redshift (z\cle
0.75) SEDs observed in the bluest UVIS filters (F225W). Further details are
given in Rutkowski \etal\ (2011).

The WFC3 IR blue-leaks are very small (\cle 0.01\%) for all zero-redshift SEDs
with \Teff\cle 10,000 K for the F098M, F105W, F125W, and F160W filters, and
remains very small (\cle 0.1\%) even for the bluest zero-redshift SEDs with
\Teff$\simeq$30,000-50,000 K (Wong \etal\ 2010). For higher redshift SEDs (z\cge
1) of any \Teff, the redshift further reduces the IR blue-leak. Similarly, the
G141 grism was made on a glass substrate with no transmission below 750nm, and is
well blocked by its coatings shortward of 1050 nm and long-ward of 1700 nm
(Baggett \etal\ 2007). Hence, it also has acceptably an small blue-leak. The same
is true for the higher resolution G102 grism.

\subsection{WFC3 Detectors and Achieved ERS Sensitivities}

The WFC3 UV---blue optimized CCDs were chosen specifically to complement those of
ACS. They were made by E2V in the UK, and are thinned, backside illuminated, CCD
detectors with 2k$\times$4k 15 \mum\ (0\arcspt 0395) pixels, covering the
wavelength range 200--1000 nm with Quantum Efficiency QE\cge 50\% throughout
(Wong \etal\ 2010; Kimble \etal\ 2010). The total WFC3 UVIS field-of-view with
these two CCDs is 162''$\times$162''.

The WFC3 near-IR detectors were Teledyne HgCdTe infrared detectors (MBE-grown and
substrate removed) with Si CMOS Hawaii-1R multiplexers and have 1k$\times$1k 18
\mum\ (0\arcspt 130) pixels, covering the wavelength range 800--1730 nm with
QE\cge 77\% throughout (Wong \etal\ 2010; Kimble \etal\ 2010). The total WFC3 IR
field-of-view is 123''$\times$136''. Further specifications of the WFC3 detectors
are listed where relevant below.

Table 1 summarizes the resulting WFC3 sensitivities from our relatively short ERS
exposures. The table lists the number of orbits per filter and the 5-$\sigma$
depths in AB magnitudes and F$_{\nu}$ units. A net exposure time of 2600--2700
seconds was available in each HST orbit for on-source ERS observations. In Fig.
1b, the equivalent depths are plotted in physical terms, by comparing with
spectral synthesis models of Bruzual \& Charlot (2003) at the three fiducial
redshifts, following Ryan \etal\ (2007, 2010). Simple stellar populations models
with single bursts or exponentially declining star-formation rates with an
e-folding time of 1 Gyr are plotted. Fig. 1b shows the predicted spectral energy
distributions for models with ages ranging from 10 Myr to 3 Gyr, along with the
5-$\sigma$ depths of the WFC3 ERS program. The three panels represent redshifts
z=1.0, 1.5 and 2.0, and models with stellar masses of M=10$^9,$ 4$\times$10$^9$,
and 10$^{10}$ \Msun, respectively. These SED tracks illustrate the intended SED
and mass sensitivity of the WFC3 ERS observations as a function of cosmic epoch.
Galaxies with ongoing star-formation, even with fairly large ages, are easily
detected in the WFC3 UV observations. At z$\simeq$2, a maximally old
$\tau$$\simeq$1 model with a mass of $\sim$0.3 M$^*$ is detectable above the WFC3
detection threshold in the F336W filter, while at z$\simeq$1, the WFC3 ERS can
detect young star forming galaxies with masses as low as a few$\times$10$^7$
\Msun, or about a M$\sim$0.01 M$^*$ galaxy in that filter.

Table 2 summarizes the HST instrument modes and the ERS filters used, the filter
central wavelength $\lambda$, its width, and the PSF-FWHM as a function of
wavelength, the AB magnitude zero-points for all 10 filters for a count rate of
1.0\emin/sec, as well as the zodiacal sky-background measured in each ERS filter.
The GOODS sky-background values in the F435W, F606W, F775W, and F850LP filters
(hereafter BViz) are from Hathi \etal\ (2008).

Fig. 2 shows that on average, the on-orbit WFC3 UVIS sensitivity is 6--18\%
higher than the predicted pre-launch sensitivity from the ground-based thermal
vacuum test (left panel), and the on-orbit WFC3 IR sensitivity is 9--18\% higher
(right panel). The red lines are best fits to the in-flight/pre-launch
sensitivity ratio. For the UVIS data, this is just a parabolic fit, as the
in-flight/pre-launch excess does not seem to follow the CCD sensitivity curve
(Kalirai \etal\ 2009a). For the IR data, a polynomial fit was folded with the IR
detector sensitivity curve, since the in-flight/pre-launch does somewhat resemble
the IR detector sensitivity curve (Kalirai \etal\ 2009b). The true cause of this
beneficial, but significant discrepancy is unknown. It possibly results from
uncertainties in the absolute calibration procedure of the optical stimulus used
in the thermal vacuum tests of WFC3 (Kimble \etal\ 2010), and/or perhaps from
slow temporal changes in the HST Optical Telescope Assembly (OTA) itself (Kalirai
\etal\ 2009a, 2009b). The cause of this discrepancy is currently being
investigated, and lessons learned will be applied to the upcoming ground-based
calibrations of the James Webb Space Telescope thermal vacuum absolute throughput
measurements.

\section{The ERS Data Collection Strategy}


The GOODS-South field was chosen for the ERS pointings, because of the large body
of existing and publicly available data. Besides the deep, four-color ACS BViz
imaging (Giavalisco \etal\ 2004; Dickinson \etal\ 2004), there are low resolution
(R$\sim$100) ACS slitless G800L grism spectra covering the wavelength range
$\sim$0.55--0.95 \mum\ (\cf Pirzkal \etal\ 2004; Malhotra \etal\ 2005; Ferreras
\etal\ 2009; Rhoads \etal\ 2009; Straughn \etal\ 2008, 2009). There is also a
wealth of ground- and space-based data, such as deep U+R-band VLT/VIMOS imaging
(Nonino \etal\ 2009), deep VLT/ISAAC $JHK_{s}$-band imaging (Retzlaff \etal\
2010), a very large number of VLT spectra (Vanzella \etal\ 2005, 2009; Popesso
\etal\ 2009; Balestra \etal\ 2010), deep Chandra X-ray images (Giacconi \etal\
2002; 2 Msec by Luo \etal\ 2010; 4Msec by Luo \etal\ 2011), deep XMM X-ray
observations (4 Msec by Comastri \etal\ 2011), GALEX UV data (Burgarella \etal\
2006), Spitzer photometry with IRAC and MIPS (Papovich \etal\ 2006; Yan \etal\
2004, 2005), Herschel FIR images at 70, 110, and 160 \mum\ (Gruppioni \etal\
2010; Lutz \etal\ 2011), and deep ATCA and VLA radio images (\cf Afonso \etal\
2006, Kellermann \etal\ 2008), respectively. 


Given the constraint on the total amount of time available in the allotted 104
HST orbits, the ERS program could survey {\it one} 4$\times$2 WFC3 mosaic
covering 10\arcm $\times$5\arcm or roughly 50 arcmin$^2$ to 5-$\sigma$ depths of
\mAB $\simeq$ 26.0 mag in the three bluest wide-band UVIS filters, and {\it one}
5$\times$2 WFC3 mosaic covering 10\arcm $\times$4\arcm or roughly 40 arcmin$^2$
to 5-$\sigma$ depths of \mAB $\simeq$ 27.0 mag in the three near-IR filters. This
angular coverage probes co-moving scales of roughly 5--10 Mpc and provides a
sample of 2000--7000 galaxies to AB$\simeq$26--27 mag in the panchromatic ERS
images. The IR images were dithered to maximally match the UVIS field-of-view. 

Fig. 3 shows the ACS z'-band (F850LP) mosaic of the entire GOODS-South field, and
the outline of the acquired WFC3 pointings, as well as the locations of the ACS
images taken in parallel to the WFC3 ERS pointings. The ACS parallels were taken
with the ACS/WFC filters F814W and F658N to search for high redshift
Lyman-$\alpha$ emitters at z=4.415$\pm$0.03, of which several were known
spectroscopically in the GOODS-South field (Vanzella \etal\ 2005; Finkelstein
\etal\ 2011). The WFC3 ERS mosaic pointings cover the Northern $\sim$30\% of the
GOODS-South field (Fig. 3). The 8 ERS pointings are contiguous with a tiling that
can be easily extended to the South in future WFC3 GO programs (see, \eg the
Faber, Ferguson \etal\ Multi Cycle Treasury HST programs 12060--12064).

The orbital F225W and F275W ERS observations were designed to minimize possible
Earth limb contamination. To guarantee the lowest possible UV sky-background in
the WFC3 images, one 1200 sec F275W and one 1200 sec F225W exposure was obtained
in each orbit. All F225W exposures were taken at the end of each orbit, in
contrast with the common practice of observing all exposures in the same filter
in rapid succession in subsequent orbits. All three 800 sec F336W exposures were
obtained during the same single orbit for a given ERS pointing. This manner of
scheduling indeed minimized the on-orbit UV sky-background away from the Earth's
limb (see Table 2), but it somewhat complicated the sub-sequent MultiDrizzle
procedure (see \S 4.2), since no ``same-orbit'' cosmic ray rejection could be
applied to the F225W and F275W images in order to find a first slate of bright
objects for image alignment. Further details on the image alignment are given in
\S 4 and Appendix A. 

In the WFC3 IR channel, 6 exposures of 800-900 sec were taken in each of the
F098M, F125W, and F160W filters, using 2 orbits for each filter, but staying away
from the Earth's limb at the end of each orbit in order to keep the near-IR
sky-background as low as possible (see Table 2). In total, 9 or 10 Fowler samples
in each IR channel integration provided good CR-rejection, while the 6 dithered
exposures providing the capability to properly drizzle the IR images, and so more
properly sample the IR PSF (see \S 4.3.3). 

HST scheduling required that this ERS program be split into several visits.
Simple raster patterns were used to fill out the WFC3 IR mosaic, and to improve
sky-background plus residual dark-current removal. Only mild constraints were
applied to the original HST roll angle (ORIENT) to maximize overlap between the
northern part of the GOODS-South field, and to allow HST scheduling in the
permitted observing interval for the ERS (mid-Sept.--mid-Oct. 2009). The IR and
UVIS images were constrained to have the {\it same} ORIENT, to ensure that a
uniform WFC3 mosaic could be produced at all wavelengths. The slight misalignment
of the Northern edge of the {\it existing} ACS GOODS-South mosaics and the {\it
new} WFC3 ERS mosaics in Fig. 3 was due to the fact that the HST ORIENT
constraints had to be slightly changed in the late summer of 2009, since the ERS
observations needed be postponed by several weeks due to a change in the HST
scheduling constraints. Since finding good guide stars for all 19 ERS pointing
was very hard, it was necessary to only slightly change the mosaic ORIENT angles
at that point, but not the actual image pointing coordinates. The Early Release
Science mosaics would have otherwise become unschedulable. 

The area of overlap between the individual WFC ERS mosaic pointings is too small
to identify transient objects (e.g. SNe and variable AGN), since only about
1--2\% of all faint field objects show point-source variability (Cohen \etal\
2006). However, it {\it is} useful in the subsequent analysis to verify the
positions of objects, and so verify the instrument geometric distortion
corrections (GDCs) used, as discussed in \S 4 and Appendix A. 

\section{WFC3 UVIS and IR Data Processing}

The WFC3 ERS data processing was carried out with the STScI pipeline {\it
calwf3}. The WFC3 ERS data set also provided tests of the STSDAS pipeline under
realistic conditions. This process was started well before the SM4 launch in the
summer of 2008 with pre-flight WFC3 thermal vacuum calibration data, and
continued through the late summer and fall of 2009, when the first ERS data
arrived. The raw WFC3 data was made public immediately, and the ``On-The-Fly''
(``OTF'') pipe-lined calibrated and Multi-Drizzled WFC3 mosaics will be made
public via MAST at STScI when the final flight calibrations --- as detailed below
--- have been applied. The specific pipeline corrections that were applied to the
WFC3 ERS images are detailed here. Unless otherwise noted below, the latest
reference files from the WFC3 Calibration web-page were used in all cases, and
are available on this $URL$. \footnote{www.stsci.edu/hst/observatory/\\
cdbs/SIfileInfo/WFC3/reftablequeryindex}

\subsection{The Main WFC3 Pipeline Corrections} 

\subsubsection{WFC3 UVIS}

All raw WFC3 UVIS data were run through the standard {\it STSDAS calwf3}
calibration program as follows: 

\sn (1) The best WFC3 UVIS super-bias that was available at the time of
processing\\ (090611120\_bia.fits) was subtracted from all images. This is an
on-orbit super-bias created from 120 UVIS bias frames, each of which was
unbinned, and used all four on-chip amplifiers (see this $URL$\footnote{
www.stsci.edu/hst/wfc3/lbn\_archive/\\ 2009\_09\_09\_new\_uvis\_superbias}). The
measured on-orbit UVIS read-noise levels are 3.1\emin in Amp A, 3.2 \emin in Amp
B, 3.1 \emin in Amp C, and 3.2 \emin in Amp D, respectively, or on average about
3.15 \emin per pixel across the entire UVIS CCD array.

\sn (2) A null dark frame was applied, since the only available darks at the time
were from the ground-based thermal vacuum testing in 2007--2008, and these were
not found to reduce the noise in the output ERS frames. Hence, the subtraction of
actual on-orbit 2D dark-frames was omitted until better, high signal-to-noise
ratio on-orbit dark-frames have been accumulated in Cycle 17 and beyond. Instead,
a {\it constant} dark-level of 1.5 \emin/pix/hr was subtracted from all the
images, as measured from the {\it average} dark-current level in the few on-orbit
dark-frames available thus far. This dark-level is about 5$\times$ higher than
the ground-based thermal vacuum tests had suggested, but still quite low enough
to not add significant image noise in an average 1200 sec UVIS exposure. 

\sn (3) A bad pixel file (tb41828mi\_bpx.fits) was created (by H. Bushouse) and
updated over the one available in the ``Office of Space Sciences Payload Data
Processing Unified System'' or ``OPUS'' pipeline at the time, and applied to all
the UVIS images.

\sn (4) All flat-fields came from the 2007--2008 WFC3 thermal vacuum ground tests
and had high signal-to-noise ratio. We used these flats, since the WFC3 data base
is not yet large enough to make a reliable set of on-board sky-superflats. (As in
the case for the WFPC2 Medium-Deep Survey, this can and will be done during
subsequent years of WFC3 usage). These thermal vacuum flats left some large-scale
gradients in the flat-fielded data, due to the illumination difference between
the thermal vacuum optical stimulus and the real on-orbit WFC3 illumination by
the zodiacal sky-background. For each passband, the mean UV-sky-background was
removed from the individual ERS images (as part of MultiDrizzle), and the
resulting images were combined into a median image in each UV filter. The large
scale gradients from this illumination difference correspond to a level of about
$\sim$5--10\% of the on-orbit zodiacal sky-background and have very low spatial
frequency. This situation will be remedied with on-orbit internal flats and
sky-flats, that will be accumulated during Cycle 17 and beyond. Since the UV
sky-background is very low to begin with ($\sim$25.5 \magarc, see Table 2 and
Windhorst \etal\ 2002), these residual 1--2\% sky-gradients affect the object
photometry only at the level of AB\cge 27--28 mag, \ie\ well below the UVIS
catalog completeness limits discussed in \S 5.4. Also, the spatial scales of
these gradients are much larger than $\sim$100 pixels, and faint objects are
small (see \S 5.5 and figures therein; see also Windhorst \etal\ 2008), so that
these gradients do not affect the faint object finding procedure, catalog
reliability and completeness significantly (see \S 5.1--5.4). 

We suspect, but have at this stage not been able to prove with the currently
available data, that this remaining low-level sky gradient is of {\it
multiplicative} and not of {\it additive} nature. Once we have been able to
demonstrate this with a full suite of sky-flats, we will re-process all the UVIS
data again, and remove these low-level gradients accordingly. For now, these
gradients are not visible in the high quality, high contrast color reproductions
of Fig. 5a--5b. Hence, they do not significantly affect the subsequent
object-finding and their surrounding sky-subtraction procedures, which assume
linear remaining sky-gradients. This is corroborated by the quality of the
panchromatic object counts discussed below, and consistency with the counts from
other authors in the flux range where these surveys overlap. In other words, any
remaining low-level sky gradients do not significantly affect the UVIS object
catalogs generated for the current science purposes to AB$\simeq$25.5--26.0 mag. 

We also checked for CCD window ghosts or filter ghosts next to the brightest
stars. These are in general very faint, or of very low surface brightness (SB)
and much larger than the galaxies we are studying here. Such window ghosts do not
affect the WFC IR images. In the WFC 3UVIS images, they only amount to 0.4\% of
the stellar peak flux in the F225W filter, and are much dimmer in the redder UVIS
filters (Wong \etal\ 2010). No obvious filter ghost-like objects were found by
the SExtractor object finder (Bertin \& Arnouts 1996) surrounding the bright
stars in the ERS.

\subsubsection{WFC3 IR}

The reduction of the ERS WFC3 IR data largely followed the procedures as
described in Yan \etal\ (2010). We used the {\it calwf3} task included in the
STSDAS package to process the raw WFC3 IR images, using the latest reference
files indicated by the relevant FITS header keywords. Additional corrections to
the calibrated images were applied as follows.

\sn (1) We removed residual DC offsets between the four detector quadrants, which
was caused by an error in the application of the quadrant-dependent gain values
in {\it calwf3} and documented in the WFC3 STAN (September 2009 issue, see this
$URL$ \footnote{www.stsci.edu/hst/wfc3/documents/\\
newsletters/STAN\_09\_01\_2009}; see also Wong \etal\ 2010). Specifically,
multiplicative gain correction factors were applied to each image quadrant using
g=1.004 for Quadrant 1 (upper left), g=0.992 for Quadrant 2 (lower left), g=1.017
for Quadrant 3 (lower right), and g=0.987 for Quadrant 4 (upper right quadrant).
Note that this quadrant issue was fixed in {\it calwf3 v1.8} and later. 

\sn (2) For each passband, the mean near-IR sky-background was removed from the
individual ERS images, and the resulting images were combined into a median image
in each near-IR filter. 

\sn (3) A smooth background gradient still persisted in the median image. This
gradient was fitted by a 5-order Spline function, and was then subtracted from
the individual near-IR images. This sky gradient is of order 1--2\% of the
zodiacal sky-background. Since the near-IR zodiacal sky-background is about
22.61, 22.53, and 22.30\magarc in \YsJH\ (see Table 2), respectively, these
remaining WFC3 IR gradients do not affect the large-scale object finding and
catalog generation at levels brighter than AB$\simeq$27.0 mag. 

We checked for persistence in the IR images left over from saturated objects in
previous exposures. Since the WFC3 IR observations just before the WFC3 ERS IR
observations didn't contain many highly saturated bright stars, very few obvious
persistence problems were found. Since the ERS filters were taken in the order
F125, F160W, F098M, persistence would have been most obvious in the highest
throughput F125W filter, leading possibly to objects with unusually high J-band
fluxes compared to H- and \Ys-band. Only very few such objects were found, and
where persistence was suspected, they were removed from the SExtractor catalogs.

\subsection{WFC3 Astrometry and MultiDrizzle Procedures}

\subsubsection{WFC3 UVIS Astrometry} 

The calibrated, flat-fielded WFC3 UVIS exposures were aligned to achieve
astrometric registration with the existing GOODS ACS reference frame (Giavalisco
\etal\ 2004; GOODS v2.0 $URL$\footnote{http://archive.stsci.edu/pub/hlsp/\\
goods/v2/h\_goods\_v2.0\_rdm.html}. from Grogin \etal\ 2009). To generate
accurate SExtractor (Bertin \& Arnouts 1996) object catalogs in the F225W, F275W
and F336W filters, the higher S/N-ratio GOODS ACS B-band images were used as the
detection image. This also provided an astrometric reference frame that was
matched as closely as possible to the wavelengths of the UVIS filters used in
these observations. Because the GOODS B-band images reach AB\cle 27.9 mag and so
go much deeper than the ERS UVIS images, they help optimally locate the objects
in the ERS UVIS mosaics (see \S 5.1). 

The F225W and F275W ERS exposures taken separately in successive orbits (see \S
3) needed to be aligned with each other individually, in addition to their
overall alignment onto the GOODS reference frame. For each ERS filter, the
relative alignment between exposures was achieved iteratively, starting with an
initial partial run of MultiDrizzle (Koekemoer \etal\ 2002) to place each
exposure onto a rectified pixel grid. These images were then cross-correlated
with each other, after median filtering each UVIS exposure and subtracting this
smooth exposure to reduce the impact of cosmic rays and to identify the brighter
real objects. This ensured a {\it relative} alignment between the sequential
orbital ERS exposures to the sub-pixel level, correcting for offsets that were
introduced by the guide-star acquisitions and re-acquisitions at the start of
each successive ERS orbit.

\subsubsection{The WFC3 UVIS MultiDrizzle Procedure}

These first-pass aligned images were then run through a full combination with
MultiDrizzle (Koekemoer \etal\ 2002), which produced a mask of cosmic rays for
each exposure, together with a cleaned image of the field. The cosmic ray mask
was used to create a cleaner version of each exposure, by substituting pixels
from the clean, combined image. These were then re-run through the
cross-correlation routine to refine the relative shifts between the exposures,
achieving an ultimate relative alignment between exposures with an accuracy of
\cle 2--5 mas. This process was limited primarily by the on-orbit cosmic ray
density, and the available flux in the faint UV objects visible in each
individual UVIS exposure. In the end, about 3 independent input pixels from 4
exposures contributed to {\it one} MultiDrizzle UVIS output pixel. Of these,
typically \cle 1--2\% were rejected in the cosmic ray rejection, leaving on
average 3 independent UVIS measurements contributing to one MultiDrizzle output
pixel in both the F225W and F275W filters. In F336W, about 2.3 independent input
pixels from the 3 F336W exposures contributed to one output pixel during the
MultiDrizzle process.

After this relative alignment between exposures was successfully achieved, each
set of exposures needed to be aligned to the {\it absolute} GOODS astrometric
reference frame. This was achieved by generating catalogs from the cleaned,
combined images for each of the three ERS UVIS filters F225W, F275W, and F336W,
and matching them to the GOODS B-band catalog (Giavalisco \etal\ 2004; Grogin
\etal\ 2009). This was done by solving for linear terms (shifts and rotations)
using typically $\sim$30--50 objects matched in each pointing, depending on the
UVIS filter used. This procedure successfully removed the mean shift {\it and}
rotational offsets for each visit relative to the GOODS astrometric frame.
MultiDrizzle also produced ``weight''-maps (Koekemoer \etal\ 2002), which are
essential for the subsequent object detection (\S 5.1), and for the computation
of the effective area, which is needed for the object counts (see the figures in
\S 6--7). 

In order to perform matched aperture photometry (see \S 5.2), our approach was to
create images at {\it all} wavelengths at the same pixel-scale. Since the IR data
was necessarily created at 0\arcspt 090 per pixel (see Appendix A), we created
UVIS mosaics at that same pixel size. This essentially ``smoothed'' over the
remaining issue of the geometric distortion solution (see Appendix A), and
created a sufficient data product for the purpose of producing matched aperture
photometric catalogs, reliable total magnitudes in all 10 ERS filters, performing
z$\simeq$1--3 dropout searches, and many other ``total magnitude applications''.
The performance of the panchromatic ERS images for photometric redshift estimates
is described by Cohen \etal\ (2011). Further details on the remaining
uncertainties from the UV geometric distortion and its corrections are given in
Appendix A and in Fig. 4a--4b. 

The current 0\arcspt 090 per pixel UVIS image mosaic is referred to as ``ERS
version v0.7''. In the future, when the UV geometric distortion correction is
well measured, and better on-orbit WFC3 flat-fields or sky-flats become
available, we will make higher resolution images (0\arcspt 030 per pixel) for
applications such as high-resolution faint-galaxy morphology, structure,
half-light radii, and other high-precision small-scale measurements of faint
galaxies, and make these end-products available to the community. 

\subsubsection{IR Astrometry and MultiDrizzle Procedure}

The WFC3 IR images processed as described in section 3 were first corrected for
the instrument geometric distortion and then projected to a pre-specified
astrometric reference grid according to the World Coordinate System (WCS)
information populated in the image headers. This was done by using the
MultiDrizzle software (Koekemoer \etal\ 2002) distributed in the STSDAS.DITHER
package. Similar to the processing of the UVIS images in \S 4.2.1, the GOODS
version 2.0 ACS mosaics were used as the astrometric reference. The only
difference is that the GOODS ACS mosaics were $3\times 3$ rebinned for comparison
with the ERS IR data, giving a spatial resolution of 0\arcspt 090 per pixel for
all ERS images. 

As usual, the projected ERS IR images show non-negligible positional offsets,
which is mainly caused by the intrinsic astrometric inaccuracies of the guide
stars used in the different HST visits. Following Yan \etal\ (2010), about 6--12
common objects were manually identified in each ERS IR input image and in the
reference ACS $z_{850}$ image. We subsequently solved for X-Y shift, rotation,
and plate scale between the two. These transformations were then input to
MultiDrizzle, and the drizzling process was re-run to put each input image onto
the pre-specified grid. We set the drizzling scale (``pixfrac'') to 0.8, so that
in the IR about 5 input pixels from 4 exposures, or 20 independent measurements
contributed to one MultiDrizzle output pixel. Of these, typically \cle 10\% or 2
pixels were rejected in the cosmic ray rejection, leaving on average 18
independent measurements contributing to one MultiDrizzle output pixel. 

\subsection{Resulting ERS Mosaics and their Properties}

\subsubsection{The panchromatic 10-band ERS mosaics}

Fig. 5a. shows the panchromatic 10-band color image of the entire ERS mosaic in
the GOODS-South field. All 10 ERS filters in Fig. 5, 6, 13, and 14 are shown at
the 0\arcspt 090 pixel sampling discussed in \S 4. All RGB color images of the
10-band ERS data were made as follows. First, the mosaics in all 10 filters were
registered to the common WCS of the ACS GOODS v2.0 reference frame to well within
one pixel. Second, all images were rescaled to \Fnu units of Jy per pixel using
the AB zero-points of Table 2. Next, the blue gun of the RGB images was assigned
to a weighted version of the UVIS images in the F225W, F275W, and F336W filters
and the ACS F435W filter. The green gun was assigned to a weighted version of the
ACS F606W and F775W images, which had the highest S/N-ratio of all available
images. The red gun was assigned to a weighted version of the ACS z-band (F850LP)
and WFC3 IR F098M, F125W, and F160W images. All weighting was done with the
typical image sky S/N-ratio, sometimes adjusted so as to not overemphasize the
deepest multi-epoch GOODS v2.0 images in the V and i-filters. This procedure thus
also rebalanced the different sensitivities per unit time in these filters, as
shown in Fig. 1a--1b, and corrected for the fact that some filters have their
central FWHM-range overlap somewhat in wavelength, so they are not completely
independent (see Table 2). This is especially noticeable for the ACS z-band
filter F850LP --- which at the long wavelength side is cut-off by the sharp
decline in the QE-curve of silicon --- and the IR F098M filter, which doesn't
have this problem at its blue side, but overlaps with F850LP for about 40\% of
its OTA$\times$T$\times$QE integral, where OTA is the net Optical Telescope
Assembly reflectivity, T in the product of the WFC3 optics reflectivities and
filter + window transmissions, and QE is the detector Quantum Efficiency as a
function of wavelength. (When the QE of the HgCdTe detectors produced by Teledyne
increased from $\sim$10--20\% in 2001 to \cge 80\% after 2005, the F098M filter
thus became almost a replacement of the ACS z-band).

In Fig 5a--5b, we used a log(log) stretch to optimally display the enormous
dynamic range of the full resolution ERS color TIFF images. Fig. 5a only displays
the overlap between the 4$\times$2 ERS UVIS mosaics, the GOODS v2.0 ACS BViz
mosaics, and the 5$\times$2 ERS IR mosaics. Each of the ERS mosaics are 8079
$\times$ 5540 pixels in total, but only about 6500 $\times$ 3000 pixels or
9\arcmpt 75 $\times$ 4\arcmpt 5 or 43.875 arcmin$^2$ is in common between the
UVIS and IR mosaics and shown in Fig. 5a--5b. The area of the individual UVIS
mosaics used in each of the UV-optical galaxy counts of \S 7 is substantially
larger than this, but the total usable area of the IR mosaics is comparable to
the area shown in Fig. 5a. 

Fig. 5b (see Appendix B.1) shows a zoom of the 10-band ERS color image,
illustrating the high resolution available over a factor of ten in wavelength,
the very large dynamic range in color, and the significant sensitivity of these
few orbit panchromatic images. Further noteworthy objects in the images are
discussed in Appendix B.1 below.

\subsubsection{Astrometric quality of the ERS mosaics }


To compare the astrometry of our WFC3 ERS catalogs to our catalogs derived from
the GOODS ACS v2.0 images, we selected the WFC3 H-band, because the geometrical
distortion correction (GDC) was measured thus far in the F160W filter only (see
Appendix A). Amongst the GOODS ACS v2.0 images, we select the z'-band filter as
the closest in wavelength to compare the ERS F160W images to astrometrically, and
because most faint ERS stars are expected to be red (see \$ 5.5). The GDC of the
ACS/WFC has been well measured and calibrated over time and as a function of
wavelength (Maybhate \etal\ 2010, Anderson 2002, 2003, 2007), and so is not a
major source of uncertainty in this astrometric comparison. The exposure-time
averaged effective epochs of the GOODS v2.0 ACS/WFC mosaics are: 2002.7796\ in
F435W, 2002.9755 in F606W, 2003.6083 in F775W and 2003.7634 in F850LP,
respectively. Due to a continued GOODS high-z SN search that lasted from mid 2002
through early 2005, the spread on these numbers is about one year, yielding
possibly somewhat elongated ACS images for very high proper motion stars in each
of the GOODS v2.0 image stacks. The effective time-averaged epoch for the WFC
UVIS ERS data is JD 2009.6918 in F225W, F275W, and F336W with a spread of 2 days,
while for the WFC3 IR channel images the effective epoch is 2009.7370 in F098M,
F125W, and F160W with a spread of about one week. For the WFC3 ERS images, image
elongation for high proper motion stars is thus not a concern. The effective (WFC
ERS--GOODS v2.0) epoch difference to be used for proper motions derived from this
comparison is thus (2009.7370--2003.7634) = 5.97$\pm$1 years, where the
dispersion is dominated by the GOODS ACS z'-band image-spread of about one year. 

Fig. 4c shows the measured residual astrometric offsets in (RA, DEC) for all 4614
ERS objects matched between our WFC3 H-band object catalog and our GOODS ACS/WFC
v2.0 z'-band catalog, as well as their histograms in both coordinates for 4511
matched ERS objects classified as galaxies (black) and 103 ERS objects classified
as stars (red). Best fit Gaussians are also shown for each of the histograms. As
discussed in \$ 4.2, the WCS coordinate system in the FITS headers of the WFC3
ERS images was by definition brought onto the well established GOODS ACS v2.0
WCS. This was done by applying WCS offsets averaged over {\it all} ERS objects to
the WFC3 FITS headers. The histograms and curves in Fig. 4c show that this could
be done with an accuracy of 0.32 $\pm$ 0.46 (m.e.) m.a.s. in RA, and 0.10 $\pm$
0.41 (m.e.) m.a.s. in DEC, respectively, i.e. in general to within 0.5 m.a.s.
both randomly and systematically. While residual errors in the WFC3 GDC are large
(see \S 4.2 and Appendix A), for a large number of objects spread over all the
ERS images these errors apparently average out well enough to establish the
overall WCS coordinate system of both the ERS UVIS and IR mosaics onto the GOODS
v2.0 ACS/WFC mosaics to within 0.4--0.5 m.a.s. on average. 

For the 4511 ERS galaxies {\it alone}, Fig. 4c shows that the residual WFC3
offsets compared to GOODS ACS v2.0 are $\Delta$RA = --0.64 $\pm$ 0.47 (m.e.)
m.a.s. and $\Delta$DEC = +0.38 $\pm$ 0.42 (m.e.) m.a.s., or at the 1.4 and
0.9-$\sigma$ level in RA and DEC, respectively. For the galaxies, these ERS
offsets are indeed statistically insignificant, although they are not exactly
equal to zero, because the matching onto the ACS WCS was done including the ERS
stars as well --- {\it before} it was known what the optimal star-galaxy
separation method would be. (Because the residual offsets for {\it all} ERS
galaxies alone are within the 0.4--0.5 m.a.s. errors quoted above, no second
iteration was done in bringing the WFC3 WCS system on top of the GOODS ACS v2.0
WCS system.)

In total, 21 out of the 103 ERS stellar candidates show proper motion at the
$\ge$3-$\sigma$ level in RA or DEC, respectively, as shown by the green asterisks
in Fig. 4c. In total, 37 out of the 103 ERS stellar candidates show proper motion
at the $\ge$2-$\sigma$ level, also shown in Fig. 4c. Only about 5 stars are
expected at $\ge$2-$\sigma$ for a random Gaussian distribution, and so the
stellar ($\Delta$RA, $\Delta$DEC) offsets have a non-Gaussian distribution, as
shown by the histograms in Fig. 4c. Hence, proper motion allows us to confirm
statistically about 32 out of the 103 stellar candidates in the ERS. As a
consequence, ERS proper motions {\it alone} cannot prove that {\it all} our ERS
objects classified as stellar are in fact Galactic stars. For this reason, we
will also consider object colors in \S 5.5 as confirmation of the stellar
classifications. 

Statistically, proper motions do cause significant offsets in the average
($\Delta$RA, $\Delta$DEC) distribution of the 103 ERS objects classified as stars
(Fig. 4c). For these stellar candidates, we find on average that $\Delta$RA =
13.71 $\pm$ 3.34 (m.e.) m.a.s., or 2.30$\pm$0.56 m.a.s./yr, and $\Delta$DEC =
--12.04 $\pm$ 2.09 (m.e.) m.a.s., or 2.02$\pm$0.35 m.a.s./yr. These constitute
4.1-$\sigma$ and 5.8-$\sigma$ detections of the statistical proper motion of all
103 ERS stellar candidates. The KS probability that the stellar $\Delta$RA values
are drawn from same distribution as the ERS galaxy population is
9.8$\times$10$^{-5}$, while for the stellar $\Delta$DEC values this probability
is 11$\times$10$^{-5}$. Hence, {\it average} stellar proper motion {\it is}
detected at high significance level for the sample of 103 ERS stellar candidates.
This is a significant result, since the star-galaxy separation of \$ 5.5) was
done {\it completely independently} of any proper motion information. A further
discussion of this result is given in \$ 5.6 and 6. 

Fig. 6c shows a log(log) color reproduction of the 20 ERS stars with the highest
(\cge 3-$\sigma$) proper motion (green symbols in Fig. 4c). The images used a
similar color balance as in Fig. 5a--5b, except that only the 2009 WFC3 UVIS
filters are shown in the Blue gun, {\it all} the 2003 ACS BViz filters were used
in the Green gun, and all the 2009 WFC3 IR filters in the Red gun. The proper
motion of these stars is best visible as significant centroid-displacements
between the Green 2003 ACS colors and the Blue+Red (or violet) 2009 WFC3 colors
(one has to magnify the PDF figure to best see the significant central
green-to-white-to-orange displacement). 

\subsubsection{The panchromatic ERS PSFs }

Fig. 6a shows a full color reproduction of a stellar image in the 10-band ERS
color images in the GOODS-South field, and Fig. 6b shows a ``double'' star. These
images give a qualitative impression of the significant dynamic range in both
intensity and wavelength that is present in the ERS images. Fig. 7a show images
in all 10 ERS filters of an isolated bright star that was unsaturated in all
filters, and Fig. 7b shows its 10-band stellar light-profiles. Table 2 lists the
stellar PSF-FWHM values in the 10 ERS bands. These include the contribution form
the OTA and its wavefront errors and the specific instrument pixel sampling.

Table 2 and Fig. 7b show the progression of the HST PSF ($\lambda$/D) with
wavelength in the 10 ERS filters. Table 2 implies that the larger pixel values
used in the multi-drizzling of \S 4.2.2 indeed add to the effective PSF diameter.
Table 2 also shows that that HST is diffraction limited in V-band and longwards,
while shortward of V-band, the PSF-FWHM starts to increase again due to mirror
micro-roughness in the ultraviolet. At wavelengths shorter than the V-band, HST
is no longer diffraction limited, resulting in wider image-wings, and a somewhat
larger fraction of the stellar flux visible outside the PSF-core. The ``red
halo'' at $\lambda$\cge 0.8 \mum\ is due to noticeable Airy rings in the stellar
images in the WFC3 IR channel, and the well known red halo in the ACS z-band
(Maybhate \etal\ 2010) has an additional component from light scattered off the
CCD substrate. Details of the on-orbit characterization of the WFC3 UVIS and IR
PSFs are given by Hartig (2009ab).

\subsubsection{The 10-band ERS Area and Depth }

\n Table 1 summarizes the exposure times and the actual achieved depth in each of
the observed ERS filters, while Table 2 also lists the effective area covered in
each filter mosaic at the quoted depth.

The histograms of Fig. 8a--8c give the cumulative distribution of the maximum
pixel area that possesses a specified fraction of total orbital exposure time.
These effective areas must be quantified in order to properly do the object
counts in \S 6--7. After Multi-drizzling the ERS mosaics in the UV and IR, these
effective areas were computed from the weight maps, which include the total
exposure time, and the effects from CR-rejection, dithering, and drizzling. Fig
8a shows that about 50 arcmin$^2$ of the UV mosaics has $\sim$80\% of the UVIS
exposure time, or $\sim$90\% of the intended UV sensitivity. For reference, one
0\arcspt 090 pixel could be composed of 5.2 native WFC3 UVIS pixels times the
number of exposures on that portion of sky. Due to overlapping dithers (see Fig.
3), some pixels have more than the total orbital exposure time contributing to
their flux measurement. The histograms of Fig. 8b give the same information as
Fig. 8a, but for the six GOODS v2.0 mosaic tiles in BViz that overlap with the
ERS. Fig. 8c give the same information as Fig. 8a, but for the ERS mosaics in the
IR. About 40 arcmin$^2$ has $\sim$80\% of the exposure time in the ERS IR
mosaics, or $\sim$90\% of the intended IR sensitivity. The overall WFC3 UVIS--IR
sensitivity is 9--18\% better than predicted from the ground-based thermal vacuum
tests (see \S 2.4 and Fig. 2), and so in essence 100\% of the intended ERS
exposure time was achieved over the 50 arcmin$^2$ UVIS images and the 40
arcmin$^2$ IR images. 

\section{Catalog Generation from the 10-band ERS mosaics }

\subsection{Object Finding and Detection }

All initial catalogs were generated using SExtractor version 2.5.0 (Bertin \&
Arnouts 1996). In general, these catalogs were generated SExtractor's {\it single
image mode} for each ERS filter separately, so that the object finding could be
done using the {\it total} object flux from each ERS filter independently, as is
required when determining the star counts (\S 6) and the galaxy counts (\S 7).
SExtractor was {\it only} used in its {\it dual image mode} to generate the {\it
additional} catalogs that were used exclusively to make the color-color diagrams
in \S 5.6 to confirm our star-galaxy separation procedure, using the H-band as
the detection image. As stated in \S 4.2.1, the ACS B-band image was used as the
detection image to get an optimal object definition in the UVIS filters for
reasons explained in detail here. 

It was necessary to change the parameters in SExtractor to handle the UVIS images
slightly differently than the ACS/WFC and WFC3/IR ones. This is due to several
factors, both cosmetic and physical. The major difference is that galaxies appear
much smoother in the IR and more clumpy in the UV, which occurs due to the
specific distribution of their young and old stellar populations as well as their
dust (see Windhorst \etal\ 2002 for a discussion of this effect for all Hubble
types). Hence, star-forming regions in well-resolved galaxies can be deblended
into separate objects by SExtractor if the deblending is overly aggressive.
Fortunately, most of these intermediate redshift galaxies are not bright in the
observed UV, causing the UVIS fields to appear rather sparse, and making
deblending a minor issue. 

The UVIS images have a less uniform background ($\sim$10--15\% of sky), which as
discussed in \S 4.1.1 can possibly be improved, when accurate on-orbit UVIS
sky-flats have been accumulated over time. For these reasons, we adopted a
seemingly low threshold for object detection with SExtractor, requiring 4
connected pixels that are $0.75\sigma$ above the background in the UV. Due to the
clumpy nature of the ERS objects in the UV, we convolved the UVIS image with a
Gaussian kernel of FWHM of 6.0 pixels for the object detection phase only. The
deblending parameter, DEBLEND\_MINCONT, was set to 0.1, to assure that real
objects were not over-deblended. It should be emphasized that object crowding is
{\it not} an issue in these medium depth UV images (\eg Windhorst \etal\ 2008),
so that the use of a larger SExtractor convolution filter, a lower SB-detection
threshold, and less object deblending is fully justified. The 6.0 pixel Gaussian
convolution kernel improves the SB-sensitivity by $\sim$ 2 mag, given that the
low-level UV sky-gradients occur over much larger spatial scales than the typical
faint galaxy sizes (see \S 4.1 and 5.5). Together with the fact that the UVIS
catalogs were made with SExtractor in dual image mode --- using the much deeper
ACS B-band image as the detection image, as described above --- avoids excessive
deblending of the very clumpy UV objects, which limits the number of spurious UV
objects detected, as shown in the discussion of the ERS catalog reliability in \S
5.3.

The SExtractor input parameters for the ACS images and WFC3/IR images were the
same. The detection threshold was set to $1.5\sigma$, again requiring 4 connected
pixels above the threshold for catalog inclusion. The convolution filter was a
Gaussian with FWHM of 3.0 pixels, and DEBLEND\_MINCONT was set to 0.06. For all
ten bands, the appropriate weight maps (see \S 4.2.2) were used in order to
correctly account for image borders (Fig. 3), as well as to properly characterize
the photometric uncertainties and the effective areas for each mosaic (see Fig.
8a--8c).

\subsection{Object Extraction }

Two post-processing steps were taken to clean the catalogs of residual artifacts.
First, due to the relatively small number of exposures per ERS filter, there were
residual cosmic rays at the borders of the dither patterns, and in the chip gaps.
These were cleaned by masking out all objects in regions where the number of
exposures was less than three. Therefore, the total area coverage in each filter
(Table 2 \& Fig. 8a--8c) is slightly different. Secondly, SExtractor will detect
the diffraction spikes of bright point sources. These are removed by searching
for bright objects with FWHM near the size of the PSF and searching for
surrounding objects in the catalog that are {\it both} highly elongated {\it and}
oriented radially outward in sets of 4 from that compact object. Since the number
of bright stars is not large (see \S 4.3.2 and figures therein, \$ 5.5, \& 6.1),
this is only a minor correction to the catalog, and only a few dozen diffraction
spikes were found by SExtractor and removed in this way.

Magnitudes were all measured on the AB-scale using the most current zero-points
available on the WFC3 website, which are listed in Table 2. Both Petrosian (1976)
magnitudes were measured using a Petrosian factor of 2.0, and Kron (1980)
magnitudes were measured using a Kron-factor of 2.5 (SExtractor parameter
MAG\_AUTO). For the galaxy counts in \S 7.2, Kron magnitudes were used, because
the correction to total magnitude --- preferably using a Sersic extrapolation ---
is less than that typically required for a Petrosian magnitude. Within the
errors, the Kron and Sersic based number counts are indistinguishable (Hill
\etal\ 2010b). The total magnitude errors in the ERS as a function of total
AB-magnitude were determined from the SExtractor errors in each filter, which are
a combination of the MultiDrizzle RMS-map errors and the image shot-noise.

\subsection{ERS Catalog Reliability }

To test the reliability of ERS object catalogs, we performed the negative image
test (\cf Yan \& Windhorst 2004). In brief, all 10 ERS mosaics were multiplied by
--1.0, and then SExtractor was run with {\it exactly the same} parameters as the
object finding procedure in \S. 5.1--5.2. This test showed that our ERS catalog
generation was conservative due to the combination of several factors. These
include drizzled the larger mosaic pixel size of 0\arcspt 090, the requirement of
{\it four} connected pixels above the detection threshold, and the sizes of the
Gaussian filters applied in the object detection phase. The choice of the
SExtractor parameters listed in \S 5.1--5.2 were determined by trial and error to
create the most reliable catalogs at each wavelength. 

In fact, these SExtractor parameter settings yielded {\it no plausible} spurious
detections when it was run on the 10 negative ERS images. In total, only 6
negative ``objects'' were found, but these were all on the image borders, and the
real ERS object catalogs were similar cleaned of such border defects. {\it If}
the noise in the ERS images were completely Gaussian, then at the \cge 5$\sigma$
detection limit we would expect that 2.9$\times$$10^{-7}$ of the
6500$\times$3000/4 independent ERS pixels (see Fig. 5a) would yield a bogus
object, or $\sim$1.4 bogus object in the combined 10-band catalog (the grand
union) of 22,000 ERS objects. This is consistent with {\it no convincing bogus
objects} found in the 10 negative ERS images after discarding the image borders.
This exercise shows that at the 5$\sigma$ 50\% completeness level (see also Fig.
9 below), our ERS object catalogs are \cge 99.97\% reliable. In astrophysical
data samples, a compromise always has to be sought between sample reliability and
sample completeness, while ideally maximizing the sum of the two (Windhorst,
Kron, \& Koo 1984b). The reason we choose here to have essentially 100\% reliable
catalogs --- at the expense of some catalog incompleteness below --- is the
existence of the deeper optical--near--IR ACS and WFC3 images in the HUDF
(Beckwith \etal\ 2006, Bouwens \etal\ 2010b, Yan \etal\ 2010) in the \BVizYsJH\
filters. These images made digging deeper into the ERS image noise unnecessary.
Instead, the faint-end of the galaxy counts at AB\cge 26--27 mag in \S 7 will be
derived in \BVizYsJH\ from these HUDF images, which is in roughly the same
direction of the sky. 

\subsection{ERS Catalog Completeness }

Fig. 9 shows the sample completeness functions of the 10-band ERS images. These
were derived by Monte Carlo insertion of faint point sources into the ERS images,
and plotting the object recovery ratio as a function of total magnitude. Most
faint galaxies observed by HST are slightly resolved at the HST diffraction limit
(\eg Cohen \etal\ 2003; Hathi \etal\ 2008; Windhorst \etal 2008), so that for
faint galaxies these limits are slightly brighter than for point sources, as
discussed in \S 7.2 and shown in Fig. 10a. 

In summary, the WFC3 UVIS images are $\sim$50\% complete for point-like objects
with \cge 5-$\sigma$ detections in total flux for AB$\simeq$26.3 mag in F225W,
AB$\simeq$26.4 mag in F275W, and AB$\simeq$26.1 mag in F336W. The WFC3 IR images
are $\sim$50\% complete to AB$\simeq$27.2 mag in F098M, AB$\simeq$27.5 mag in
F125W, and AB$\simeq$27.2 mag in F160W, as listed in Table 1. The multi-year
GOODS ASC/WFC v2.0 images in BViz are substantially deeper than this, and reach
AB\cge 27.5 mag in general. The actual star counts in \S 6 appear indeed complete
to roughly these limits in the UV, but to slightly brighter limits in the IR due
to the increasing object confusion of red stars with faint red background
galaxies (see the figure in \S 5.5). 

The actual UV galaxy counts in \S 7.2 turn over at a level rather close to the
above UVIS point-source sensitivity limits. This is due to the generally more
compact nature of the ERS objects, and the extremely faint zodiacal
sky-background in the UV. In the IR, the galaxy counts turn over at levels a
little brighter than the 5-$\sigma$ point-source sensitivity limits listed in
Table 2. This is due to the higher zodiacal sky-background values in the
near-IR (see Table 2), combined with their somewhat larger sizes at the longer
wavelengths (\S 5.5 and figures therein; see also Windhorst \etal\ 2002), even
though faint red field galaxies selected in \YsJH\ are in general still rather
compact. Object confusion at these flux levels is relatively modest, especially
at the HST diffraction limit, although natural confusion --- which due to the
intrinsic object sizes and {\it not} the instrumental PSF-FWHM --- becomes
increasingly important at fainter fluxes (AB\cge 25 mag; \eg Windhorst \etal\
2008). 

In summary, the point-source sensitivity of our ERS catalogs is described by the
simulations in Fig. 9. The corresponding 50\% point-source completeness limits
are listed in Table 1. The actual observed panchromatic star-counts (\S 6.1
and figures therein) are complete to roughly the 5-$\sigma$ limits of Fig. 9 in
the WFC3 UVIS filters, and to the 10-$\sigma$ limits of Fig. 9 in the ACS BViz
and WFC3 IR filters. 

Rather than repeating the object simulations of Fig. 9 to assess the actual
completeness limits for the {\it slightly extended} galaxies (see \S 5.5 and
figures therein), in \S 7 below we take a different and more empirical approach
of determining the ERS incompleteness for faint galaxies by directly comparing
the panchromatic ERS galaxy counts to AB\cle 25.5--27 mag to the galaxy counts in
the much deeper, but adjacent HUDF field. Since the effects of cosmic variance
are relatively small over the 4\arcm separation between the ERS and HUDF fields
(see \S 7 and Fig. 3), this comparison directly shows that the ERS catalogs are
approximately \cge 50\% complete for samples of compact galaxies (see \S 5.5) to
flux levels that are approximately 0.5--1.0 mag brighter (from the UVIS to the
IR) than the ERS stellar samples in the same filters. 

\subsection{Star Galaxy separation in the ERS Mosaics }

Stars and galaxies were separated as follows. In each filter, the stellar locus
was defined as shown in the usual plot of the object FWHM versus total
AB-magnitude (\eg Cohen \etal\ 2003). Fig. 10a illustrates our star-galaxy
separation procedure used for the 10-band ERS images. Objects with
FWHM$<$PSF-FWHM (from Table 2) are image defects, bad pixels, and some remaining
CRs. Hence, objects to the left of the black vertical lines plotted at
FWHM$\simeq$0\arcspt 07--0\arcspt 15 are {\it smaller} than the PSF, and are
discarded. Objects in the thin vertical filaments immediately larger than this
are stars (plotted as thin reds dots in Fig. 10a), and they in general have
PSF-FWHM\cle FWHM\cle (1+$\epsilon$)$\times$PSF-FWHM, where 0\cle $\epsilon$$<$1.
Fig. 10a shows that ERS stars straddle the instrumental PSF-FWHM to AB$\simeq$27
mag. Objects to the right of the black dashed slanted lines are galaxies, and
they in general have FWHM\cge (1+$\epsilon$)$\times$PSF-FWHM, where
``1+$\epsilon$'' is a weak function of total S/N-ratio, as shown by the two solid
slanted lines in each of the panels in Fig. 10a.

Next, all ten independent ERS object catalogs --- each with their independent
star-galaxy separation and photometry --- were merged by astrometric
cross-matching. In order to maximize the 10-band information, we considered an
object to be a star if it was classified as stellar in {\it at least 3 out of 10
ERS filters}. This provides robust stellar classifications, perhaps to fainter
limits than can be done in single filter alone. 

From Fig. 10a, we can see that the stars separate out well from the galaxies to
about AB$\simeq$25.5--26.0 mag, and somewhat fainter in the ACS BViz filters and
the WFC3 IR filters. To these flux levels, the star counts follow a power-law
rather well, as shown in \S 6.1. The choice to require an object to be classified
as a star in three filters --- as opposed to all ten filters --- was made because
stars can be either very red or sometimes very blue, and so stars are often not
detected in all 10 ERS filters. Also, we are merging the data from three
different HST instruments and detectors (WFC3 UVIS CCD, ACS/WFC CCD, and the WFC3
IR detector), so subtle PSF and sampling differences between these instruments
can lead to the mis-classification of a faint star in some of the 10 ERS filters,
depending on its color. Since brighter stars are always seen at least in three
(adjacent) ERS filters, we require all of the fainter stars to be classified as
stellar in at least three of the ERS filters as well. 

\subsection{Confirmation of Star-Galaxy Separation: Proper Motion and Stellar
Colors }


As shown in \S 4.3.2 and Fig. 4c, all ERS objects together --- which are \cge
97.8\% galaxies --- allowed us to bring the WFC3 UVIS and IR coordinate system on
average on top of the ACS WFC WCS to within 0.4--0.5 m.a.s. in both RA and DEC.
After having done this, the 4511 ERS objects classified as galaxies show indeed
no significant residual motion over 5.97 years. However, for all 103 ERS objects
classified as stars, an {\it average} proper motion of 2.30--2.02 m.a.s./yr was
detected at the 4.1--5.8-$\sigma$ level in RA and DEC, respectively (see \S
4.3.2). For at least 32 out of the 103 ERS stars, this proper motion was
discovered {\it individually} at the $\ge$3-$\sigma$ level (see Fig. 4c). The
total {\it average} proper motion of all 103 ERS stars on the sky is 3.06
m.a.s./yr, summed in quadrature over the RA and DEC proper motions. The mean
error is 0.66 m.a.s./yr, which amounts to a 4.6-$\sigma$ detection of the overall
ERS stellar proper motion. The dispersion on this number in Fig. 4c is 6.70
m.a.s./yr. We will interpret these numbers in more detail below. In conclusion,
while ERS proper motions {\it alone} cannot confirm that {\it all} our ERS
objects classified as stellar are in fact Galactic stars, {\it average} stellar
proper motion {\it is} detected at high significance level for the {\it entire
sample} of 103 ERS stellar candidates, making it in retrospect likely that most
of these are in fact Galactic stars. 

The ERS colors in Fig. 10b are more conclusive in supporting our ERS star-galaxy
separation procedure a posteriori. Fig. 10b shows the (i--z) vs. (B--V)
color-color diagram in the GOODS filters, where all ERS objects have the highest
S/N ratio (see Fig. 9). ERS objects classified as galaxies as shown as black
dots, and ERS objects classified as stars with green or red asterisks. Green
asterisks mark ERS objects that show proper motion at the $\ge$3-$\sigma$ level
in Fig. 4c (see \S 4.3.2). The (i--z) vs. (B--V) colors are compared to the
Pickles (1998) model library SEDs (blue squares), which range in spectral type
from O5V--M7V, and also show giant branch models up to M7III. The observed ERS
stellar colors are shown with their 1-$\sigma$ error bars. Only stars with
combined color errors along each axis \cle 0.5 mag are plotted. It is clear that
within the errors, most stellar ERS objects have colors of K--M type main
sequence stars, although there are a few ERS stars with color as blue as BV--GV
types. 

There are about 20 ERS stars in the color range 0\cle (B--V)\cle 1.4 mag with
(i--z) colors \cge 0.3 mag redder than the Pickles (1998) main sequence. At least
8 of these also have significant proper motion at the \cge 3-$\sigma$ level, and
are thus real Galactic stars. All 20 objects significantly redder than the
Pickles (1998) main sequence were inspected by eye and correlated with other
data. Three have VLT redshifts: one is a Seyfert galaxy at z=1.031 in a group
with similar spectroscopic redshifts, the other a compact elliptical at z=0.251
next to a very faint stellar object, and the third is a stellar object next to an
object at z=2.217 with a neighbor at a similar redshift. The latter two stellar
objects also have significant proper motion, and are likely real Galactic stars
that are very close to an extragalactic object whose VLT redshift was actually
measured.

Thus at least 17 --- and likely 19 --- of these 20 ERS stars with colors
significantly redder than the Pickles (1998) main sequence have either
significant proper motion and/or clear stellar morphology. Many of these appear
to be stars that were unavoidably saturated in the ERS images. Because both their
colors are \cge 0.0 mag --- they have higher ADU in i-band than in z-band, and in
V-band than in B-band --- given the ACS/WFC QE curve. Hence, their measured
(i--z) flux will be too red, and their (B--V) color slightly too blue. Since
these stars are saturated in nearly every ACS exposure, more accurate colors
cannot be determined with the current mosaics, and their appearance above the
Pickles (1998) main sequence does thus not rule them out as stars. Some of the
reddest stars well away from the Pickles (1998) main sequence could be AGB stars
with surrounding dust shells that could considerably redden their (B--V) and
(i--z) colors significantly.

We also need to address how many of the stellar objects that are within the
errors consistent with the Pickles (1998) stellar main sequence could be compact
background galaxies. Fig. 10b shows that the (i--z) vs. (B--V) color space of
zero redshift stellar SEDs overlaps somewhat with the colors of intermediate
redshift galaxies, which largely occupy the region (B--V) \cle 1 mag {\it and}
(i--z) \cge 0.15 mag. This is a limitation of the current color confirmation of
the stellar ERS candidates. Only multi-parameter 10-band SED fitting of {\it all}
ERS objects can address this issue more fully, which we will address in a future
paper (Cohen \etal\ 2011). However, the available VLT spectra can give us some
idea. 

Among the 103 objects classified as ERS stars, many would have been bright enough
to have been included in the VLT redshift survey, which is complete to AB\cle 24
mag, but covered objects as faint as AB\cle 25 mag. Among the 103 stellar ERS
objects, 11 appear to be compact extragalactic objects confirmed to have VLT
redshifts z$>$0. One of these is a quasar with an underlying early-type galaxy at
z=0.734. 

In summary, this detailed study in Fig. 10b confirms that at least 90--92 of the
103 stellar ERS objects have either significant proper motions (Fig. 4c), stellar
colors consistent with the Pickles (1998) main sequence (Fig. 10b), and/or are
not extragalactic objects with VLT redshifts z$>$0. Conservatively, we can thus
conclude that our ERS stellar objects are indeed Galactic stars with \cge 87\%
reliability. This provides a posteriori confirmation that our ERS star-galaxy
separation method of \S 5.5 --- which was based on measured effective image
diameters {\it only} (Fig. 10a) --- is indeed valid. This then also means that we
may consider the {\it statistical proper motions of all} ERS stars in Fig. 4c, to
investigate if the implied {\it average} proper motion shows us something
interesting about Galactic structure of very faint stars at high Galactic
latitude (the GOODS-South field has \bII=--54\degree). We discuss this in section
6 in more detail. 

As a final remark, ideally one would like the star-galaxy separation procedure to
also include accurate fits of a library of (redshifted) SEDs to the 10-band ERS
object photometry, and so help decide whether an object is most likely a
zero-redshift star or a redshifted galaxy. This is beyond the scope of the
current paper, and is pursued by Cohen \etal\ (2011) for the ERS stars and
galaxies separately. This paper will also attempt to do a stellar + power-law AGN
SED fits to {\it all} ERS objects, including on a pixel-to-pixel basis. This
procedure will also provide better estimate of the AGN contamination of the ERS
sample, rather than relying on the presence of weak point-like components in the
ERS images alone. Suffice to say here that in Fig. 10b most ERS stars have colors
consistent with the stellar main sequence, so that contamination of our {\it
stellar} ERS objects by weak AGN or other compact extragalactic objects is likely
small. This is confirmed by the high proper motion stars in Fig. 10b, which have
statistically the same color-distribution as the stellar ERS objects without
measurable proper motion. 

\section{Results on ERS Stars }

\subsection{The Panchromatic ERS Star Counts to AB$\simeq$26 mag}

Fig. 11a shows the differential panchromatic star counts in the 10-band ERS
images. The star-galaxy separation breaks down at fluxes AB\cge 26 mag in
\BVizYsJH. However, for AB\cge 22 mag galaxies outnumber stars by a large margin,
so the more uncertain star-galaxy separation for AB\cge 26 mag will not affect
the quality of the galaxy counts. The star counts, of course, become
correspondingly harder to do for AB\cge 25 mag, and so in section 6.2 and Fig.
10b additional criteria such as colors and proper motion were used to confirm
faint stars at higher confidence. 

The solid black line in the F850LP panel of Fig. 11a represents the spectroscopic
star counts to AB$\simeq$25 mag from the HST Cycle 14 ACS grism survey ``PEARS''
(Probing Evolution And Reionization Spectroscopically) of Pirzkal \etal\ (2009).
Since for these objects we can use the low resolution ACS G800L spectroscopic
data as independent confirmation of their stellar nature, they have higher
accuracy than the image-based star-counts alone. The southern PEARS ACS grism
survey was done over a similar sized area to the ERS, covering an overlapping
portion of GOODS-S. Hence, it is encouraging that the Pirzkal \etal\ (2009) grism
star counts, which are for spectral types of $M0$ and later, have the same {\it
slope} as our direct-imaging ERS star counts in F850LP. The slightly different
star count {\it amplitude} for the flux range 18\cle AB\cle 25 mag is likely due
to the fact that the spectroscopic sample was limited to later spectral types,
where those grism classifications were most reliable. The good agreement in star
counts between the two different survey methods confirms that --- as far as faint
stars are concerned --- our star-galaxy separation is accurate to about
AB$\simeq$25 mag. 

Fig. 11b shows the ERS star count slope versus observed wavelength in the flux
ranges AB$\simeq$19--25.5 mag for the 3 UV WFC3 filters, AB$\simeq$16--26 mag for
the GOODS/ACS BViz filters, and AB$\simeq$15--25 mag for the 3 WFC3 IR filters,
respectively. The GALEX FUV and NUV points of Xu \etal\ (2005) are also plotted
at 153 and 231 nm, and cover the flux range AB$\simeq$17--23 mag, using primary
color criteria to accomplish star-galaxy separation at the 4--6'' FWHM GALEX
resolution. The statistics of the star counts in the three bluest ERS filters
(F225W, F275W and F336W) are rather sparse (Fig. 11a). As a consequence, our
three bluest best-fit ERS power-law slope values are rather uncertain, as
indicated by the larger error bars in Fig. 11b. The GALEX FUV and NUV star count
points of Xu \etal\ (2005) are complete to somewhat brighter levels
(AB$\simeq$23--24 mag), but have much better statistics. As a consequence, the
GALEX FUV and NUV star count slope points in Fig. 11b have smaller error bars,
and at the common wavelength of $\sim$230 nm they therefore carry larger weight
than the ERS star count slope in the mid--UV. 

Given these uncertainties, the Galactic star count slope at the faint-end is
remarkably flat at all wavelengths from the Balmer break to the near-IR, although
it is rather poorly determined from the ERS data alone below the Balmer break.
The values of the best-fit ERS power-law star count slope is in general of order
0.03--0.05 dex/mag. Including the more accurate GALEX star count slope values at
150 and 230 nm, the panchromatic faint-end of the star-count slope appears to be
always in the range 0.03--0.20, i.e. at all ERS wavelengths the faint star-count
slope remains well below Euclidean value of 0.6 dex/mag. This shows that to a
depth of AB$\simeq$25--26 mag at all wavelengths in the range 0.2--2 \mum, the
ERS images readily penetrate through most of the thick disk of our Galaxy at
these latitudes (\bII$\simeq$--54\degree) in the direction of the Galactic
anti-center (\lII=224\degree; Ryan \etal\ 2005; Beckwith \etal\ 2006). This trend
was also seen in the red ACS imaging parallels of Ryan \etal\ (2005), and the red
ACS GRAPES and PEARS grism observations of Pirzkal \etal\ (2005, 2009),
respectively, each of which constrained the scale-height of Galactic L \& T
dwarfs to about 300--400 pc. 

In the bluest three WFC3 filters F225W, F275W, and F336W, the WFC3 ERS finds only
22, 18, and 27 faint stellar candidates, respectively. Most of the full sample of
ERS stars have 10-band colors red enough to indicate Galactic halo K--M type
stars and possibly a few L \& T dwarfs, but a few of the faint stars are blue
(see \S 5.6 and Fig. 10b). The latter could be faint Galactic white dwarfs,
though there were no UV detected stars that were not also seen in the ACS
$B$-band. The effective ERS UV stellar detection limit of AB$\simeq$25 mag could
trace a T=2$\times$$10^{4}$K white dwarf with an absolute magnitude in the range
$M_{bol}$$\simeq$+15 to +10 mag (Harris \etal\ 2006) out to a distance of 1--10
kpc, respectively, assuming modest bolometric corrections. The faint stellar
objects in the 10-band ERS data are obviously of great interest by themselves,
and will be studied in more detail in a future paper. An important lesson learned
for UV observations of high-latitude fields is that there are so few UV bright
stars, that one has to ensure that individual exposures are long enough in order
to align and drizzle (or stack) them using faint galaxies. In the ERS UV images
we did this using the cross-correlation technique, as explained in more detail in
Appendix A.

\subsection{The Nature of Faint ERS Stars from Proper Motion and Stellar Colors}

Fig. 10b suggest that the majority of ERS stars have red colors characteristic of
faint red main sequence stars. Their typical spectral type is in the range
K0V--M5V with an average spectral type of $\sim$K5V. For these stars, the
expected \MV ranges from +5.8 to +12.3 mag with an average of \MV$\sim$+8 mag.
With the flat ERS star count slope of Fig. 11a, the {\it average} ERS star has an
AB magnitude of $\sim$23 mag, while the ERS stellar detection limit is AB\cle 26
mag. Hence, the {\it average} faint red ERS star will be at a distance of about
R$\simeq$10 kpc from the Sun. The direction of the GOODS-South field at
RA=03$^{h}$ 32$^{m}$ 39$^{s}$, DEC=--27\degree 47\arcm 29\arcspt 1 is in the
Southern Galactic hemisphere, about 54\degree\ below the plane of the Milky Way,
and in the general direction of the Galactic anti-center. Stars in this direction
at $\sim$10 kpc distance from the Sun will thus be bulge and disk stars $\sim$8
kpc below the Galactic plane, \ie well below the outer Orion arm of the Galaxy.
At this distance, the above {\it average} ERS proper motion of 3.06 $\pm$ 0.66
m.a.s./yr amounts to an average stellar {\it tangential} velocity of \cle 145
$\pm$ 31 km/sec (m.e.). These stellar velocities and their dispersion are in part
due to the projected solar motion in the sky --- compared to the Galactic Local
Standard of Rest (LSR) --- and in part due to the true space velocities of ERS
stars in the Galactic bulge and disk. While a detailed discussion of the proper
motion of ERS stars and its implication for Galactic structure is beyond the
scope of the current paper, we will here briefly discuss how our ERS proper
motion results compare to previous proper motion and radial velocity observations
of the Galactic bulge and disk at intermediate to high Galactic latitude. 

A proper motion study to V\cle 22.5 mag at the North Galactic Pole by Majewski
(1992) shows a systemic velocity of --120 km/sec at the edge of the thick disk
($Z$\cge 5.5 kpc). Clarkson \etal\ (2008) show similar proper motion values at
R$\sim$5 kpc, but in Baade's window from multi-epoch HST/ACS images that are
effectively complete to about I\cle 21 mag given the more crowded images at low
Galactic latitudes. Kinman \etal\ (2007) find that Galactic Blue Horizontal
Branch (BHB) stars with an average distance to the Galactic plane of $Z$\cle 8
kpc have zero galactic rotation and roughly isotropic velocity dispersions of
93--81 km/sec in the tangential and $Z$ directions, respectively, or $\sim$123
km/sec in the plane of the sky. Theoretical disk+bulge+halo models (Battaglia
\etal\ 2005, Klypin \etal\ 2002) predict velocity dispersions at R$\sim$5-10 kpc
of 150 km/sec for the Galactic bulge+thick disk components.


These values are very similar to the 145 km/sec value implied from the {\it
average} proper motion observed for our faint red ERS stars at relatively high
Galactic latitude (\bII=-54\degree). In other words, the Galactic (thick
disk+bulge) velocity dispersion {\it implied} from our faint red ERS stellar
proper motions is consistent with these previous observations and theoretical
model predictions. It is worth noting, however, that these previous observations
are based on high resolution radial velocities {\it and} proper motions of
Galactic stars with V\cle 16 mag, or proper motions alone for stars with V\cle
22.5 mag. Our ERS observations push this substantially fainter for proper motions
at high Galactic latitudes, to flux levels (AB\cle 25--26 mag) where {\it high}
resolution spectroscopy cannot be done with current ground based facilities.
Systematic proper motion studies of very faint stars at HST resolution at high
Galactic latitudes thus have the potential to sample part of Galactic structure
that is otherwise not easily accessible. This topic will be studied in further
detail in a subsequent paper.

\section{Results on Galaxies in the ERS }

\subsection{The Panchromatic ERS Galaxy Counts to AB$\simeq$26--27 mag}

Fig. 11a also shows the differential panchromatic galaxy number counts in the
10-band ERS images. {\it No} completeness corrections --- as implied by Fig. 9
and 10a --- were applied to the faint-end of the panchromatic ERS galaxy counts
in Fig. 11a. Instead, to be conservative, the data points are not plotted in all
panels of Fig. 12 at the flux level where the counts clearly start to deviate
from a best-fit faint-end power-law (see Fig. 11a \& 11c) by more than 50\%. This
in general occurs 0.5--1.0 mag brighter than the 50\% point-source completeness
limits listed in Table 1, as derived from Fig. 9. 

Fig. 11c shows the ERS galaxy count slope versus observed wavelength in the flux
ranges AB$\simeq$19--25 mag for the 3 UV WFC3 filters, AB$\simeq$18--26 mag for
the GOODS/ACS BViz filters, and AB$\simeq$17--25 mag for the 3 WFC3 IR filters,
respectively. The GALEX FUV and NUV points of Xu \etal\ (2005) are also plotted
at 153 and 231 nm, and cover AB=17--23 mag, again using primary color criteria to
accomplish star-galaxy separation at the 4--6'' FWHM GALEX resolution. The
statistics of the galaxy counts in the three bluest ERS filters (F225W, F275W and
F336W) are sparse (Fig. 11a), and at the BRIGHT-END the ERS object definition and
deblending is more complex given the sizes of the brighter objects (Fig. 10a). As
a consequence, the best-fit ERS galaxy count power-law slope values become
somewhat more uncertain at the bluer ERS wavelengths (Fig. 11c). The GALEX FUV
and NUV galaxy counts of Xu \etal\ (2005) are complete to somewhat brighter
levels (AB$\simeq$25 mag), but have much better statistics than the ERS. Given
the much larger size of the GALEX beam (4--6'' FWHM), they do however suffer from
more uncertain star-galaxy separation, which for GALEX needed to be done by using
multi-color information as well (Baldry \etal\ 2010). At the longer GALEX
wavelength, it may be harder to separate Galactic stars from low redshift older
galaxies using colors alone. As a consequence, the GALEX NUV count slope point in
Fig. 11c has a somewhat larger error bar at the common wavelength of $\sim$230
nm. 

Given the above uncertainties in the GALEX and ERS UV galaxy count slopes, the
galaxy counts show the well known trend of a steepening of the best-fit power-law
slope at the bluer wavelengths, which is caused by a combination of the more
significant K-correction {\it and} the shape of the galaxy redshift distribution
at the selection wavelength. The galaxy count slope changes significantly from
$a$\cge 0.44 dex/mag at 0.23 \mum\ wavelength to $a$$\simeq$0.26 dex/mag at 1.55
\mum\ wavelength. Hence, the galaxy counts have a slope flatter than the
Euclidean value of $a$$\simeq$0.6 dex/mag. However, at wavelengths below 0.4
\mum\ the UV galaxy counts are steeper than the $a\le$0.4 dex/mag value required
for their sky-brightness integral to converge to a finite value, {\it if} they
were to continue with this power-law slope for AB\cge 27 mag. That is, the UV
galaxy counts will have to turn over with a slope flatter than $\le$0.4 dex/mag
for AB\cge 27 mag. In \S 7.3 we will confirm that they indeed do so. 

\subsection{The Panchromatic Galaxy Counts for 10\cle AB\cle 30 mag}

Fig. 12a--12j show the galaxy number counts in the 10-band ERS images compared to
a number of other panchromatic surveys at the bright end and at the faint end. At
the bright end, we added the counts from the Galaxy and Mass Assembly (GAMA)
survey (Driver \etal\ 2009) in NUV+ugrizYJH, which cover AB=10--21 mag (Xu \etal\
2005; Hill \etal\ 2010a). Details on the ground-based data obtained in the GAMA
survey are given by Driver \etal\ (2009) and Hill \etal\ (2010a). The GAMA survey
also uses matched aperture Kron magnitudes, which are comparable to our total
magnitudes used for the ERS. 

The GAMAs star-galaxy separation procedure is quite different than for the ERS,
since there is a significant potential for contamination by stars which dominate
the number counts at brighter magnitudes (Fig. 11a). The GAMA star-galaxy
separation therefore uses the ($r_{PSF}$--$r_{model}$) parameter to determine how
extended an object is, and also uses the well-defined stellar locus in the (g--i)
vs. (J--K) color-color diagram. Details of the star galaxy separation methods
used in the GAMA survey are given by Baldry \etal\ (2010). 

At the faint end in Fig. 12d--12j, we added the HUDF counts in BViz from Beckwith
\etal\ (2006), which cover AB=24--30 mag, and the HUDF counts from the Bouwens
\etal\ (2010b) data in the F105W, F125W, and F160W filters (hereafter \YJH), as
compiled by Yan \etal\ (2010), which cover AB=24--30 mag. In the UV, we added the
WHT U-band counts and the HDF-North and South F300W counts as compiled by
Metcalfe \etal\ (2001), the deep LBT U-band counts of Grazian \etal\ (2009), as
well as the deep HST STIS UV counts of Gardner \etal\ (2000). 

We note that the filter systems in the comparisons in each of Fig. 12a--12j are
{\it not} quite identical, although they were chosen in every case to be as close
as possible in wavelength (typically to within 0.1 dex). Hence where necessary,
small but appropriate AB-mag offsets or color transformations were made between
the filters and models used in each plot, following Metcalfe \etal\ (2001) or
Windhorst \etal\ (1991). These corrections generally contain some color
dependence. When applying them as a single AB-flux scale correction to a given
survey can introduce uncertainties of order 0.1 mag in the flux-scale used. Over
the entire AB=10--30 mag range shown in Fig. 12a--12j, such $\sim$0.1 mag offsets
would not be noticeable. However, for a detailed set of modeling in subsequent
work, more subtle flux and color dependent corrections may need to be applied.
This will be the subject of a future paper. 

The error bars in each of the counts in Fig. 12a--12j are Poisson, and therefore
do not include effects from cosmic variance. To increase the statistics where
necessary at the bright-end of each survey, bins were combined logarithmically
using the local slope of the galaxy counts, following the prescription of
Windhorst, van Heerde \& Katgert (1984a). At the faint-end, the counts are not
plotted in Fig. 12a--12j fainter than the flux level where these are deemed
complete. Based on Fig. 9 and 11a, this occurs in general where the counts turn
over significantly within the statistical error bars. Hence, in principle, the
panchromatic counts of Fig. 12a--12j should be comparable over the entire flux
range AB$\simeq$10--30 mag, except for the effects of cosmic variance. The latter
is small for the panchromatic $\sim$116 deg$^2$ GAMA survey, but it could be
important for the other panchromatic survey areas plotted, such as the ERS and
the HUDF which are both (disjoint) parts of GOODS-S, as well as the HDF-North (in
GOODS-North) plus the HDF-South and various other ground-based surveys in the
U-band. 

In the flux range where the panchromatic GAMA and HUDF counts in Fig. 12a--12j
overlap with the panchromatic ERS counts, the agreement is in general quite good,
except at the bright-end of the ERS, which suffers from cosmic variance, and from
the fact that the GOODS/ERS field was chosen to be devoid of objects much
brighter than AB=18 mag. As a consequence, the bright end of the ERS galaxy
counts also suffers somewhat from incompleteness. The same is true for the bright
end of the HUDF, which suffers similarly from cosmic variance and the avoidance
of objects much brighter than AB$\sim$21--22 mag when that field was selected in
early 2003 for the ultradeep HST survey. The HUDF is adjacent to, but does not
overlap with the ERS area (see Fig. 3). The HUDF galaxy counts in \BVizYsJH\ are
in good agreement with the ERS galaxy counts to the flux levels where the
panchromatic ERS counts are considered complete (AB$\simeq$26--27 mag in Fig.
11a). 

The good agreement between these various surveys also implies that the flux
scales of the panchromatic ERS counts are approximately correct. Had one of the
ERS filter zero-points been off significantly (\eg by more than 0.1--0.2 mag), we
would have noticed this as a significant offset between the counts. Note this
argument only holds for the counts at AB\cge 20 mag, as these cover the same
general area of sky in the GOODS-South field. Hence, cosmic variance likely
affects these counts similarly over such scales (\cle 30\%, Somerville \etal\
2004). 

\subsection{Modeling the Panchromatic Galaxy Counts for 10\cle AB\cle 30 mag}

Fig. 12a--12j over-plots simple galaxy evolution models by Driver \etal\ (1995,
1998), which have been updated for the current comparison with the ERS. Driver
\etal\ (1995, 1998) and Cohen \etal\ (2003) provide further details of these
models. In short, they use the best available local LF as a function of galaxy
type, and K-corrections from the Bruzual \& Charlot (2003) spectral evolution
models, also as a function of galaxy type. The local galaxy LF then gets
integrated with the WMAP-year 7 (Komatsu \etal 2010) cosmological volume element
to predict the number density of galaxies observed to the total-flux limit in
each ERS filter. These models also include internal reddening as a function of
galaxy type. 

These models have been updated with the best available local panchromatic LFs in
ugrizYJHK from the MGC survey by Hill \etal\ (2010b), and in the UV they use the
GALEX LFs of Robotham \& Driver (2010). These LFs will be updated in future work
with GAMA redshifts from the 3-year AAT survey that was concluded in 2010.
Robotham \etal\ (2010) provide further details on the GAMA redshift survey. This
GAMA data set is currently 99\% redshift-complete for $r_{Petro}$$\simeq$19.4
mag, and provides high-fidelity, matched-volume panchromatic LFs, which are
essential to interpret the higher redshift panchromatic ERS work. 

Evolution of the faint-end LF-slope is not yet included in these simple models,
but the panchromatic local GAMA LFs automatically includes the well-known
steepening of the local LF towards shorter rest-frame wavelengths, where the
local LF is dominated by the more rapidly star-forming late-type galaxies (see
Driver \etal\ 1998; Robotham \& Driver 2010; Hill \etal\ 2010b). The
epoch-dependent galaxy redshift distribution is also {\it not} an input to these
models, but can instead be inferred from the local LF as a function of type (\eg
Driver \etal\ 1996), the total constraints provided by the panchromatic galaxy
counts (fig. 12a--12j), the adopted K-corrections as a function of galaxy type,
and the adopted evolutionary model parameters. These simple evolution models used
are either a pure luminosity evolution model (PLE) with
e(z)~=~2.5~log[(1+z)$^{\beta}$] and PLE exponent $\beta$, {\it or} a number
density evolution (NDE) model with n(z)~=~n$_o$ [(1+z)$^{\gamma}$] and NDE
exponent $\gamma$, or a combination thereof. 

Based on the currently available, best panchromatic local GAMA LFs, Fig. 12a--12j
show for AB=10--30 mag the pure luminosity evolution model prediction with PLE
exponent $\beta$=1 (dotted lines), the number density evolution model prediction
with NDE exponent $\gamma$=1 (short-dashed lines), {\it or} a combination of both
models indicated by {\it both} parameters $\beta$=$\gamma$=1 (long-dashed
lines). The solid lines show in all cases the non-evolving model
($\beta$=$\gamma$=0). 

Over the entire flux range AB$\simeq$10--30 mag, Fig. 12a--12j show that {\it no
single PLE model} fits all the available galaxy count data, since there is not
enough volume at high redshift to make any PLE model fit the high observed
counts, especially in the visual to near-IR (Fig. 12d--12j). This means that one
has to invoke number density evolution (NDE) as well, \ie $\beta$\cge 1 and some
modest amount of luminosity evolution for $\gamma$\cge 0. The best models using
these PLE or PDE prescriptions are either the $\beta$$\simeq$1 or the
$\gamma$$\simeq$1 model. {\it Fig. 12a--12j also shows that no single combined
PLE+NDE model fits the panchromatic galaxy counts in all 10 bands simultaneously
over the entire observed range AB$\simeq$10--30 mag,} not even with the best
available panchromatic local LF from the GAMA survey (Hill \etal\ 2010b). 

In particular, Fig. 12d--12j show that at the faint-end of the galaxy counts in
YJH there exists a significant excess in observed object numbers compared to
these simple models. This could be due to either an additional population of
objects at high redshifts, or --- more likely --- due to lower luminosity objects
at lower redshifts, and/or because some essential aspects are missing in these
models. A detailed discussion of these possibilities is beyond the scope of the
current paper, but the reader is referred to \eg Bouwens \etal\ (2010), Yan
\etal\ (2010), and Wyithe \etal\ (2011) for a discussion of some of these
possibilities.

More realistic galaxy evolution models would be ones in which the faint-end
Schechter LF-slope $\alpha$ {\it also} evolves with redshift, as has been
suggested by \eg Ryan \etal\ (2007) and Khochfar \etal\ (2007). In particular,
the faint-end LF-slope $\alpha$ steepens significantly at higher redshifts, which
was suggested by \eg Yan \& Windhorst (2004), Ryan \etal\ (2007), Hathi \etal\
(2010), and Oesch \etal\ (2010). One would like to compare our panchromatic ERS
galaxy counts from 0.2--2 \mum\ over the entire observed range AB=10--30 mag
(Fig. 12a--12j) to state-of-the art hierarchical simulations that reach from z=8
to z=0, and properly fold star-formation and stellar population evolution models
with these $\Lambda$CDM simulations. Such models have been made, \eg by Croton
\etal\ (2006), de Lucia \& Blaizot (2007), and Nagamine \etal\ (2005a, 2005b),
and are discussed in further detail by \eg Kitzbichler \& White (2007), Monaco,
Fontanot, \& Taffoni (2007), and Nagamine (2006). It is difficult for such models
to predict reliable panchromatic galaxy counts from 0.2--2 \mum\ that cover the
entire observed range AB=10--30 mag, because such simulations take a very large
amount of computing time to cover the low redshift regime (z\cle 3), to cover the
entire galaxy LF, and to let and them run all the way to z=0. In addition, such
models need to include increasingly complex physics with cosmic time, such as
hierarchical merging and/or downsizing, feedback processes from supernovae and
AGN, and how these processes changed the faint-end and bright-end of the galaxy
LF, respectively, over cosmic time, as well as include the physics of dust. 
This will be the subject of a future paper that compares the observed 
panchromatic galaxy counts from 0.2--2 \mum\ over the entire flux range
AB=10--30 mag to available hierarchical models. 

For now, we conclude that the simple PLE+PDE models with the best available
observed panchromatic local LF from GAMA can explain each of the 10-band counts
for 10\cle AB\cle 30 mag {\it individually}, but no single one of the simple
PLE+NDE models can explain the counts over this entire flux range in all 10
filters {\it simultaneously.}

\section{Summary and Conclusions}

In this paper, we presented the new HST WFC3 Early Release Science (ERS)
observations in the GOODS-South field. We presented the scientific rationale of
this ERS survey and its data, the data taking plus data reduction procedures of
the panchromatic 10-band ERS mosaics. We described in detail the procedure of
generating object catalogs across the 10 different ERS filters.

The new WFC3 ERS data provide calibrated, drizzled mosaics in the UV filters
F225W, F275W, and F336W, as well as in the near-IR filters F098M (\Ys), F125W
(J), and F160W (H) in 1--2 HST orbits per filter. Together with the existing HST
Advanced Camera for Surveys (ACS) GOODS-South mosaics in the BViz filters, these
panchromatic 10-band ERS data cover 40--50 square arcmin at 0\arcspt 07--0\arcspt
15 FWHM resolution and 0\arcspt 090 Multidrizzled pixels to depths of
AB$\simeq$26.0--27.0 mag (5-$\sigma$) for point sources and AB$\simeq$25.5--26.5
mag for compact galaxies. The galaxy samples are \cge 50\% complete to these
limits, and \cge 99.97\% reliable. 

We also described the high quality star-galaxy separation made possible by the
superb resolution of HST/WFC3 and ACS over a factor of 10 in wavelength to
AB$\simeq$25--26 mag from the UV to the near-IR, respectively, using proper
motion and BViz colors to confirm the star-galaxy separation procedure, which has
an overall reliability of \cge 87\% for stars. Our main science results are: 

\sn 1) We present the resulting Galactic star counts and galaxy counts in 10
different filters. From the ERS data, these could be accurately determined from
AB$\simeq$19 mag to AB$\simeq$26 mag over a full factor of 10 in wavelength from
the mid-UV to the near-IR. 

\sn 2) Both the Galactic stars counts and the galaxy counts show mild but
significant trends of decreasing count slopes from the mid--UV to the near-IR
over a factor of 10 in wavelength: 

\sn 2a) The faint-end of the Galactic star count slope is remarkably flat at all
wavelengths from the Balmer break to the near-IR. The values of the best fit
power-law star count slope is in general of order 0.03--0.05 dex/mag above the
Balmer break, i.e. well below Euclidean value of 0.6 dex/mag. This shows that to
a depth of AB$\simeq$25--26 mag at all wavelengths in the range 0.2--2 \mum, the
ERS images are looking through most of the thick stellar disk of our Galaxy at
intermediate latitudes. 

\sn 2b) We measured the proper motion for 103 ERS stars, most of which have
colors consistent with the stellar locus in a (i--z) vs. (B--V) color-color
diagram. Their ensemble {\it average} proper motion is $\sim$3.06$\pm$0.67
m.a.s./year. This is a 4.6$\sigma$ measurement of the average proper motion of
faint stars with 16\cle AB\cle 26 mag. At the typical distance of these faint
Galactic K--M type stars ($\sim$ 10 kpc), the average ERS proper motion
correspond to a velocity of $\sim$145\kms, consistent with Galactic structure
models and previous stellar radial velocity and proper motion observations at
much brighter levels. 

\sn 3) The galaxy count slope changes significantly from $a$$\simeq$0.44 dex/mag
at 0.23 \mum\ wavelength to $a$$\simeq$0.26 dex/mag at 1.55 \mum\ wavelength,
showing the well known trend of a steepening of the best-fit power-law slope at
the bluer wavelengths. This is caused by a combination of the more significant
K-correction and the shape of the galaxy redshift distribution at the selection
wavelength. 

\sn 4) We combine the 10-band ERS counts with the panchromatic GAMA counts at the
bright end (10\cle AB\cle 20 mag), and with the ultradeep HUDF counts in
\BVizYsJH\ and other available HST UV counts at the faint end (24\cle AB\cle 30
mag). {\it The galaxy counts are now well measured over the entire flux range
10\cle AB\cle 30 mag over nearly a factor of 10 in wavelength. }

\sn 5) We fit simple galaxy evolution models to these panchromatic galaxy counts
over this entire flux range 10\cle AB\cle 30 mag, using the best available
10-band local galaxy luminosity functions (LFs) from the GAMA survey (Hill \etal
2010a), as well as simple prescriptions of luminosity and/or density evolution.
{\it While these models can explain each of the 10-band counts for 10\cle AB\cle
30 mag individually, no single one of the simple PLE+NDE models can explain the
counts over this entire flux range in all 10 filters simultaneously.} Any more
sophisticated models of galaxy assembly, including hierarchical merging and/or
downsizing with or without feedback, need to at least reproduce the overall
constraints provided by the current panchromatic galaxy counts for 10\cle AB\cle
30 mag over a factor of 10 in wavelength. 

These surveys also emphasize the need for extending this work to levels fainter
than AB\cge 30 mag and at longer wavelengths, which the James Webb Space
Telescope will do for wavelengths in the range 0.6--28 \mum\ after its launch in
2015. The unique abilities of WFC3 to do deep UV imaging will need to be explored
fully before that time.

\sn 6) We show examples of interesting panchromatic faint galaxy structure in
intermediate redshift objects, including early-type galaxies with nuclear
star-forming rings and bars, and/or weak AGN activity, and some objects of other
unusual appearance. 

The panchromatic ERS database is very rich in structural information at all
rest-frame wavelengths where young or older stars shine during the peak epoch in
the cosmic star-formation rate (z$\simeq$1--2), and constitutes a unique new HST
data base for the community to explore in the future. 

\acknowledgements 

This paper is based on Early Release Science observations made by the WFC3
Scientific Oversight Committee. We are grateful to the Director of the Space
Telescope Science Institute, Dr. Matt Mountain, for generously awarding
Director's Discretionary time for this program. We thank Drs. Neill Reid, Ken
Sembach, Ms. Tricia Royle, and the STScI OPUS staff for making it possible to
have the ERS data optimally scheduled. We also thank WFC3 IPT and the GSFC WFC3
Project for their very hard dedicated work since 1998 to make this wonderful
instrument work, and for their timely delivery of the essential hardware and
software to make the acquisition and reduction of the new WFC3 data possible. We
also thank the entire GOODS team, and in particular Drs. Norman Grogin and Mauro
Giavalisco for making the exquisite GOODS v2.0 mosaics available. We thank Drs.
Norman Grogin and Paul Schmidtke for helpful discussions. We thank the referee
for very thoughtful suggestions, which helped improve the presentation of this
paper, and for prompting a more detailed study of the ERS stellar proper motions
and colors, which we believe illustrates the multi-faceted promise of the ERS
data. 

Support for HST program 11359 was provided by NASA through grant GO-11359 from
the Space Telescope Science Institute, which is operated by the Association of
Universities for Research in Astronomy, Inc., under NASA contract NAS 5-26555.
RAW also acknowledges support from NASA JWST Interdisciplinary Scientist grant
NAG5-12460 from GSFC. HY is supported by the long-term fellowship program of the
Center for Cosmology and AstroParticle Physics (CCAPP) at The Ohio State
University.

We thank the STS-125 astronauts for risking their lives during the Shuttle
Servicing Missions SM4 to Hubble, and for successfully installing WFC3 into HST
in May 2009. We dedicate this paper to the memory of the STS-107 Columbia Shuttle
astronauts and of Dr. Rodger Doxsey, who during their lives contributed so much
to the Space Shuttle and the Hubble Space Telescope projects.

\ve 

\bn {\bf APPENDIX A. The Wavelength-Dependent UVIS Geometric Distortion
Correction}

\mn Given the issues related to the GDC discussed in \S 4.2.2, it was apparent
that some astrometric residuals remained --- at the level of \cle 2--3
Multidrizzled pixels --- that were a function of position across both UVIS
detectors, after the shifts and rotations had been removed. A {\it worst case} is
shown in the lower panel of Fig. 4a. These residuals were apparent between the
three UVIS filters, {\it as well as} relative to the GOODS astrometric frame. At
the time of the data reduction, the best available Instrument Distortion
Coefficients (IDCtab's; file t982101i\_uv\_idc.fits) --- which are used for the
GDC --- had been created using the WFC3 SMOV $F606W$ data {\it alone} (see this
$URL$\footnote{http://www.stsci.edu/hst/wfc3/idctab\_lbn}). This GDC did {\it not
yet} include any wavelength-dependent distortion terms, though these will be
measured during Cycle 18 or beyond (see this
$URL$\footnote{http://www.stsci.edu/hst/wfc3/STAN\_09\_2009}), and distributed to
the community by the WFC3 instrument team. Since our WFC3/UVIS data is all taken
at wavelengths below 3500 \AA, it is likely that the remaining astrometric
residuals are indicative of the absence of these color-dependent terms in the
currently existing GDC tables. The wavelength-dependent terms of the GDC come
from the different index of refraction of the glass optics and filters in the
UVIS channel of WFC3. This is in addition to the first--third order geometric
distortion coefficients, which results from the CCD tilt (which was necessary to
minimize internal reflections), and from other reflective optics, and are
therefore largely wavelength-independent.

The ultimate goal of the UVIS part of the WFC3/ERS is to create images in F225W,
F275W, and F336W filters {\it at the same pixel scale} (0\arcspt 030 per pixel)
and astrometric grid as the GOODS ACS images (Giavalisco \etal\ 2004) of the same
portion of the GOODS-South field. As discussed in \S 2, the eight ERS UVIS
pointings were designed to overlap with each other, in order to achieve the best
possible astrometric precision. After the extensive astrometric work described
above, it turned out that we had reached the limit of what was possible with the
present {\it wavelength-independent} geometric distortion solution. We were able
to align all UVIS images such that the {\it average} object was well aligned with
the GOODS B-band image. However, objects near the individual UVIS field edges
were not well aligned (lower panel of Fig. 4a). Fig. B.1 of the WFC3 Instrument
Handbook (Wong \etal\ 2010) also suggests that in these locations the geometric
distortion is the largest. This was most noticeable on the few stars in the
overlap regions between pointings, as Fig. 4a demonstrated.

Some of these stars also showed significant proper motion between the original
2004--2007 GOODS observations and the 2009 ERS observations, complicating their
use as astrometric fiducials (see \S 4.3.2 and 6). This is quite visible as
significant color offsets in some of the ERS bright star images in the very high
resolution versions of Fig. 5a--5b and Fig. 6c, which are available on the $URL$
quoted in the figure caption. In attempting to create a 0\arcspt 030 per pixel
UVIS mosaic, the stars in the overlap regions of the 8 UVIS pointings were
somewhat elongated (Fig. 4a), even though the exposures from the same UVIS orbit
were well aligned --- \ie having round stars and astrometric residuals to a
fraction of a pixel --- {\it and} even though each pointing was well matched to
the GOODS frame. In other words, even after correcting the WCS in the FITS
headers for {\it both} the relative {\it and} the absolute position deviations,
there still remained some astrometric residuals primarily at the field edges,
that we suspect are likely due to the uncertain wavelength dependence of the GDC
(Fig. 4b). 

Fig. 4b shows the astrometric residuals for four of our ERS pointings in the
F336W filter (visits 25, 26, 29, 30). Each frame within a visit was registered
using the MultiDrizzle scripts of Koekemoer \etal\ (2002), which CR-cleaned and
cross-correlated the images, to revise the flat-fielded (``FLT'') image WCS
header keywords, as described in \S 4.2.1. These images were then drizzled with
PIXFRAC=1.0 to a scale of 0\arcspt 030 per pixel and to a position angle of 0
degrees (final\_rot=0). The PA(V3) was 111 degrees. Catalogs were then created
with SExtractor (Bertin \& Arnouts 1996), which were in turn matched in
equatorial coordinates to a similar catalog made from the GOODS-South B-band
image (GOODS v2.0; Giavalisco \etal\ 2004; Grogin 2009, priv. comm.). Note that
{\it all four} visits in Fig. 4b show similar bimodal residuals, suggesting that
this is a systematic error. We suspect that this is due to the {\it
wavelength-dependent} geometric distortion in the UV, since the only distortion
solution available at the time of the data processing was measured in the F606W
filter. Since two distinct blobs of points occur in {\it similar locations in all
four panels of Fig. 4b,} the MultiDrizzle images of objects seen in {\it only
one} pointing --- which includes 80--90\% of the total ERS area (see Fig. 3 and
8a) --- {\it are} round at 0\arcspt 090 pixel sampling. However, the images of
the brighter objects in the overlap areas between mosaic pointings --- 10--20\%
of the total area --- are not always completely round, as shown in Fig. 4a.
(Visual inspection of the F225W and F275W mosaics showed similar trends for
bright stars as seen for F336W in Fig. 4a--4b --- confirming our conjecture that
this is due to the geometric distortion correction only having been calibrated in
the F606W filter. However, these trends are seen at lower S/N ratio than in F336W
--- since faint stars are red, see \S 4.3.2 and figures therein --- and so Fig.
4a--4b are not reproduced here for F225W and F275W.)

When Multi-drizzling the four exposures of each pointing, each individual object
is well aligned with itself. However, in the WFC3 field corners, an object can be
offset from the GOODS WCS by $\pm$0\arcspt 1--0\arcspt 2 (Fig. 4a--4b). In the
area of field overlap, an object in the ERS image corners or at a visit's field
edges can thus be misaligned with itself by $\sim$0\arcspt 1--0\arcspt 2, and so
be elongated (see Fig. 4a), but still be {\it on average} on top of the GOODS
v2.0 WCS. In other words, with the {\it current} wavelength-independent GDC and
MultiDrizzle, we cannot have it both ways for a given object in the image borders
of the ERS mosaic, until the wavelength-dependent GDC has been fully measured in
the UV, {\it and} a sufficient larger number of dither-points is available to
take full advantage of the full 0\arcspt 0.0395 pixel sampling of the WFC3 UVIS
images. This is at the moment not the case. Therefore, in this paper we use the
total-flux measurements from the UVIS images that were Multidrizzled at 0\arcspt
090/pixel, and use large enough apertures that the total flux measurements are
reliable. 

For the WFC3 IR Channel, the GDC has only been measured in the F160W filter (see
this $URL$\footnote{http://www.stsci.edu/hst/wfc3/STAN\_09\_2009}), and so the
GDC in the other two ERS IR filters F098M and F125W is correspondingly more
uncertain. However, because the WFC3 IR Channel GDC is much smaller to begin with
(see Wong \etal\ 2010) --- due to the IR Channel design --- and because of the
larger IR channel pixels, the wavelength dependence of the IR Channel GDC is much
less of an issue than for the WFC3 UVIS channel. As for the WFC3 UVIS channel,
the wavelength dependence of the IR Channel GDC will be measured in more detail
in Cycle 17 and beyond, and will be applied to a future version of the ERS data
release and discussed in subsequent ERS papers. 

\bn {\bf APPENDIX B. Other Applications of the ERS Data: }

\mn {\bf B.1. Interesting Classes of Objects in the Panchromatic ERS mosaics }

\sn Fig. 13 shows several panchromatic postage stamps of lower redshift
early-type galaxies in the ERS with nuclear star-forming rings, bars, or other
interesting nuclear structure. Each postage stamp is displayed at a slightly
different color stretch, that best brings out the UV nuclear structure. Rutkowski
\etal\ (2011) present examples of ellipticals at z$\simeq$0.3--1.5 with various
amounts of UV-excess flux. These galaxies are red and relatively featureless in
the optical, but show considerable emission in the UV, which is not all
point-like. Hathi \etal\ (2010) and Ryan \etal\ (2010) present ERS samples of
faint UV-dropouts and very high redshift red galaxies, respectively. Many more
such studies can be done with the ERS data. 

Fig. 14 shows panchromatic postage stamps of objects with interesting
morphological structure in the 10-band ERS color images of the GOODS-South field,
yielding high signal-to-noise ratio detections of galaxies resembling the main
cosmological parameters \Ho, $\Omega$, $\rho_{o}$, $w$, and $\Lambda$,
respectively. The panchromatic galaxy morphology and structure in the ERS images
is so rich that a patient investigator can find galaxies of nearly any appearance
in the epoch z$\simeq$1--3. This is because galaxy merging was in full swing in
this epoch.

\mn {\bf B.2. Examples of Uses and Future Potential of WFC3 IR Grism Data }

\sn Fig. 15a--15d displays the WFC3 G102 and G141 grism images of the same region
in the GOODS-South field (green box in Fig. 3), together with short finder
exposures in the WFC3 grism continuum filters F098M and F140W. For a detailed
discussion of the ACS G800L and WFC ERS G102 and G141 grism spectra, please see
Straughn \etal\ (2009, 2011). All brighter object WFC3 grism spectra in Fig.
15a--15d show a 0$^{th}$ order image to their left, displaced by about twice the
spectral image length, which should not be confused with real emission lines.
When one zooms in on the full-resolution version of this image (see the caption
of Fig. 15a--15d), many faint object spectra are visible to a continuum flux as
faint as AB$\sim$25--25.5 mag in two HST orbits, including many faint emission
line galaxies. 

Fig. 15e--15f show examples of some specific WFC3 G102 and G141 spectra for
emission line galaxies in the ERS at z=0.738 and z=0.610. For the emission line
galaxy at z=0.610, the available lower-resolution ACS G800L grism spectrum is
also shown in green. For further details, see Straughn \etal\ (2011). Fig. 15f
shows that WFC3 near-IR grisms G102 and G141 have a spectral resolution of
R$\simeq$210 and R$\simeq$130, respectively, while the ACS G800L has a lower
resolution of R$\simeq$100. It is clear from Fig. 15f that R\cge 200 is far
superior for faint emission-line (and by implication also for weak absorption
feature) detection than an R$\sim$100 grism. This should be a critical
consideration for the design of the JDEM/WFIRST missions that are now being
planned (Blandford \etal\ 2010) to carry out wide-field near-infrared imaging and
grism spectroscopy from space for this coming decade. 

In conclusion, the WFC3 broad-band IR filters and its G102 and G141 grisms
provide low-resolution imaging and faint object spectroscopy over the entire
0.80--1.70 \mum-range, unimpeded by the ground-based OH-forest. They have shown
the tremendous power of the WFC3 IR grism data above the Earth's atmosphere,
which are now used by a number of different WFC3 GO programs in HST Cycle 17 and
beyond. They also illustrate the discovery power of the new frontier that will be
opened up by the near--mid infrared 6.5 meter James Webb Space Telescope (JWST)
after its launch in 2015 (Gardner \etal\ 2006, Windhorst \etal\ 2006), and by
future wide-field near-infrared space missions such as JDEM/WFIRST and Euclid.

\baselineskip=10pt

\ve 

\begin{verbatim}
Table 1. Filters, Exposure Times, & Depths of the WFC3 ERS & GOODS-South ACS 
data.

============================================================================
Channel    Filter1 Filter2 Filter3 Filter4  GRISM  GRISM   TOTAL ORBITS
----------------------------------------------------------------------------

WFC3/UVIS   F225W   F275W   F336W            --             
Orbits      2       2       1                --             5x8 = 40 Direct
Depth (AB)  26.3    26.4    26.1             --             
nJy         110     100     132              --             

ACS/WFC     F435W   F606W   F775W   F850LP   G800L          
Orbits      3       3       4       9        80            15x19=285 Direct
Depth (AB)  27.9    28.1    27.5    27.3     27.0           2x20= 40 Grism
nJy         14      14      27      44       58             

WFC3/IR     F098M   F125W   F160W            G102   G141    
Orbits      2       2       2                2      2       6x10= 60 Direct
Depth (AB)  27.2    27.55   27.25            25.2   25.5    1x 4=  4 Grism
nJy         48      36      48               303    230     

----------------------------------------------------------------------------
TOTAL WFC3 ERS ORBITS             (Aug. 2009--Oct. 2009)          104
TOTAL ACS GOODS ORBITS inside ERS (Jul. 2002--Mar. 2005)          325 
TOTAL HST ORBITS                  (Jul. 2002--Oct. 2009)          429
----------------------------------------------------------------------------
\end{verbatim}

\sn Notes to Table 1:

\sn Note 1: The orbital integration times listed are those as achieved for the
WFC3 UVIS and IR ERS observations, as well as for the GOODS ACS v2.0
observations. For the ACS grism, 40 out of 200 orbits ACS G800L observations from
the HST Cycle 14 PEARS project 10530 (PI: S. Malhotra; Malhotra \etal\ 2005;
Pirzkal \etal\ 2004, 2005, 2009; Straughn \etal\ 2008, 2009) that reside inside
the ERS mosaic are listed (see Fig. 3). 

\sn Note 2: The listed depth is the 50\% completeness limit for 5-$\sigma$
detections in total SExtractor AB-magnitudes for typical compact objects at this
flux level (circular aperture with 0\arcspt 4 radius; 0.50 arcsec$^2$ aperture),
as derived from Fig. 9. 

\sn Note 3: For spectral continuum detection in the G102 and G141 grisms, these
flux limits are about 2.0--1.8 mag brighter, respectively. 

\sn Note 4: We note that the pre-flight WFC3 ETC sensitivity values were
0.15--0.2 mag more conservative in both the UVIS and the IR than the in-flight
values quoted here in Table 1, as also shown in Fig. 2. 

\sn Note 5: Due to the limited observing time available, and its poorer prism
performance, UV-prism observations in P280 were not taken as part of the ERS.

\ve 

\begin{verbatim} 

Table 2. ERS Filters, PSFs, Zero-points, Sky-background, and Effective Area

==============================================================================
HST-        ERS      Central   Filter    PSF   Zeropoint  Sky-back   Effective
Instrument  Filter   Lambda    FWHM      FWHM   (AB-mag@  ground     Area
/Mode                (mum)     (mum)     (")    1e-/sec)  mag/(")^2  (arcm^2)
------------------------------------------------------------------------------

WFC3/UVIS   F225W    0.2341    0.0547    0.092    24.06     25.46     53.2
WFC3/UVIS   F275W    0.2715    0.0481    0.087    24.14     25.64     55.3
WFC3/UVIS   F336W    0.3361    0.0554    0.080    24.64     24.82     51.6
ACS/WFC     F435W    0.4297    0.1038    0.080    25.673    23.66     72.4
ACS/WFC     F606W    0.5907    0.2342    0.074    26.486    22.86     79.2
ACS/WFC     F775W    0.7764    0.1528    0.077    25.654    22.64     79.3
ACS/WFC     F850LP   0.9445    0.1229    0.088    24.862    22.58     80.3
WFC3/IR     F098M    0.9829    0.1695    0.129    25.68     22.61     44.8
WFC3/IR     F125W    1.2459    0.3015    0.136    26.25     22.53     44.7
WFC3/IR     F160W    1.5405    0.2879    0.150    25.96     22.30     44.7

------------------------------------------------------------------------------
\end{verbatim}

\n Notes to Table 2:

\sn Note 1: The panchromatic PSF FWHM was measured from ERS stars as in Fig.
7a--7b, and includes the contribution from the OTA and its wavefront errors, the
specific instrument pixel sampling or Modulation Transfer Function (MTF).

\sn Note 2: The WFC3 zero-points are in AB magnitudes for 1.0 \emin/sec from this
$URL$\footnote{http://www.stsci.edu/hst/wfc3/phot\_zp\_lbn}.

\sn Note 3: The GOODS BViz sky-background values are from Hathi \etal\ (2008).

\sn Note 4: The effective areas used in this paper are in units of arcminutes
squared for the effective number of WFC3 or ACS tiles available and used.

\sn Note 5: The GOODS v2.0 BViz data release is from this $URL$\footnote{
http://archive.stsci.edu/pub/hlsp/goods/v2/h\_goods\_v2.0\_rdm.html}. 

\sn Note 6: The panchromatic effective ERS area was derived from Fig. 8a--8c for
the WFC3 mosaics and the relevant GOODS tiles, and indicates the total area over
which at least half of the total ERS exposure time was available. (Fig. 8 shows
that \cge 80\% of the total ERS exposure time was available for 50 square arcmin
in the WFC3 UVIS filters, and for 40 square arcmin in the WFC3 IR filters).

\ve 

\vspace*{-1.0cm}
\n\cl{
\includegraphics[width=1.0\linewidth]{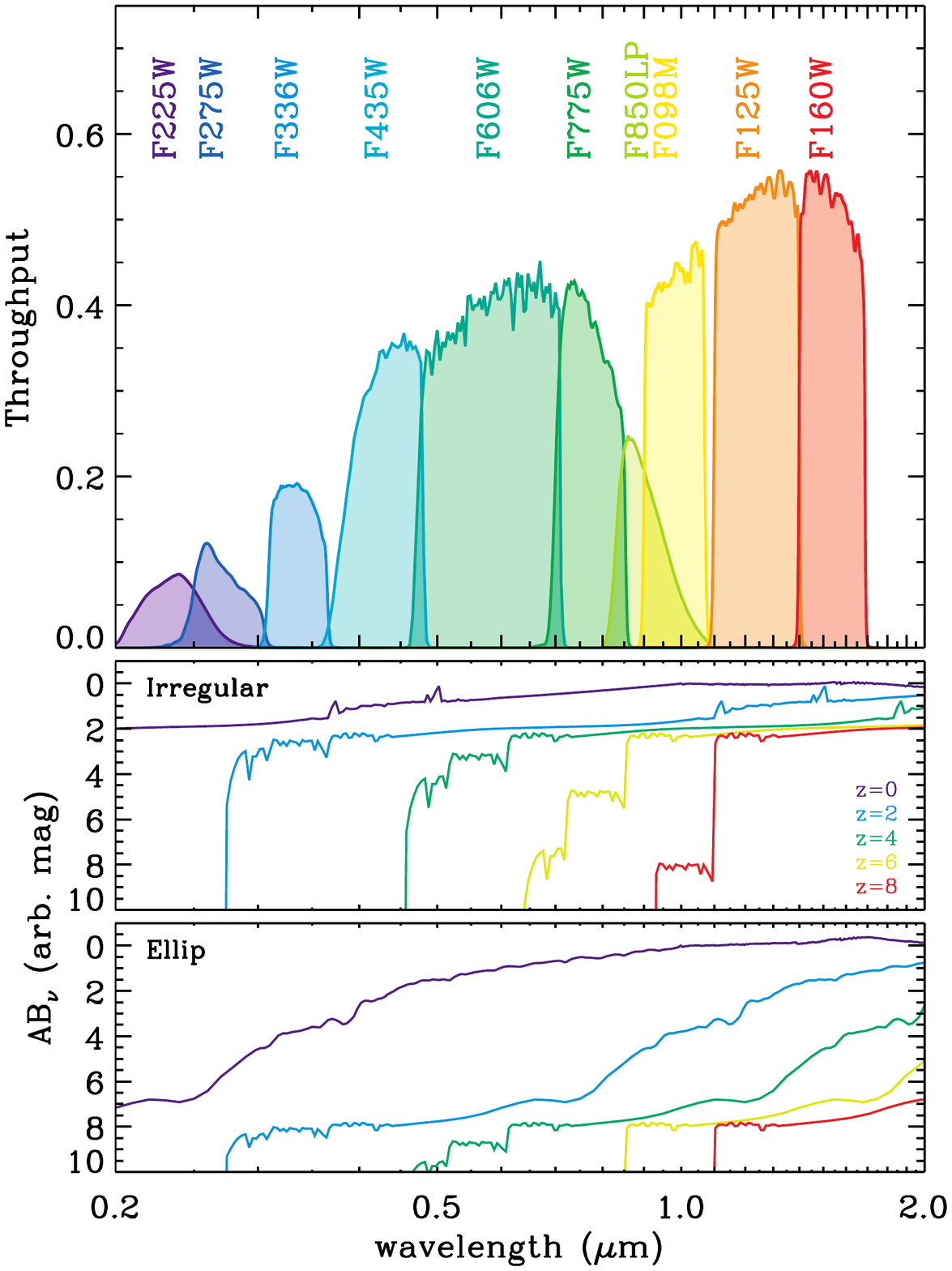}
}

\vspace*{-2.0cm} 
\bn{\textbf Fig. 1a. (Top panel)\ The full panchromatic HST/WFC3 and ACS filter
set used in the ERS imaging of the GOODS-South field. Plotted is the overall
system throughput, or the HST Optical Telescope Assembly throughput OTA $\times$
filter transmission T $\times$ detector QE. (Middle and bottom panels)\ Spectral
energy distributions for two single burst model galaxies with ages of 0.1 and 1
Gyr and redshifts of z=0, 2, 4, 6, 8, are shown as black, blue, green, yellow,
and red curves, respectively. IGM absorption shortward of Lyman-$\alpha$ was
applied following Haardt \& Madau (1996). The UVIS filters sample the
Lyman-$\alpha$ forest and Lyman continuum breaks, while the near-IR filters probe
the 4000 \AA\ and Balmer breaks at these redshifts. Additional photometry is
available from ground-based VLT K-band imaging, and Spitzer/IRAC imaging at 3.5,
4.5, 5.6 and 8.5 micron. }

\ve 

\n\cl{
\includegraphics[width=0.80\linewidth]{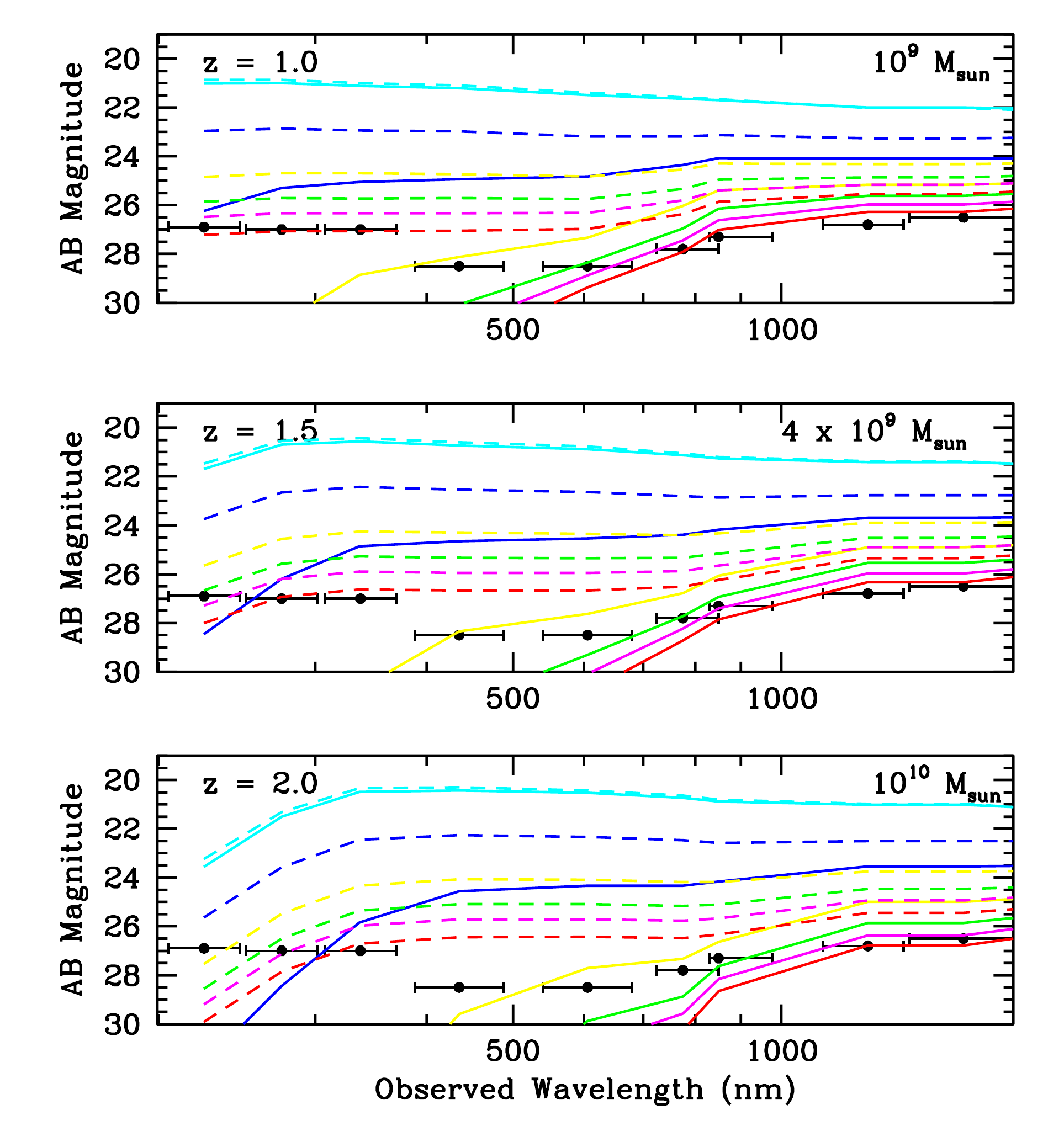}
}

\vspace*{-0.0cm}
\bn{\textbf Fig. 1b.\ Mass sensitivity of the WFC3 ERS filters used. The three
panels show the depth of the WFC3 ERS images as horizontal bars, compared to
evolving galaxy models of different masses at three redshifts. The top panel is
for z=1 and a stellar mass of 10$^9$ \Msun. The solid lines represent nearly
instantaneous bursts, and the dashed lines are for declining star-formation with
a 1 Gyr e-folding time. The colors refer to ages of 10, 100, 500, 1000, 1400,
2000, and 3000 Myrs from cyan to red. The middle panel shows similar models at
z=1.5 for 4 $\times$ 10$^9$ \Msun, while the bottom panel is for z=2 and masses
of 10$^{10}$ \Msun. At z=1, the WFC3 ERS reaches masses of 0.02 M$^*$ for typical
star-formation histories. Both the flux scale and the wavelength scale are
logarithmic, illustrating WFC3's exquisite panchromatic coverage and sensitivity
in probing the stellar masses of galaxies through its IR channel, and the
star-formation in galaxies through its UVIS channel. }

\ve 

\n\cl{
\includegraphics[width=0.50\linewidth,angle=-0]{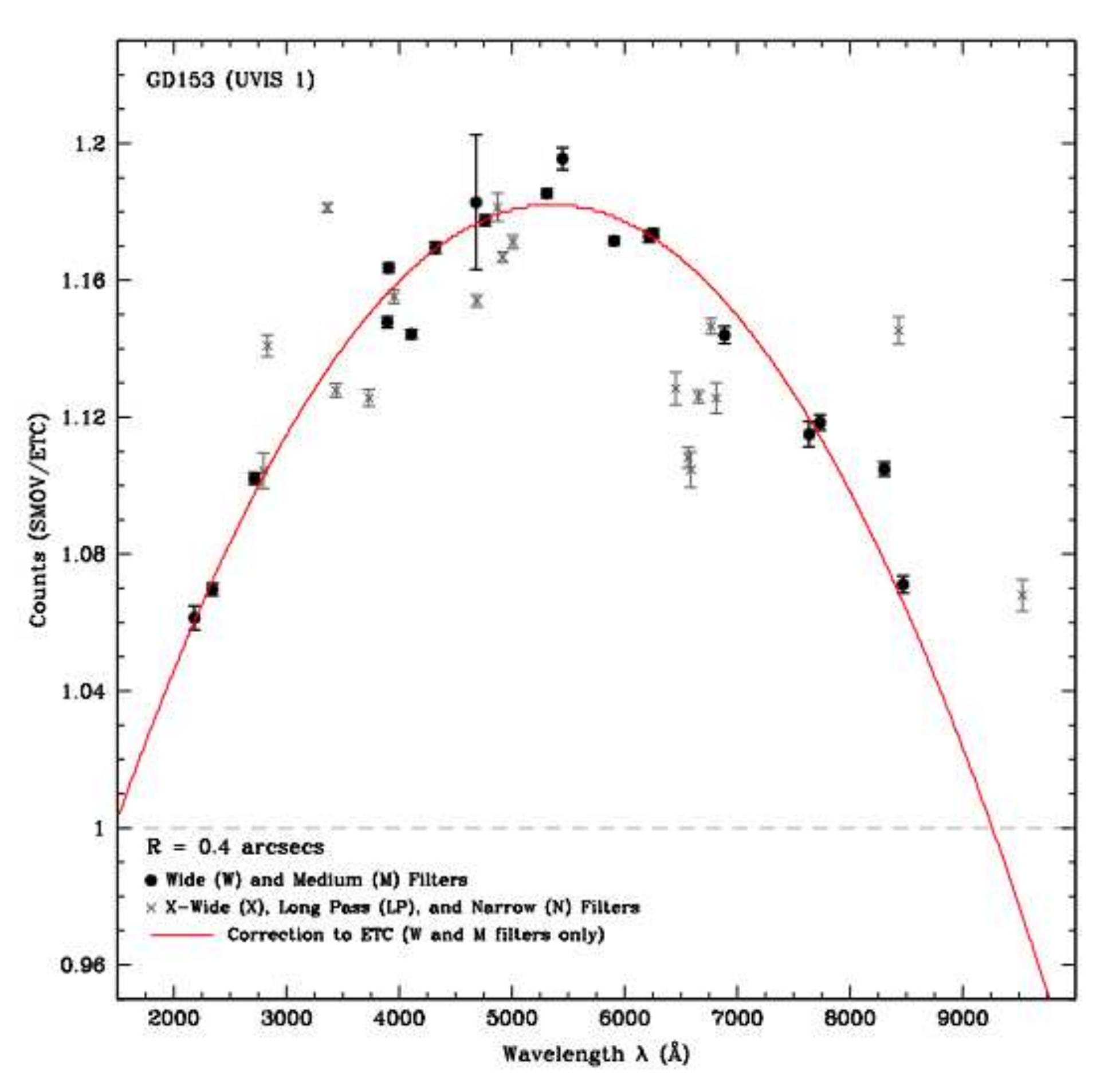}
\includegraphics[width=0.50\linewidth,angle=-0]{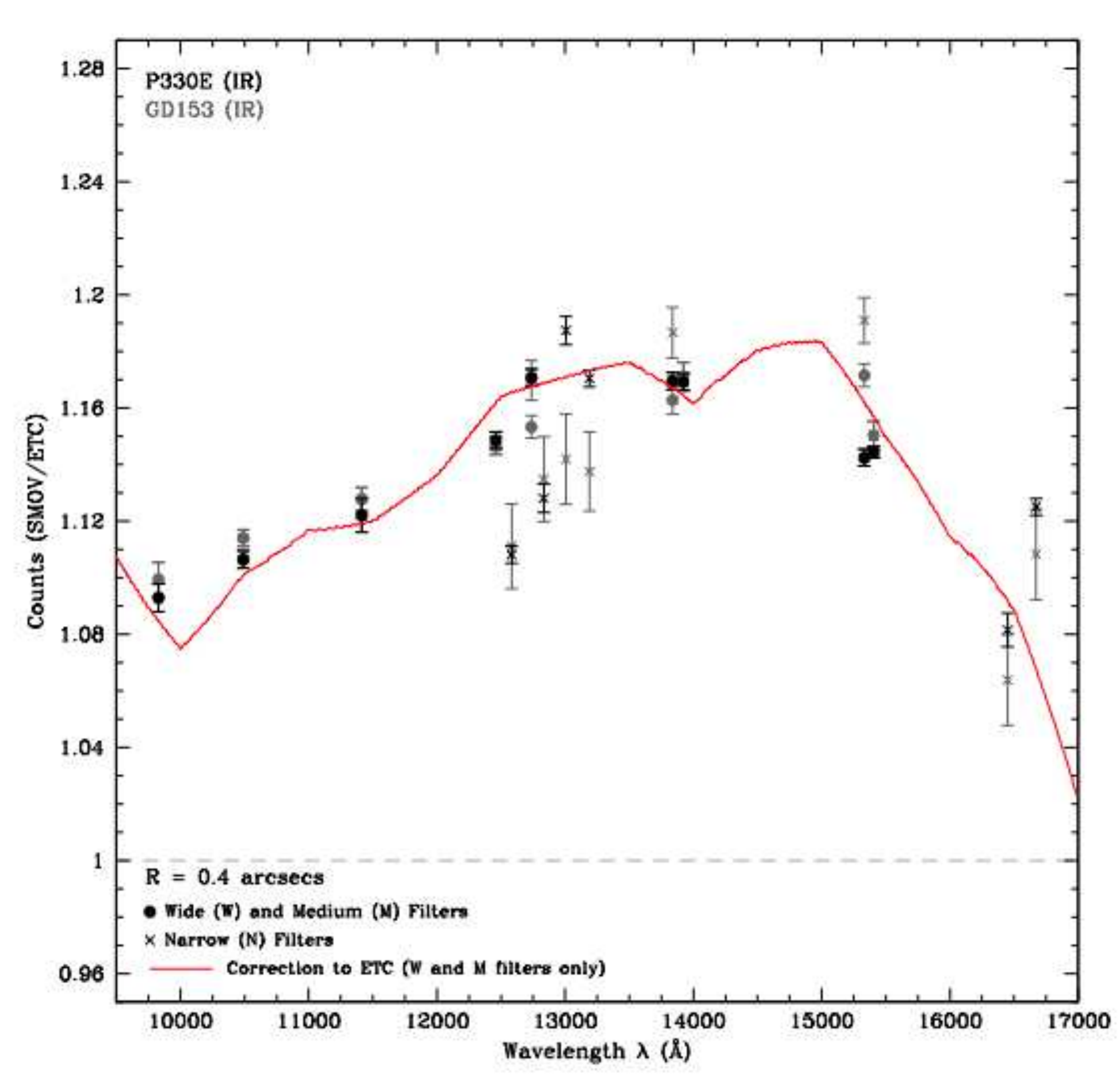}
}

\vspace*{-0.0cm}
\bn{\textbf Fig. 2.\ Ratio of on-orbit count rate to that obtained in
ground-based thermal vacuum tests for various WFC3 filters in the UVIS (left
panel) and IR (right panel), respectively. On-orbit rates are significantly
higher than in ground tests. See text for details. }

\ve 

\vspace*{-0.0cm}
\n\cl{
\includegraphics[width=0.95\linewidth]{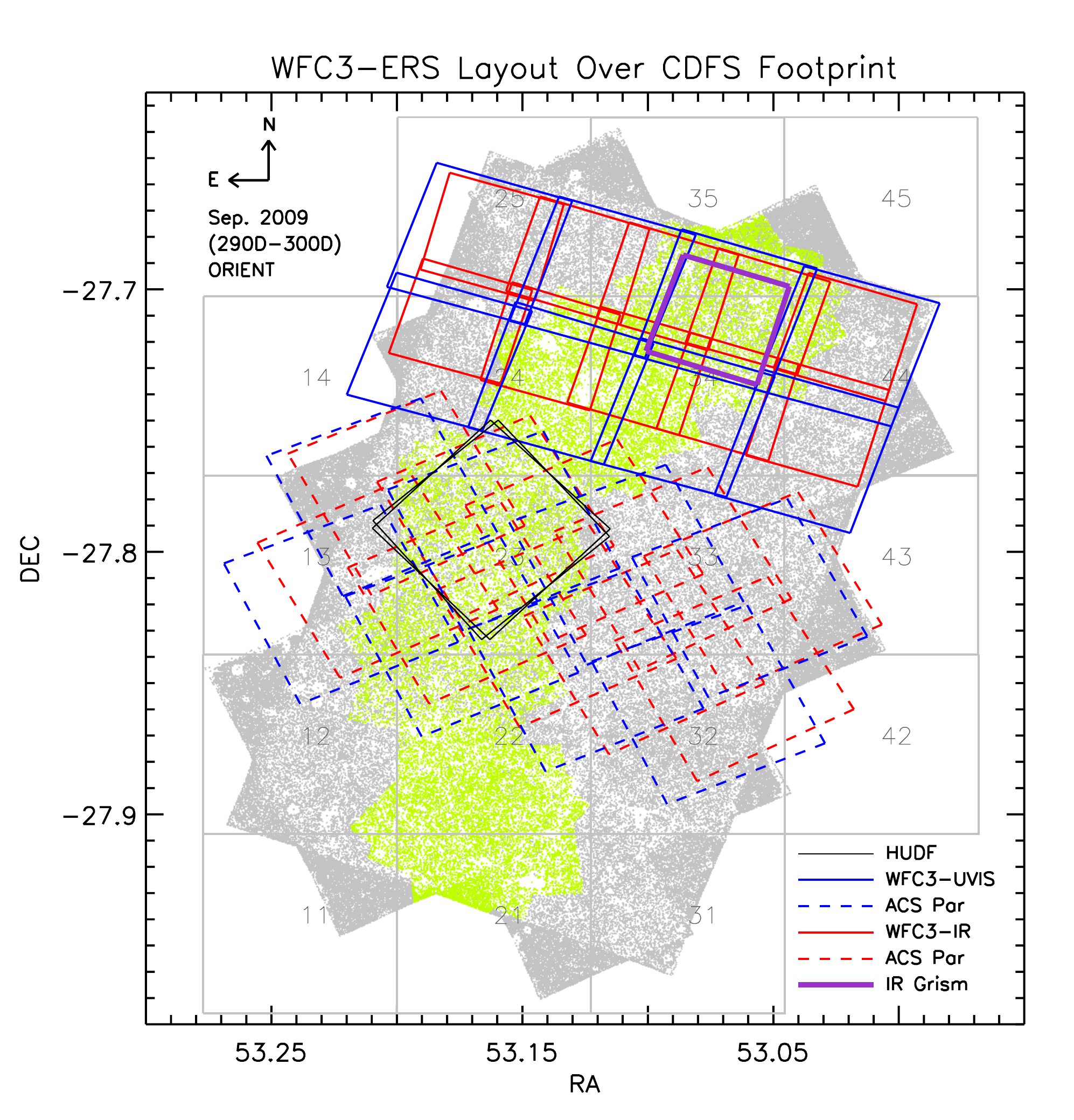}
}

\vspace*{+0.3cm} 
\n{\textbf Fig. 3.\ Layout of the GOODS-South field and its WFC3 ERS visits
footprint. The light-grey area indicates the part covered by GOODS v2.0 data, and
the numbered grey tiles are those of the GOODS-South survey. The green tiles
indicate the GOODS-South area with ACS G800L grism data from the PEARS survey.
The 4$\times$2 ERS UVIS mosaic is superposed in blue, and the 5 $\times$2 ERS IR
mosaic is superposed in red. The UVIS fields are numbered from left to right,
with UVIS fields 1--4 in the top row and UVIS fields 5--8 in the bottom row. The
IR fields are numbered from left to right, with IR fields 1--5 in the top row and
UVIS fields 6--10 in the bottom row. The dashed blue and red boxes indicate the
location of the ACS parallels to the ERS WFC3 images (Finkelstein \etal\ 2010).
The ERS IR G102 and G141 grism field is shown by the purple box (see Fig.
15a--15d), and overlaps with the northern most of the PEARS ACS G800L grism
fields in the GOODS-South region. The black boxes show the ACS Hubble Ultra Deep
Field pointings in the GOODS-South field. The ERS program was designed to image
the Northern $\sim$30\% of the GOODS-South field in six new WFC3 filters: F225W,
F275W, and F336W in the UVIS channel, and F098M, F125W, and F160W in its IR
channel. Further details are given in Tables 1--2. The exact pointings
coordinates and observing parameters for all pointings in HST ERS program 11359
can be obtained from this
$URL$\footnote{www.stsci.edu/observing/phase2-public/11359.pro}.

\ve 

\n\cl{
\includegraphics[width=0.50\linewidth,angle=-0]{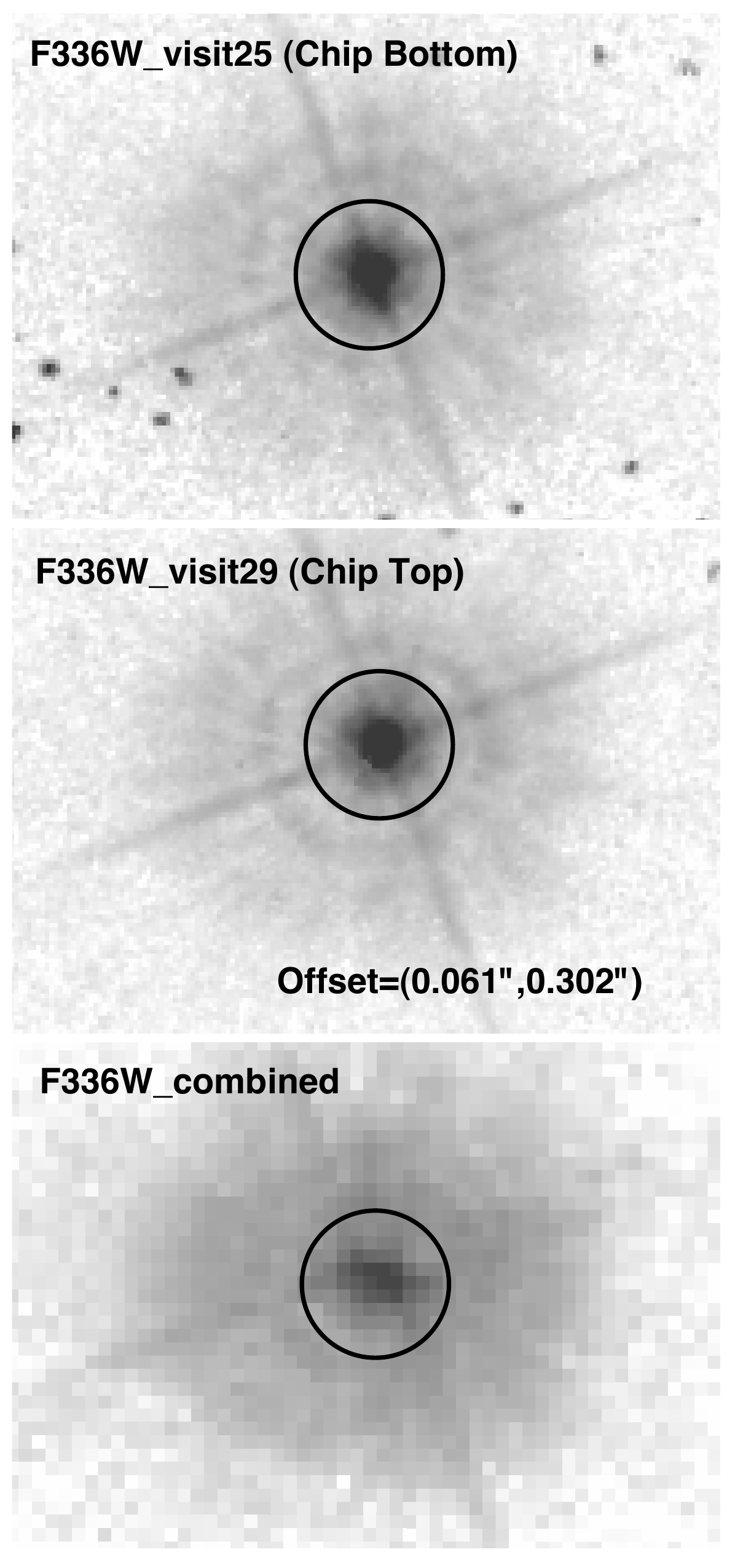}
}

\vspace*{-0.0cm}
\bn{\textbf Fig. 4a.\ An ERS star imaged in the overlap region between UVIS
visit 25 (upper panel) and visit 29 (middle panel) in the F336W filter. While the
star is properly processed by MultiDrizzle in these {\it individual} visits (\ie
its images are round in the upper and middle panels), it is clearly displaced by
(0\arcspt 061, 0\arcspt 301) in its WCS location between these two visits, as
shown in the combined image (bottom panel). This is due to the
wavelength-dependent geometric distortion correction (GDC). The GDC causes this
star --- and other objects in the mosaic {\it border} regions --- to be elongated
by approximately this amount when a MultiDrizzle is done of {\it all available}
pointings, as can be seen in the bottom panel. A full wavelength-dependent GDC
will make the images in the border regions round as well across the full
MultiDrizzle mosaic. A proper measurement and continued monitoring of the
wavelength-dependent GDC is scheduled for Cycle 18 and beyond. }

\ve 

\n\cl{
\includegraphics[width=1.10\linewidth,angle=0]{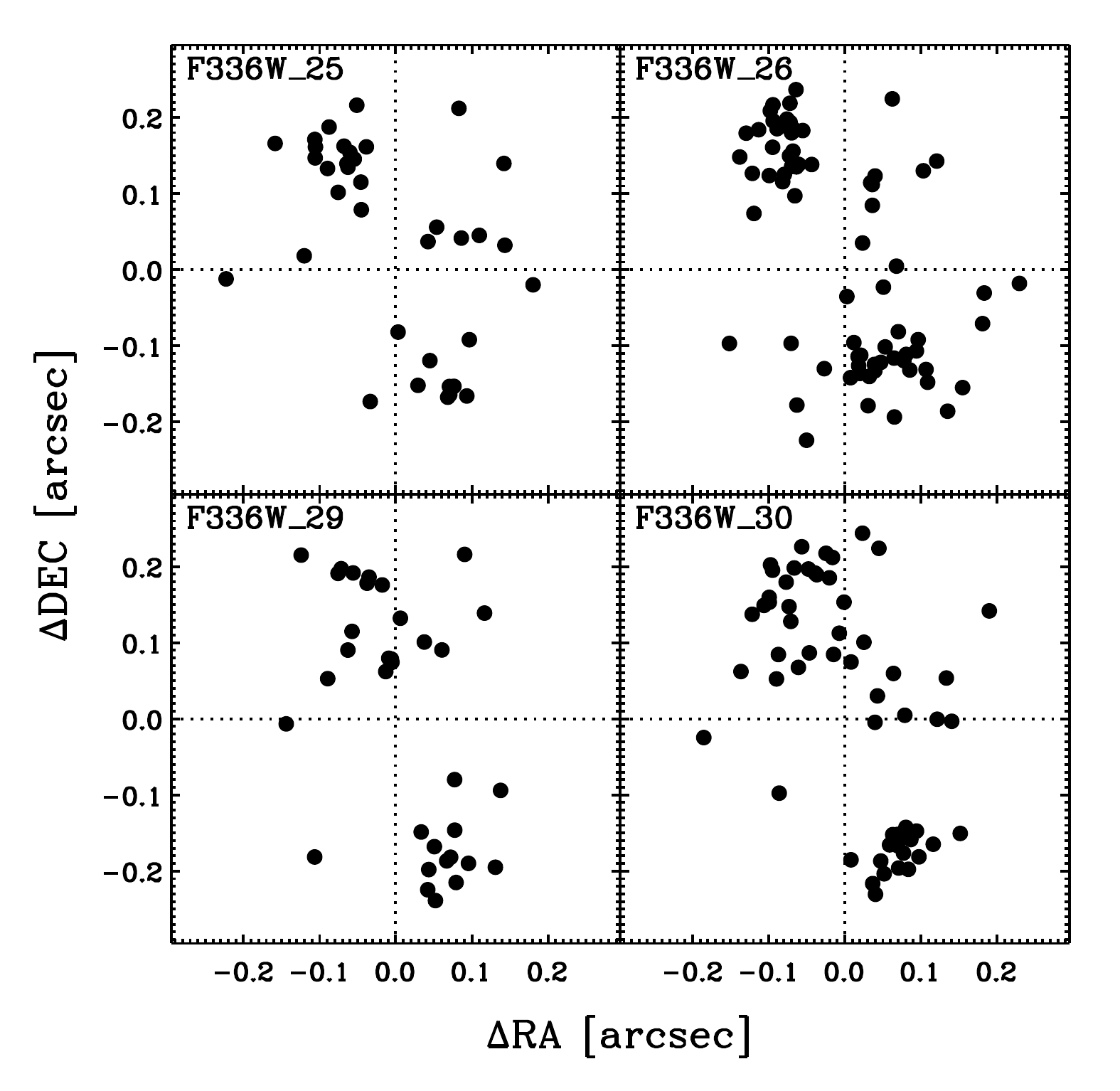}
}

\vspace*{-0.0cm}
\bn{\textbf Fig. 4b.\ The astrometric residuals for four of our ERS pointings in
the F336W filter, defined as the differences between visits 25, 26, 29, or 30 and
the image mosaic that was Multidrizzled using these four visits. Note that {\it
all four} visits show similar bimodal residuals, suggesting that this is a
systematic error. We suspect that this is due to the wavelength-dependent
geometric distortion in the UV, since the only currently available distortion
solution was measured in the F606W filter. Since two distinct groups of points
occur in {\it similar locations in all four panels,} the MultiDrizzle images of
objects seen in only one pointing --- which includes 80--90\% of the total ERS
area (see Fig. 3 and 8a) --- {\it are} round at 0\arcspt 090 pixel sampling.
However, the images of the brighter objects in the overlap areas between mosaic
pointings --- 10--20\% of the total area --- are not always round, as can be seen
in Fig. 4a. (We confirmed this by visual inspection of the F225W and F275W
mosaics, where this trend is seen at lower S/N ratio (see Appendix A), since
faint stars are red (see \S 4.3.2).) }

\ve 

\vspace*{-0.0cm}
\hspace*{+0.0cm} 
\n\cl{
\includegraphics[width=1.00\linewidth]{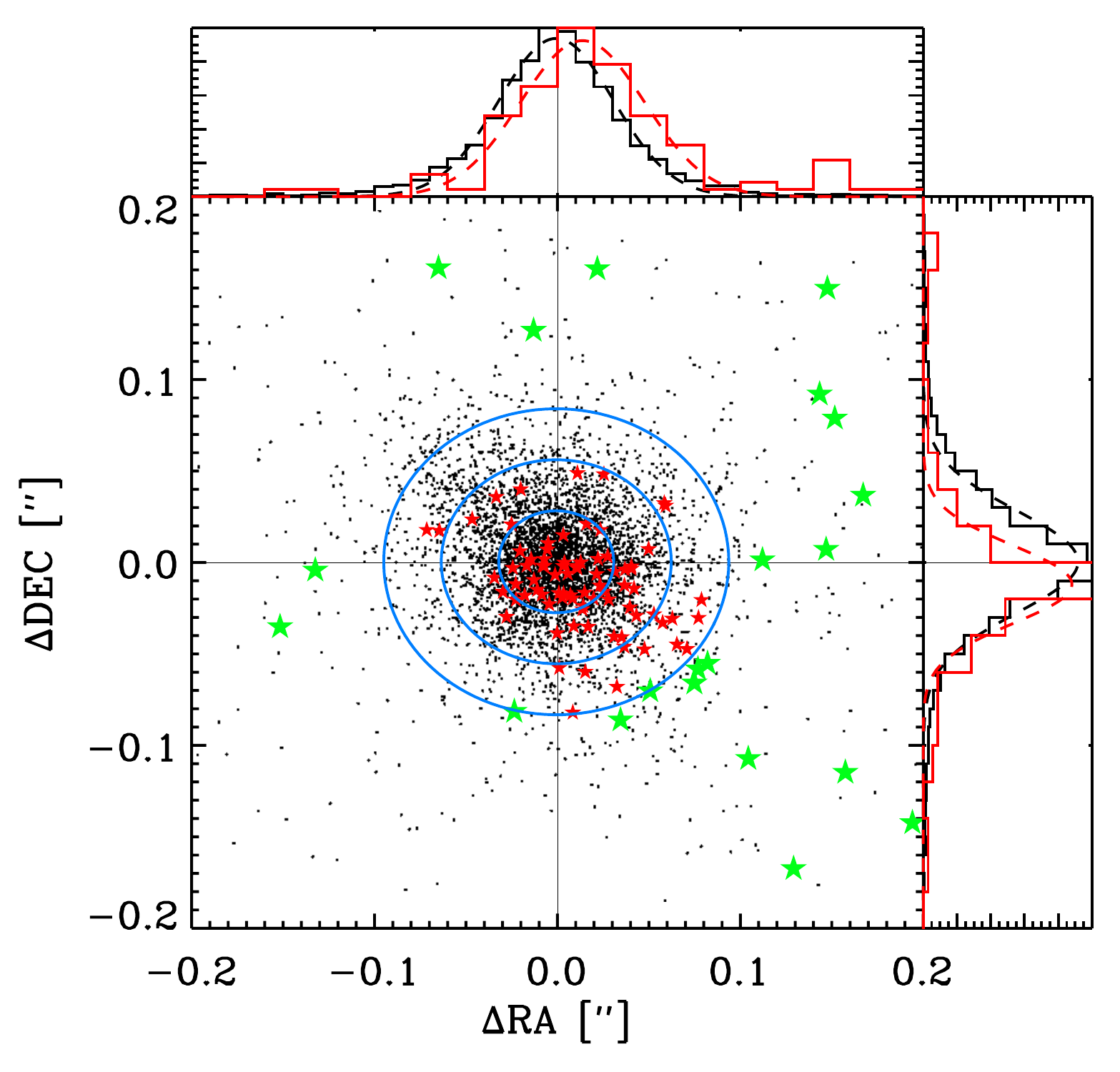}
}

\vspace*{-0.2cm} 
\bn{\textbf Fig. 4c.\ The measured residual astrometric offsets in (RA, DEC) for
all 4614 ERS objects in our WFC3 IR H-band object catalogs relative to our object
catalogs based on the GOODS ACS/WFC v2.0 z'-band images. The 4511 ERS objects
classified as galaxies are shown as small black dots, and the 103 ERS objects
classified as stellar are shown as red or green asterisks. Histograms normalized
to unity are also shown in each coordinate: black histograms indicate ERS
galaxies and red histograms ERS stellar objects. Best fit Gaussians are also
shown for each of these histograms. Objects classified as ERS galaxies have a
nearly Gaussian error distribution centered around ($\Delta$RA, $\Delta$DEC) =
(0, 0), while objects classified as ERS stars have {\it on average (red
asterisks)} --- {\it and} in a significant number of individual cases (green
asterisks) --- significant evidence for proper motion at the \cge 3$\sigma$
level. Further details are given in \S 4.3.2 \& 5.6. }

\ve 

\vspace*{-1.6cm}
\n\cl{
\includegraphics[width=1.25\linewidth,angle=90]{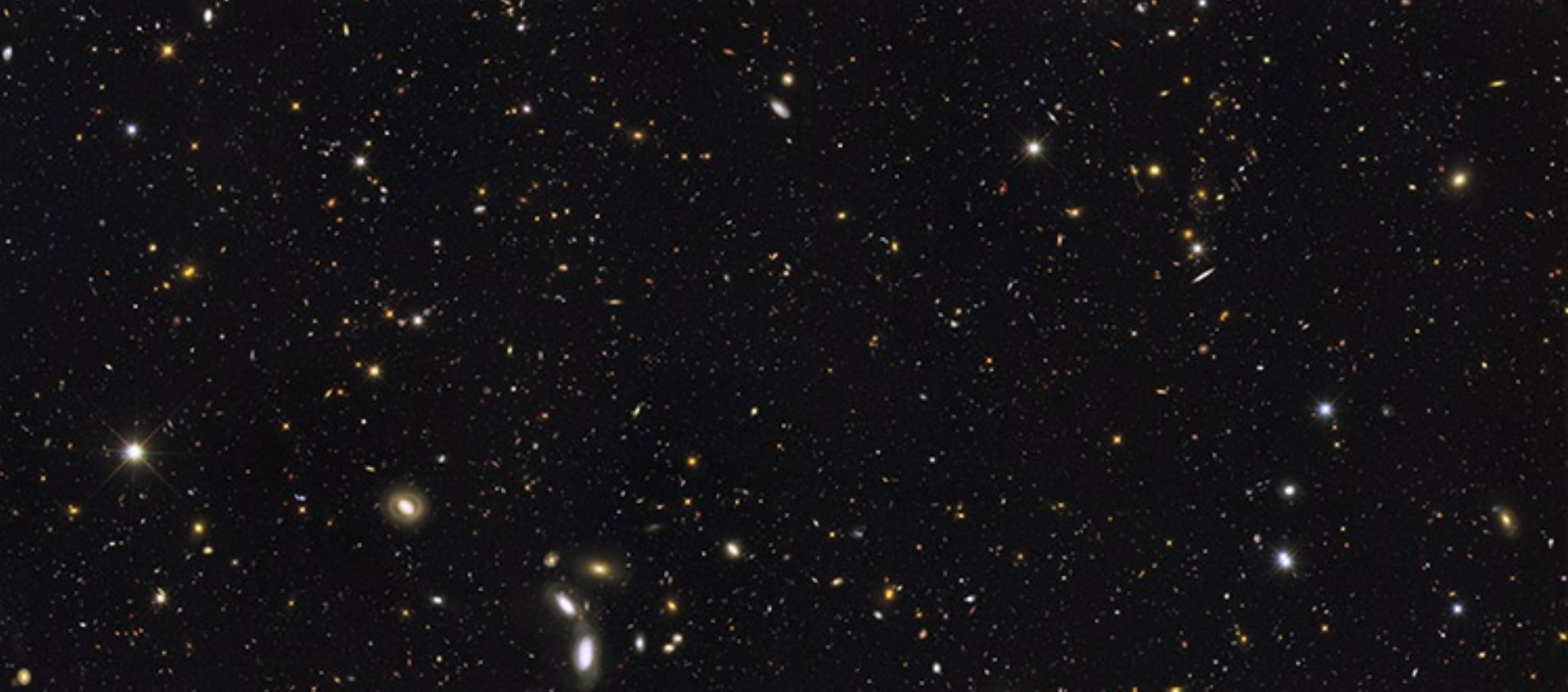}
}

\vspace*{+0.1cm}
\n{\textbf Fig. 5a.\ Panchromatic 10-band color image of the entire ERS mosaic in
the GOODS-South field. [This image is only displayed in the electronic version of
this journal paper.] Shown are the common cross sections between the 4$\times$2
ERS UVIS mosaics, the GOODS v2.0 ACS BViz, and the 5$\times$2 ERS IR mosaics. The
total image shown is 6500 $\times$ 3000 pixels, or 9\arcmpt 75 $\times$ 4\arcmpt
5. We used color weighting of the 10 ERS filters as described in the text, and
log(log) stretch. [For best display, please zoom in on the full-resolution
version of this image, which is available on this $URL$
\footnote{http://www.asu.edu/clas/hst/www/wfc3ers/ERS2\_loglog.tif}]. Note that
in these color images the reddest objects are {\it not} necessarily at z$>$7, due
to the way the colors were combined. }

\ve 

\n\cl{
\includegraphics[width=0.49\linewidth,angle=0]{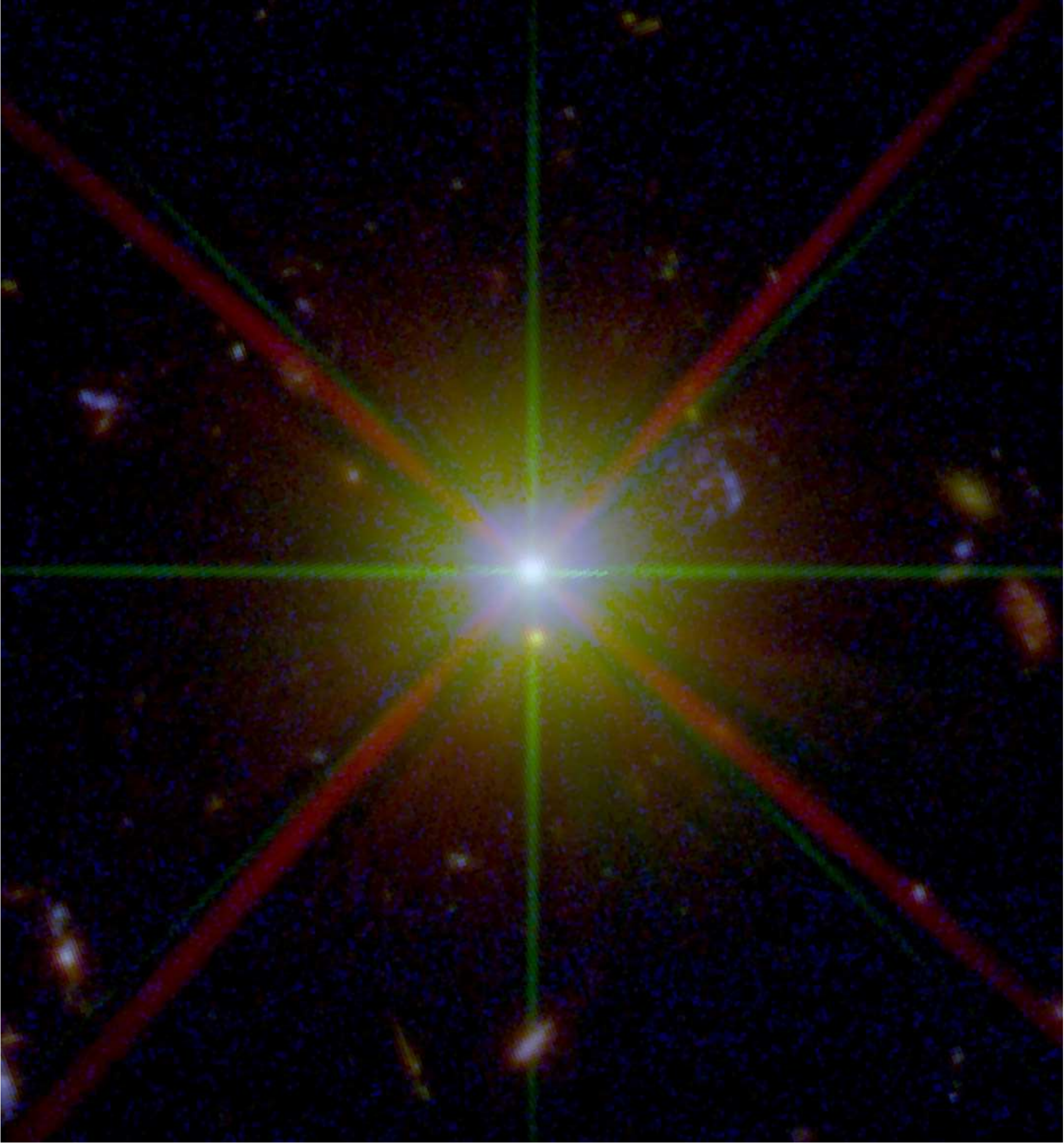}
\includegraphics[width=0.475\linewidth,angle=0]{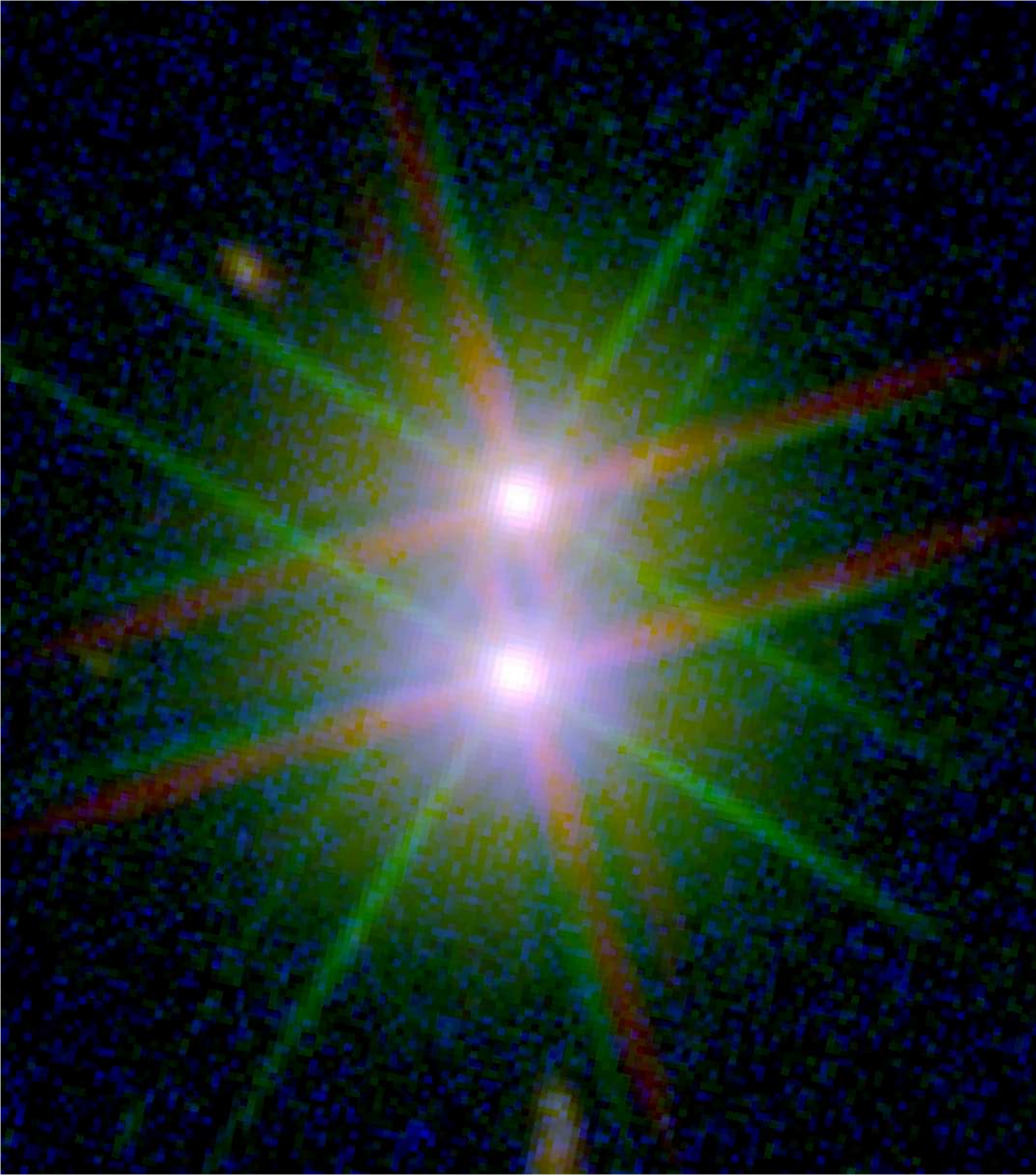}
}

\vspace*{-0.0cm}
\bn{\textbf Fig. 6a.\ Log(log) color reproduction of a bright but unsaturated
star image in the 10-band ERS color images in the GOODS-South field, using the
same color balance prescription as in Fig. 5a--5b. Fig. 6b.\ Same for a
``double'' star. These images give a qualitative impression of the significant
dynamic range in both intensity and wavelength that is present in these
panchromatic ERS images. }

\ve 

\vspace*{-0.8cm}
\n\cl{
\includegraphics[width=0.23\linewidth,angle=-90]{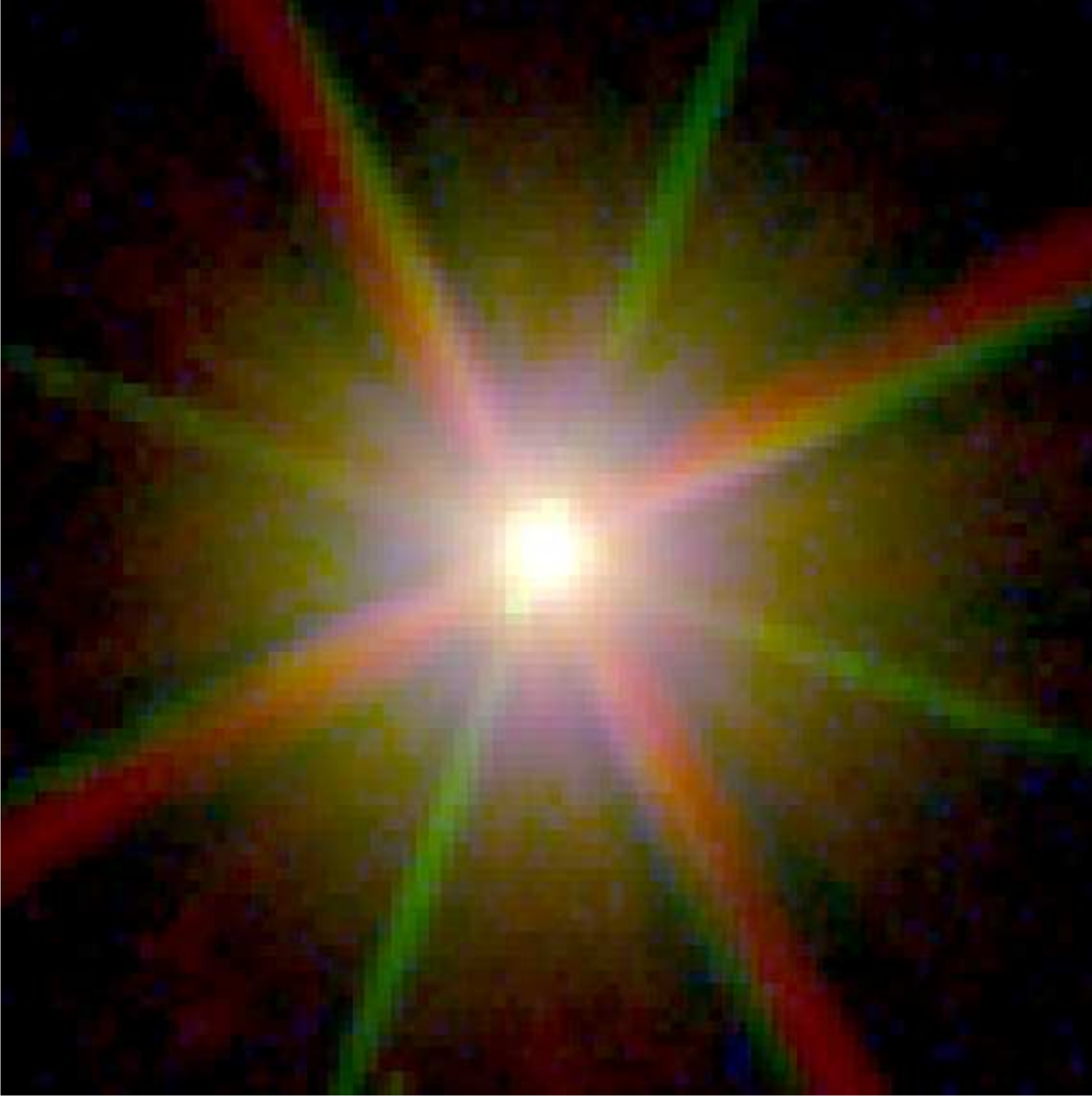}
\includegraphics[width=0.23\linewidth,angle=-90]{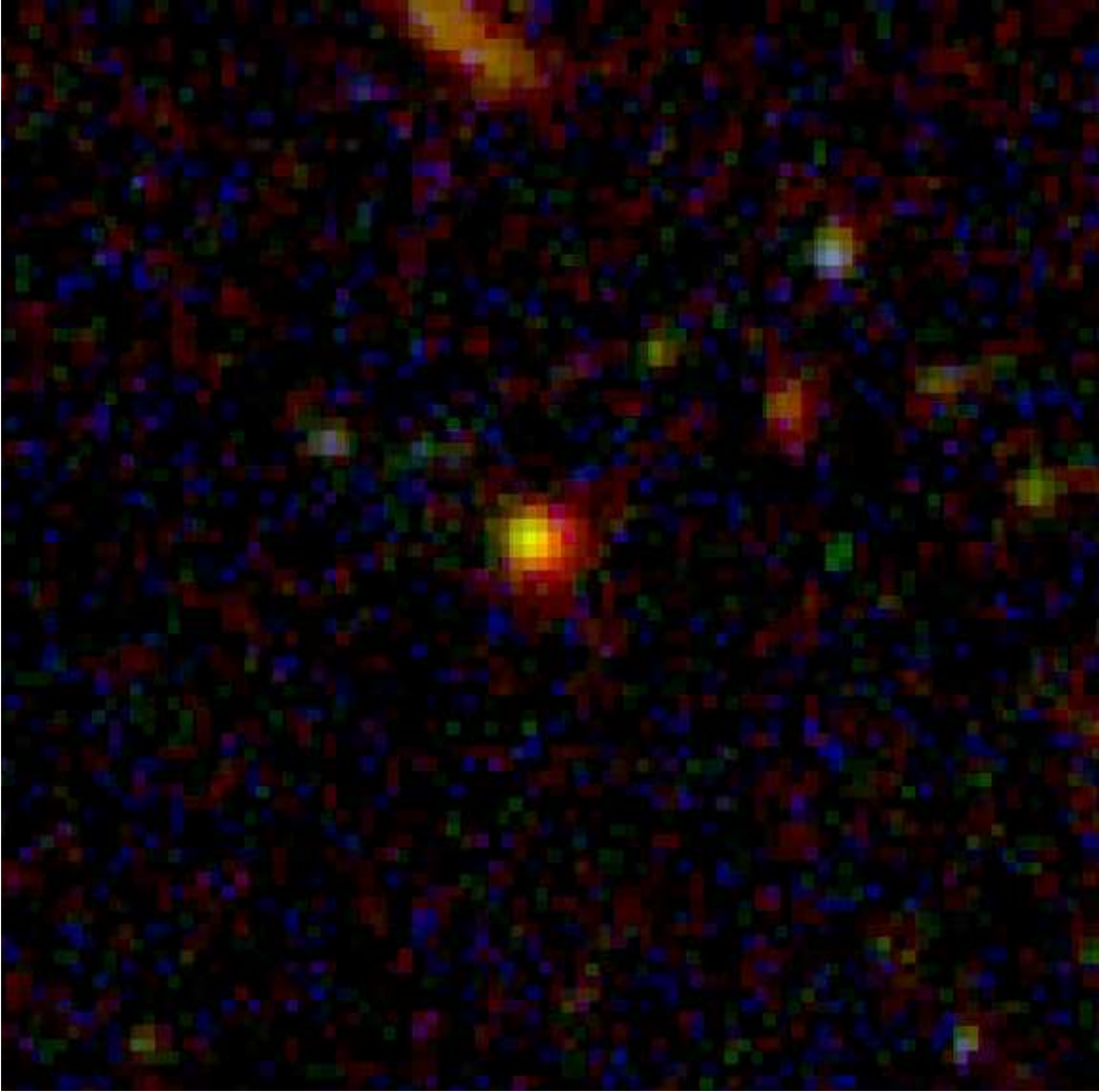}
\includegraphics[width=0.23\linewidth,angle=-90]{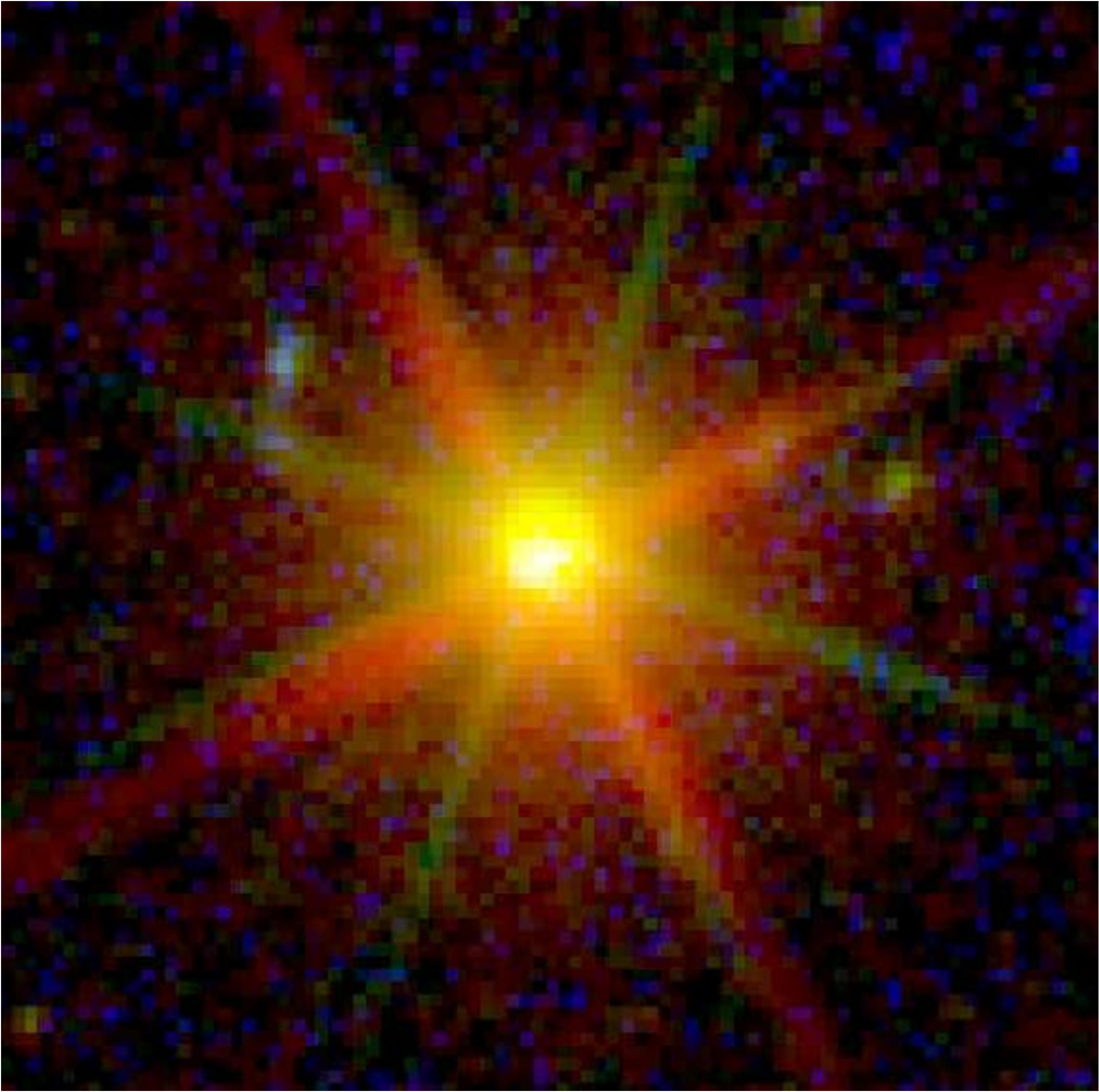}
\includegraphics[width=0.23\linewidth,angle=-90]{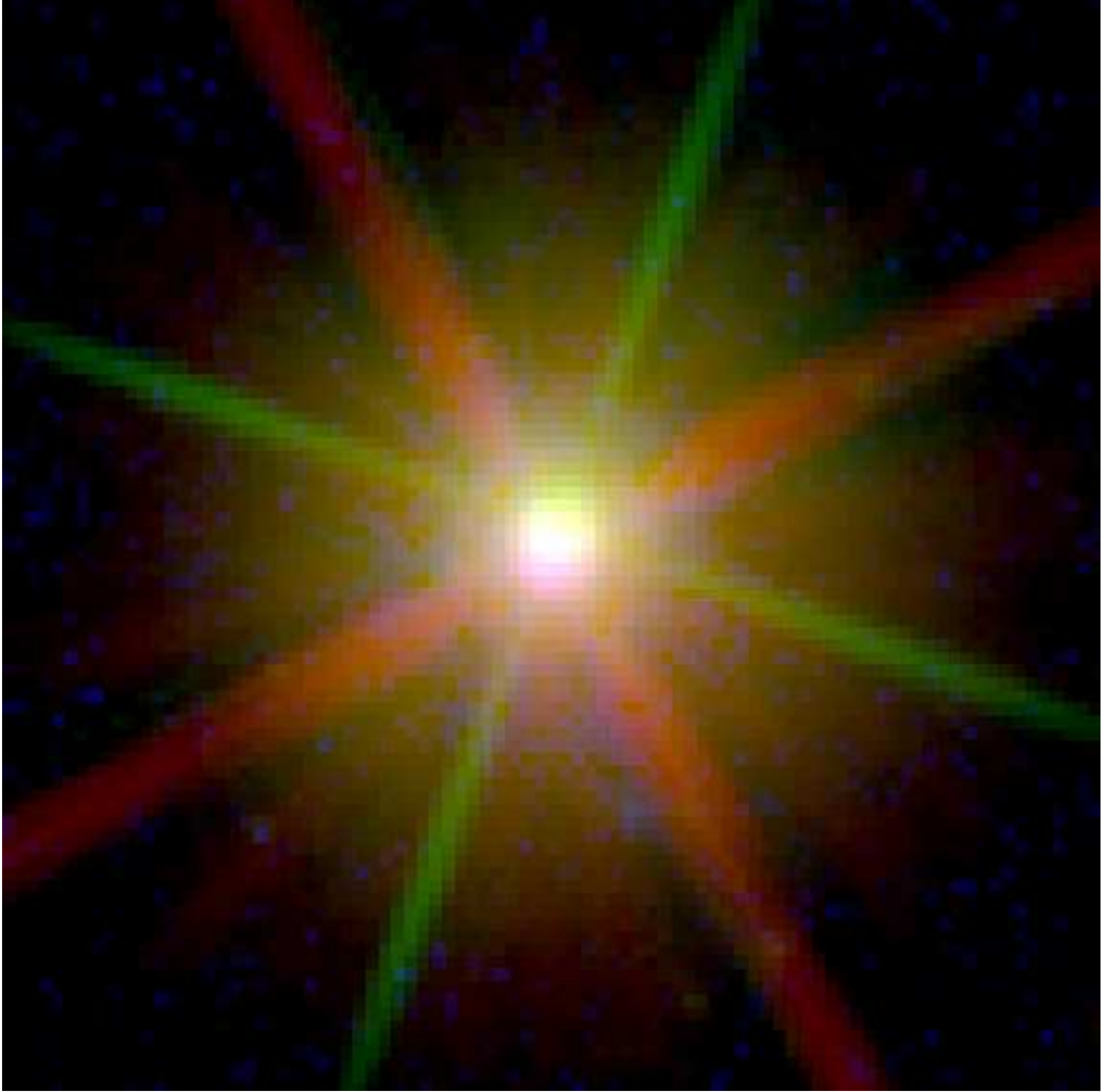}
}

\sn\cl{
\includegraphics[width=0.23\linewidth,angle=-90]{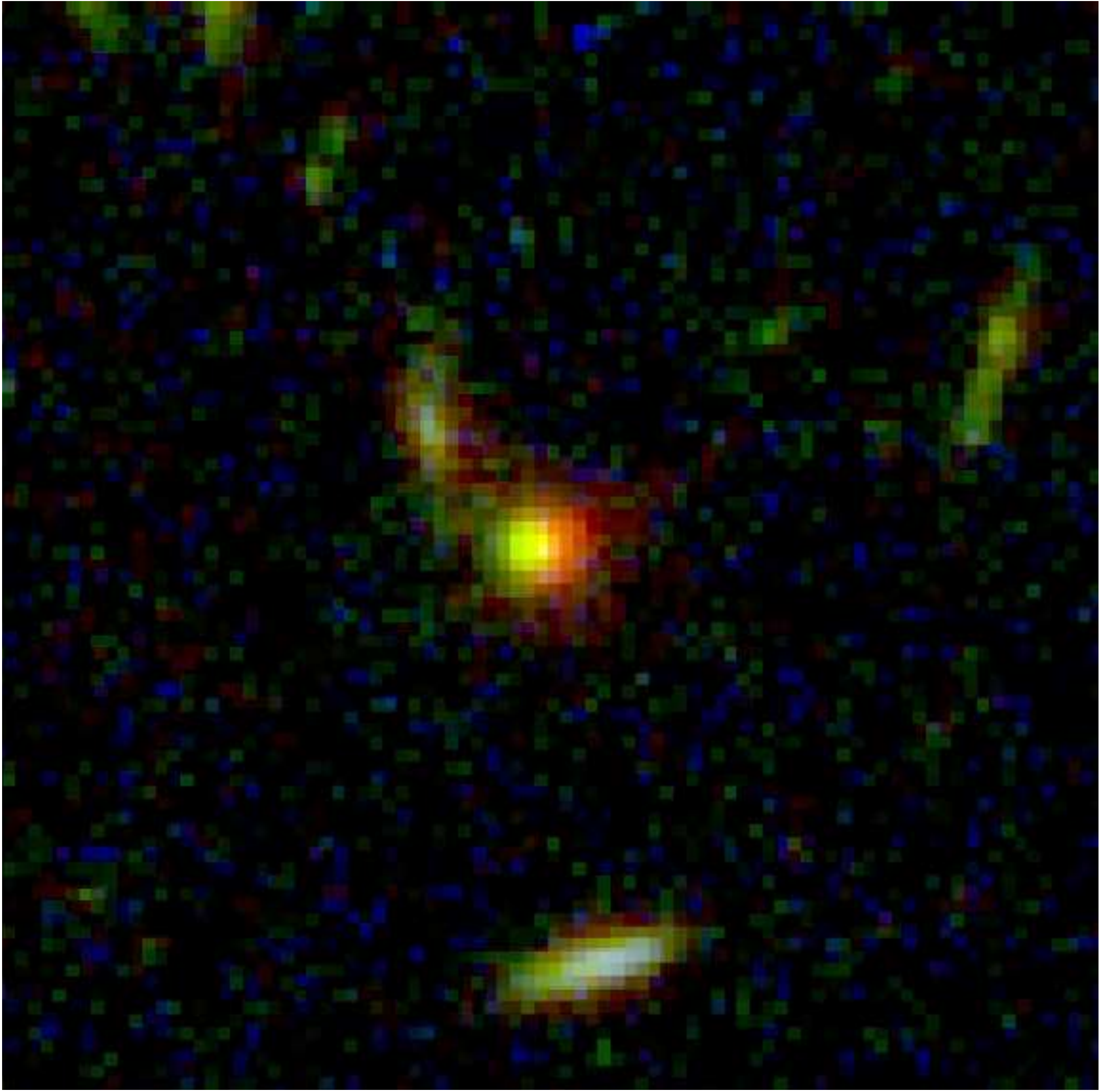}
\includegraphics[width=0.23\linewidth,angle=-90]{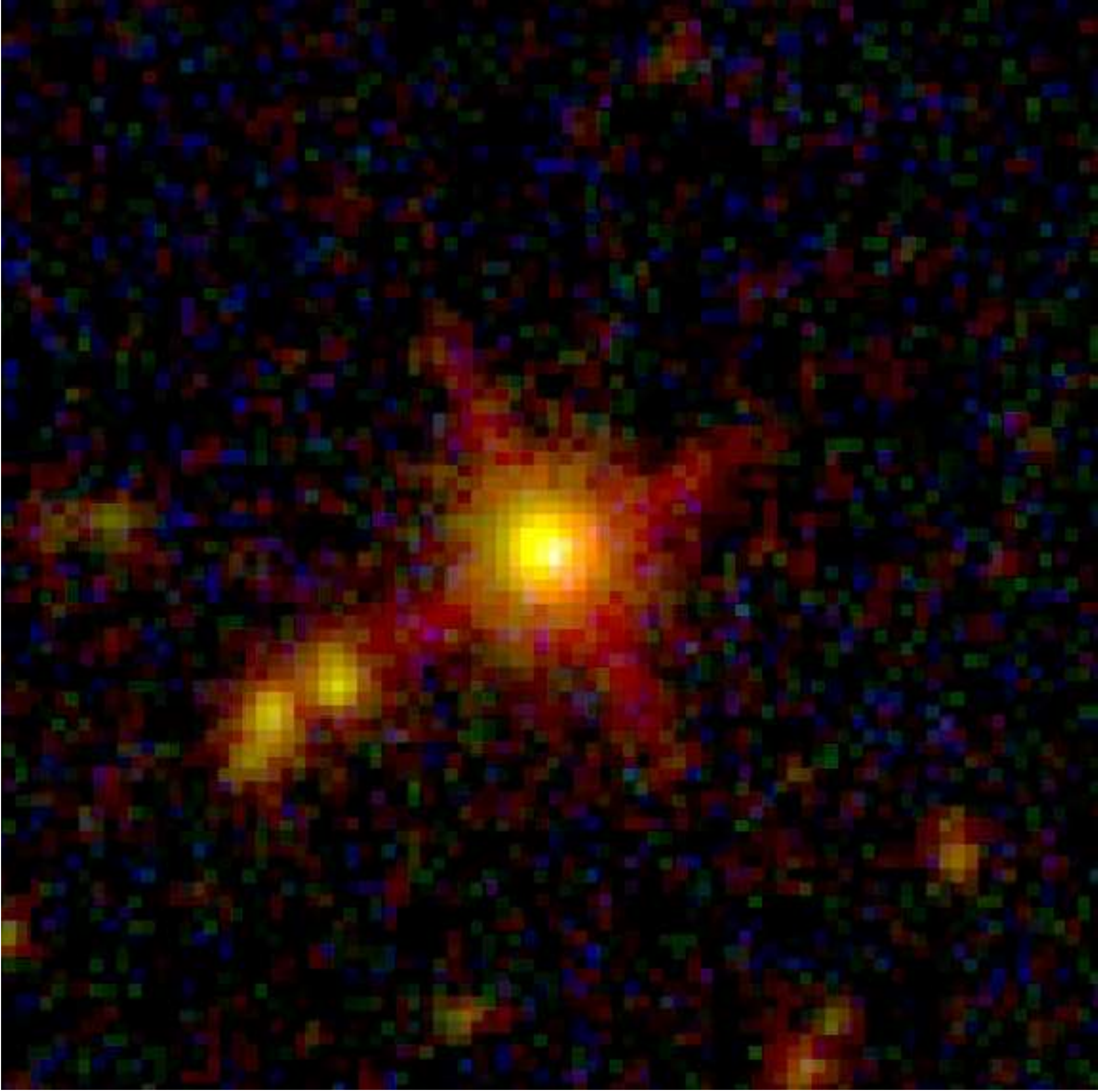}
\includegraphics[width=0.23\linewidth,angle=-90]{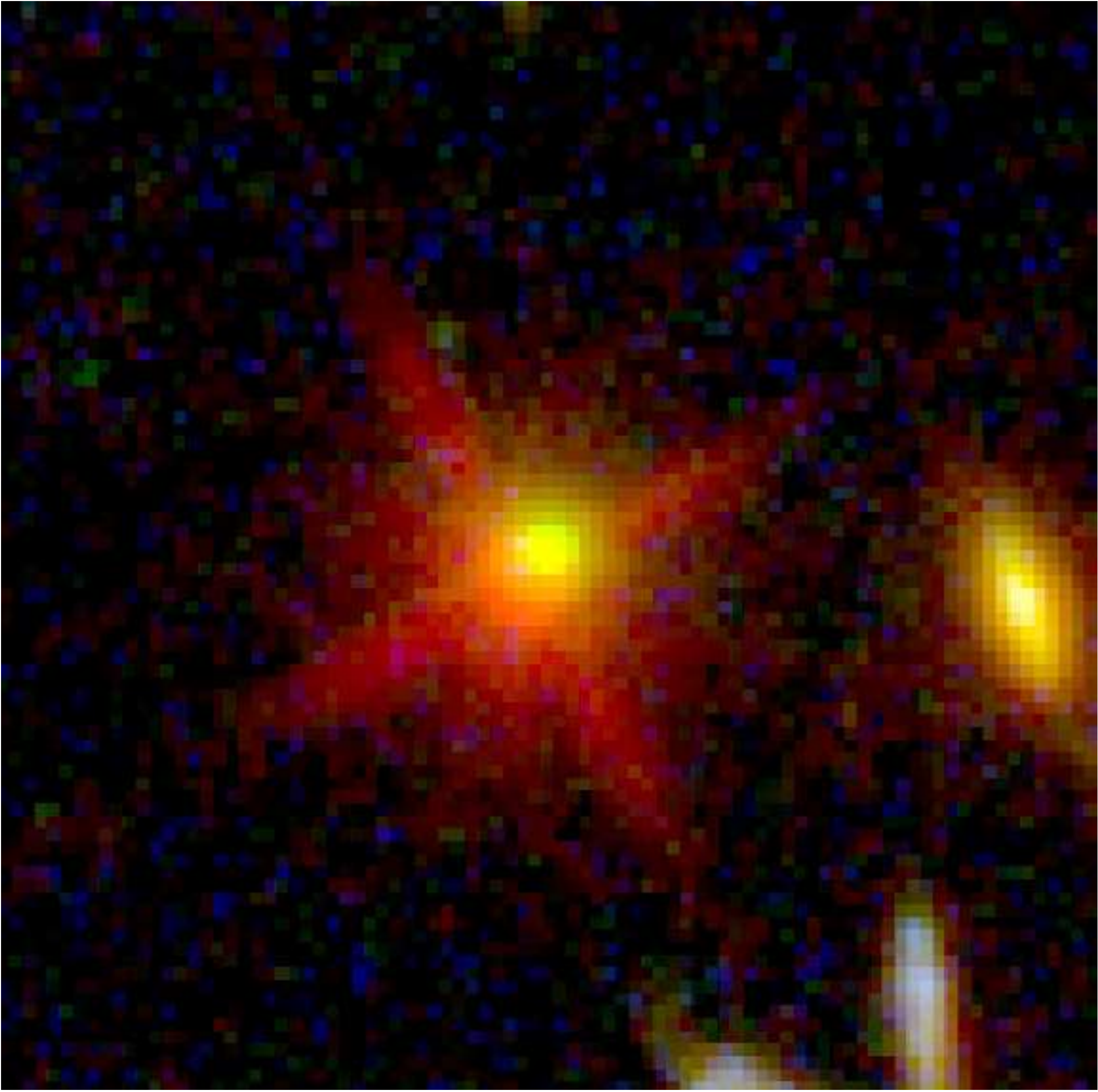}
\includegraphics[width=0.23\linewidth,angle=-90]{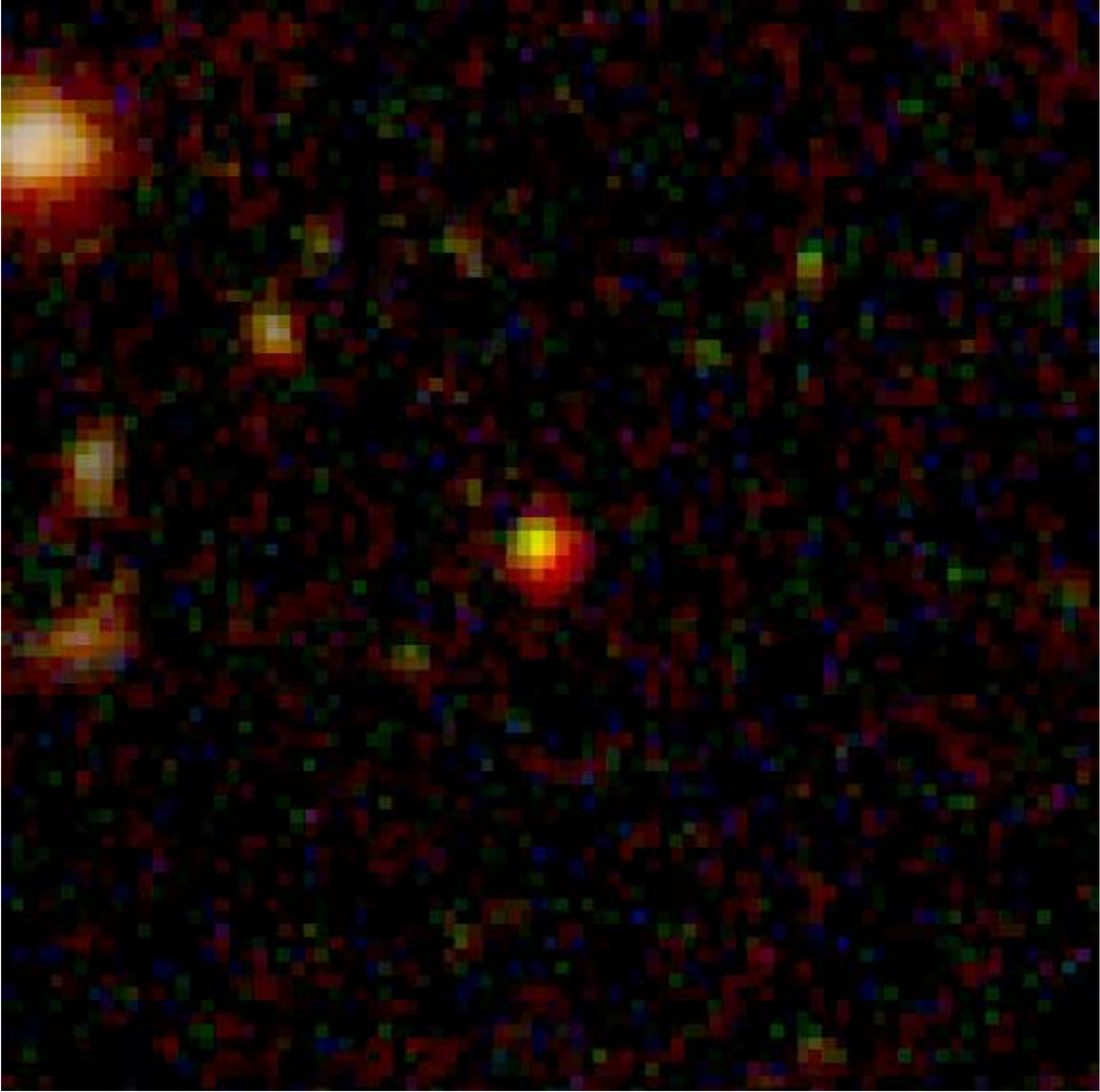}
}

\sn\cl{
\includegraphics[width=0.23\linewidth,angle=-90]{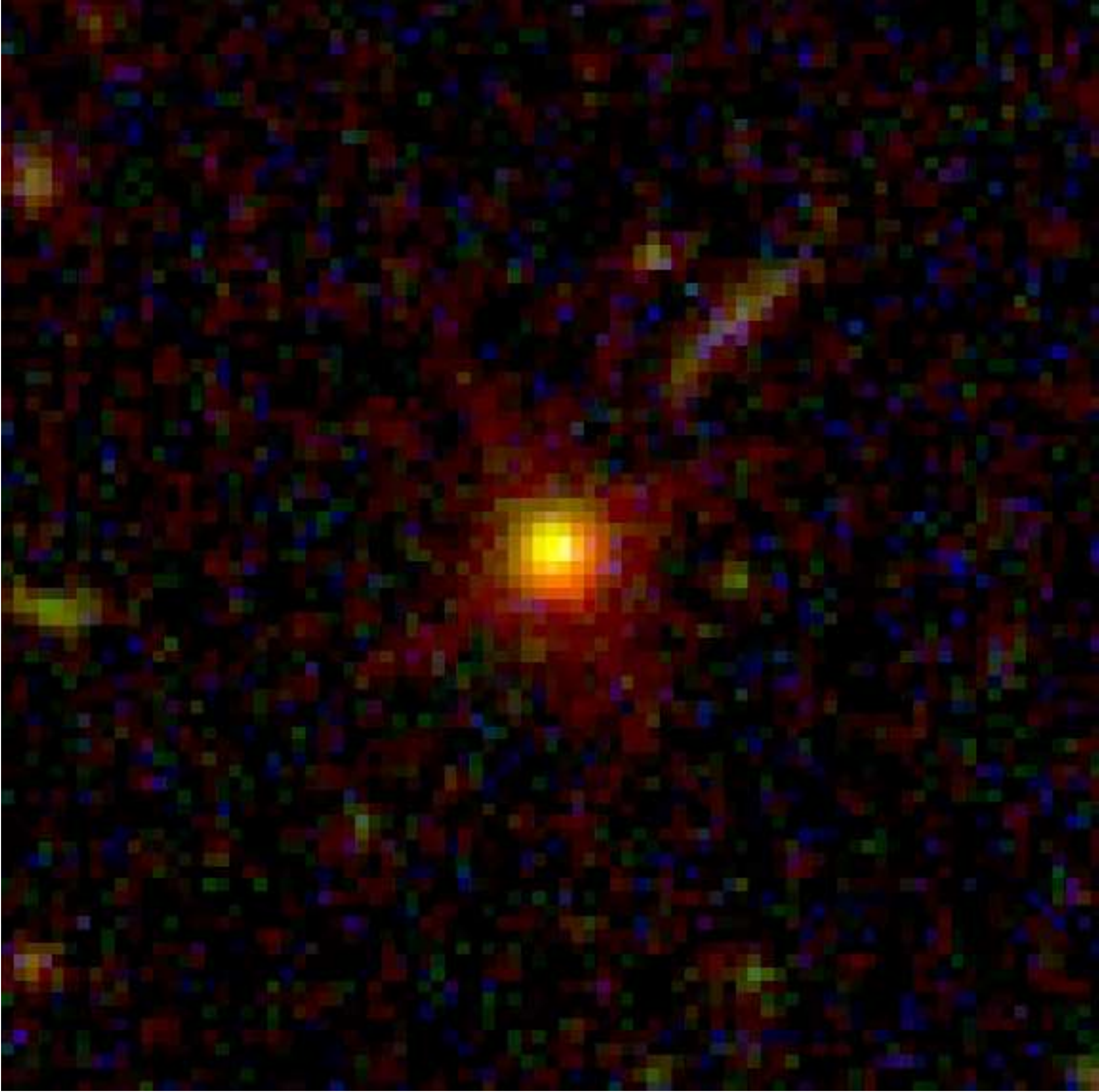}
\includegraphics[width=0.23\linewidth,angle=-90]{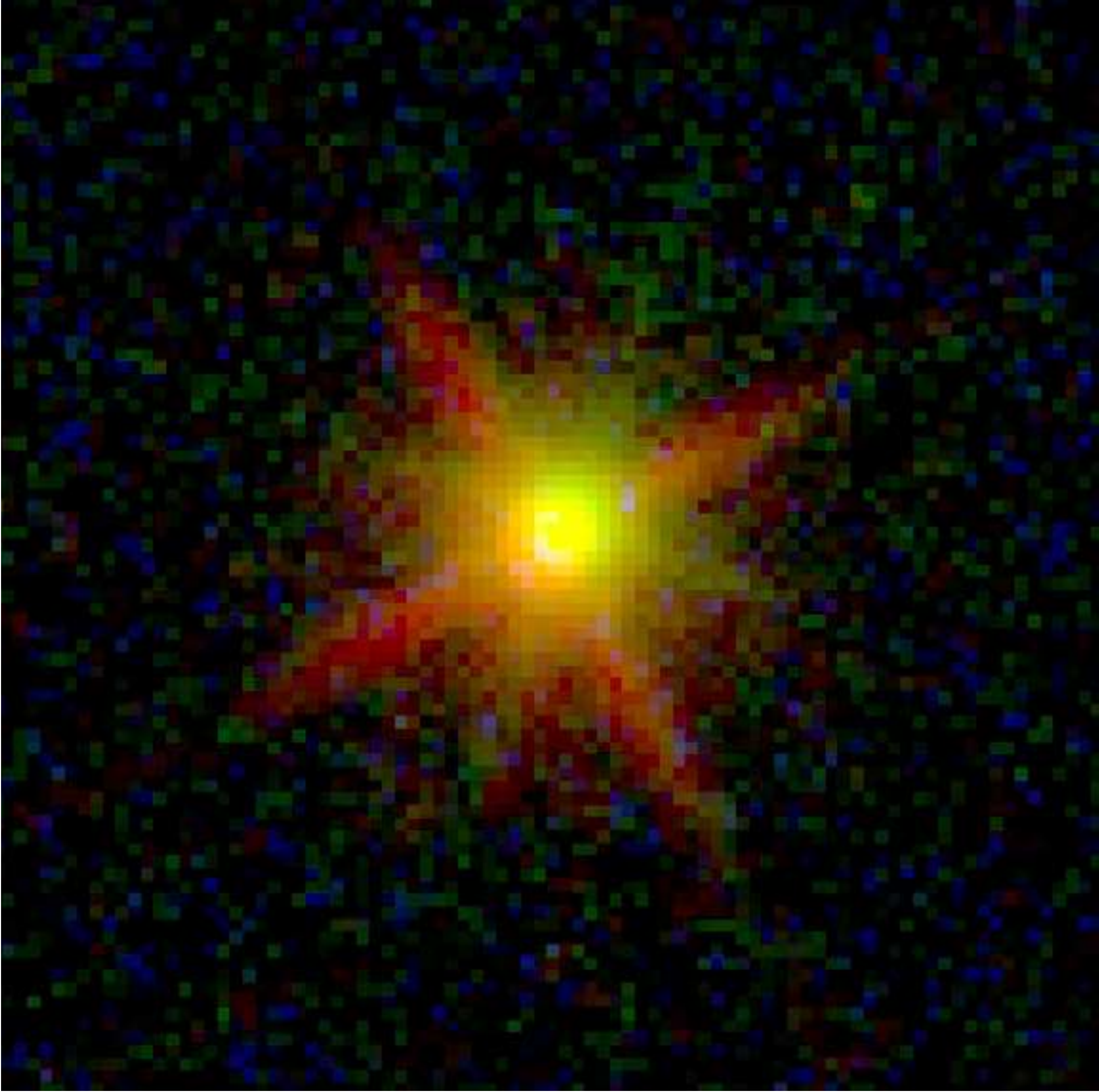}
\includegraphics[width=0.23\linewidth,angle=-90]{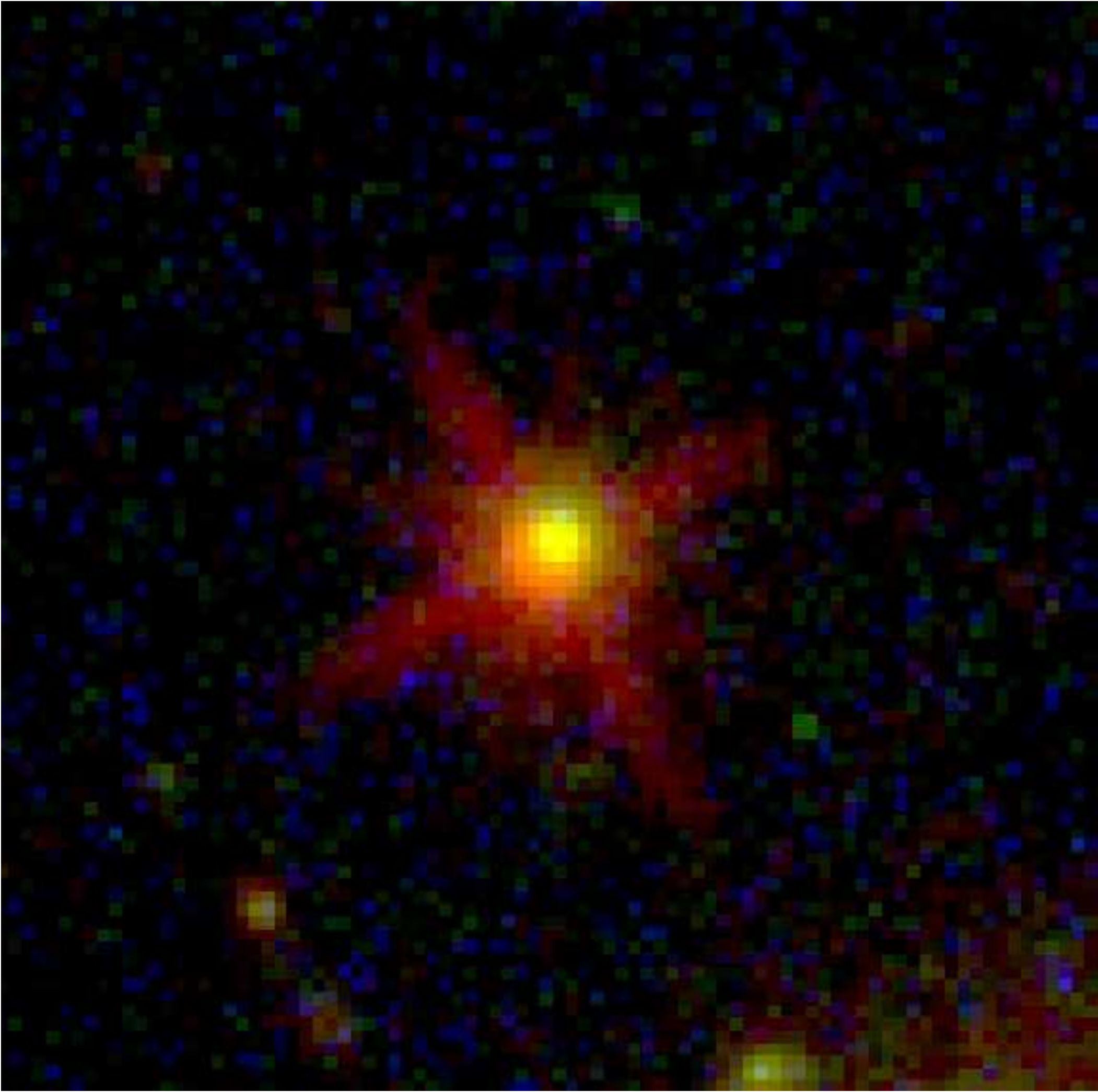}
\includegraphics[width=0.23\linewidth,angle=-90]{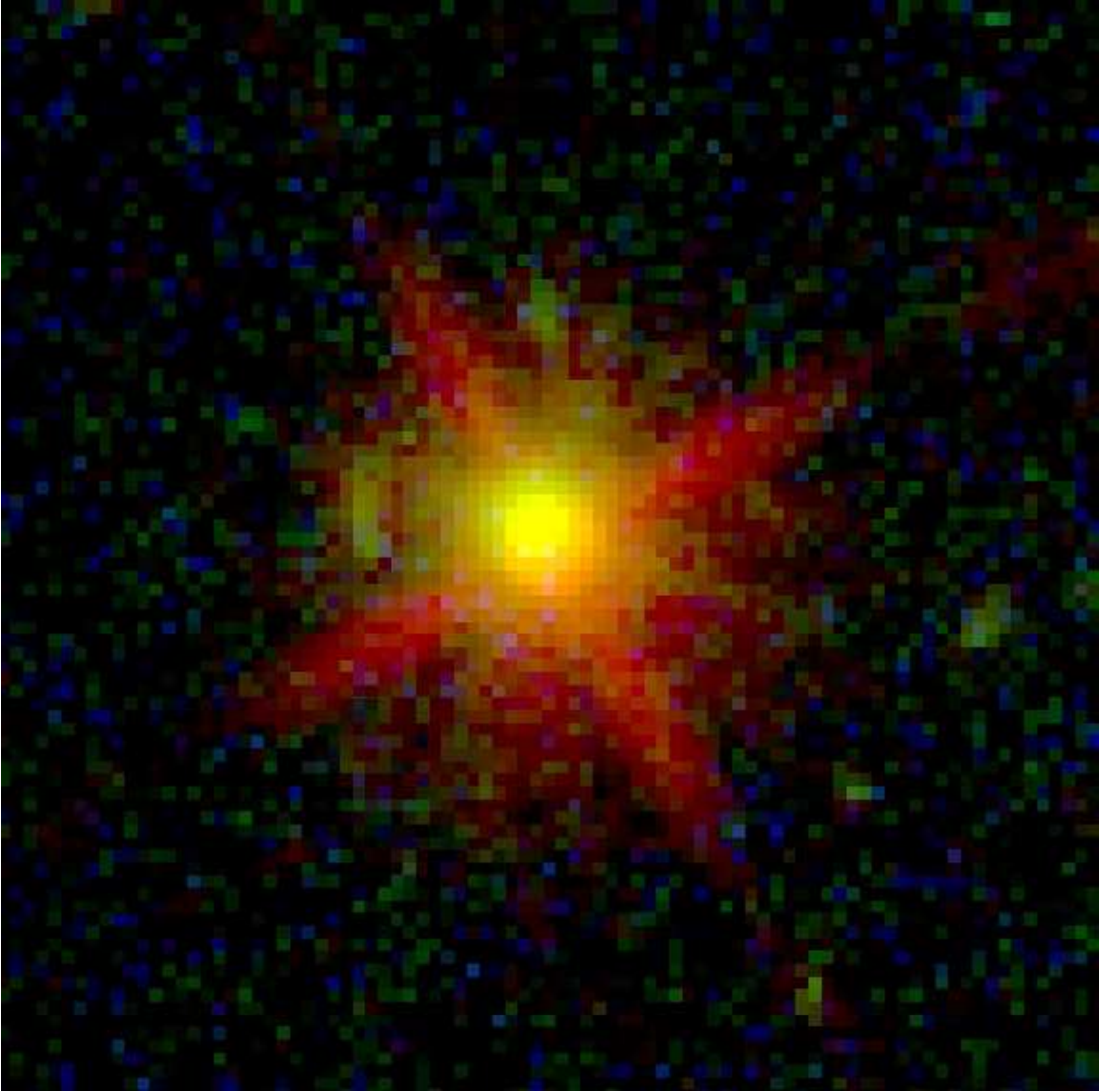}
}

\sn\cl{
\includegraphics[width=0.23\linewidth,angle=-90]{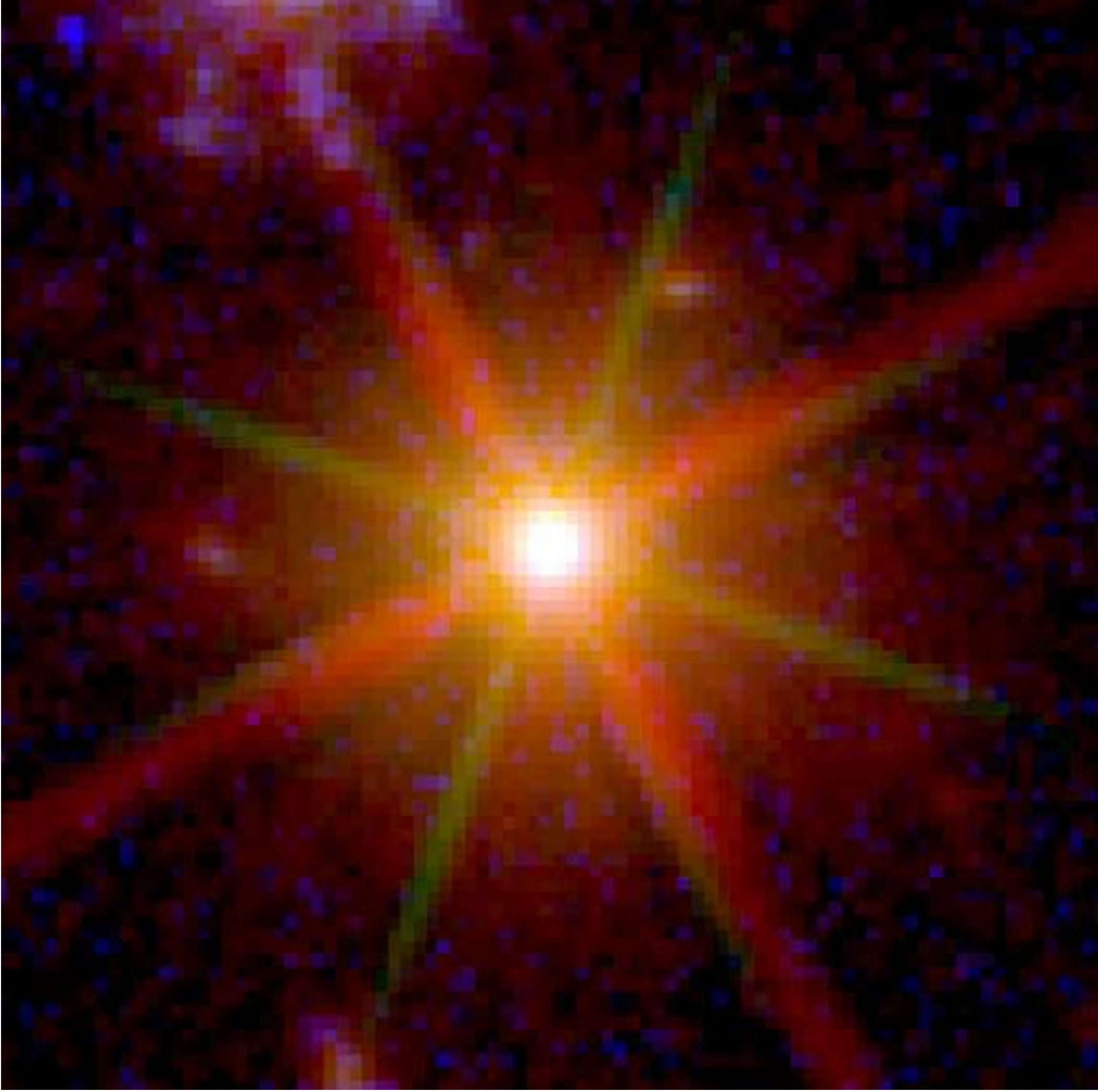}
\includegraphics[width=0.23\linewidth,angle=-90]{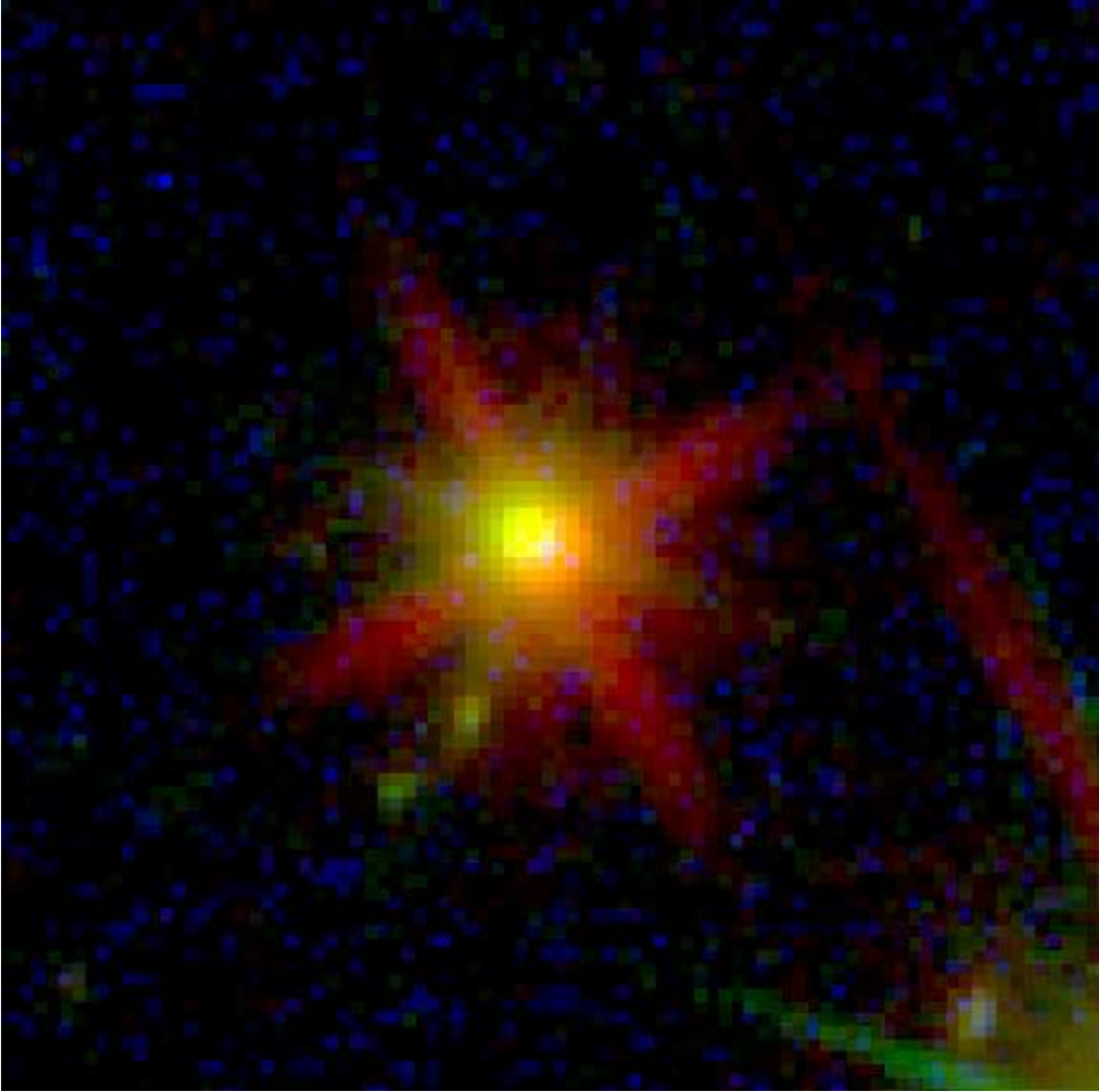}
\includegraphics[width=0.23\linewidth,angle=-90]{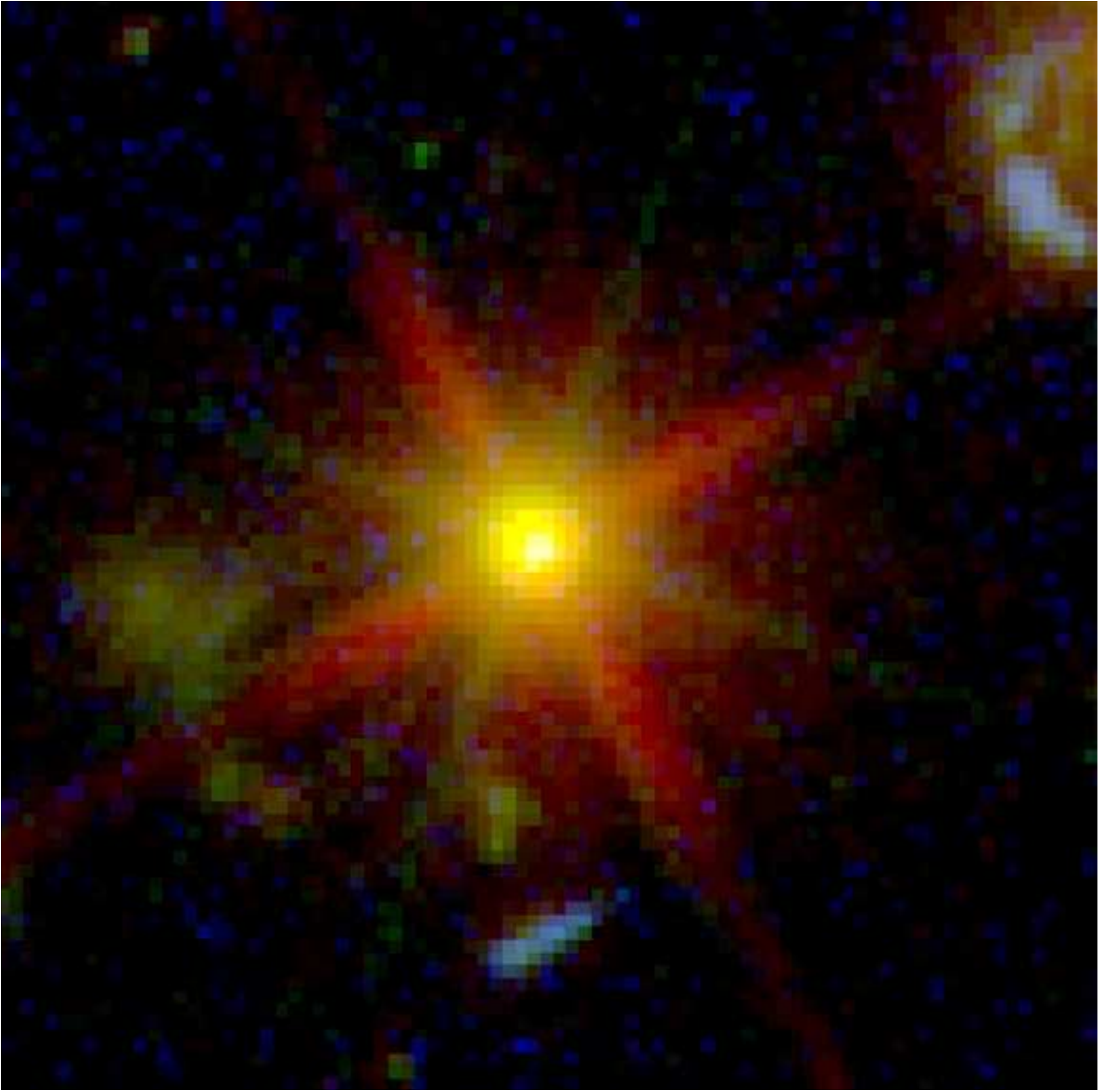}
\includegraphics[width=0.23\linewidth,angle=-90]{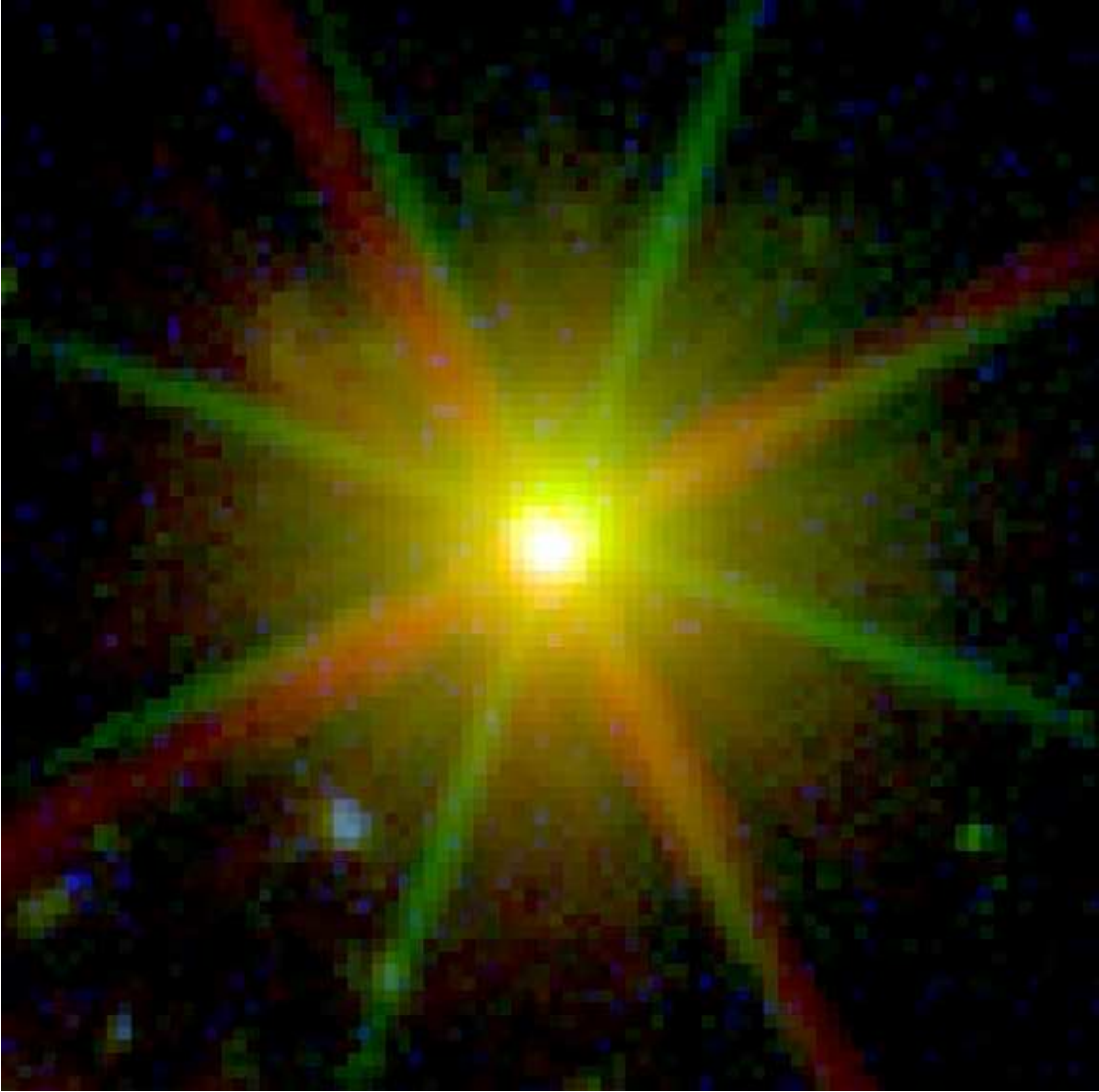}
}

\sn\cl{
\includegraphics[width=0.23\linewidth,angle=-90]{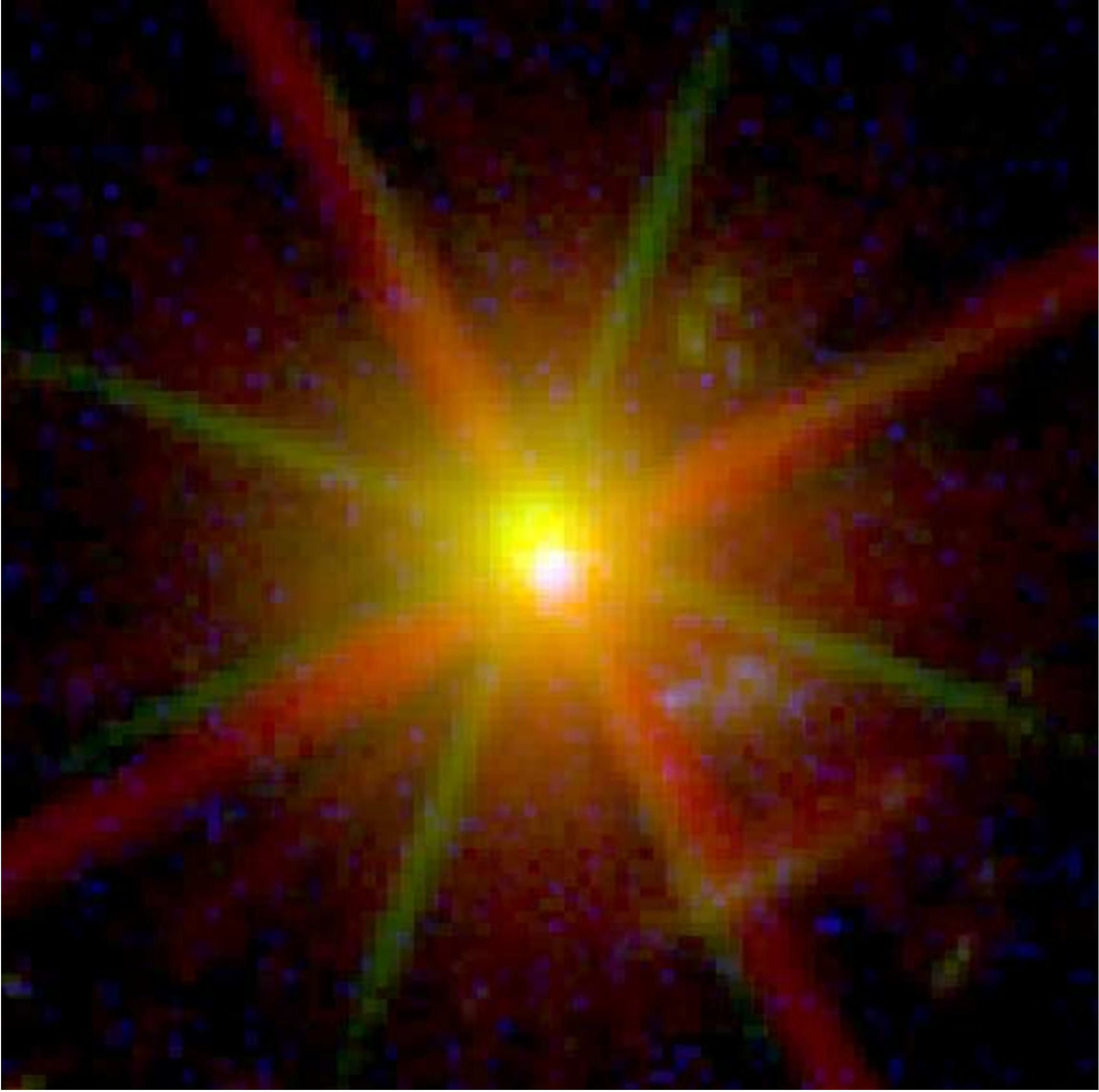}
\includegraphics[width=0.23\linewidth,angle=-90]{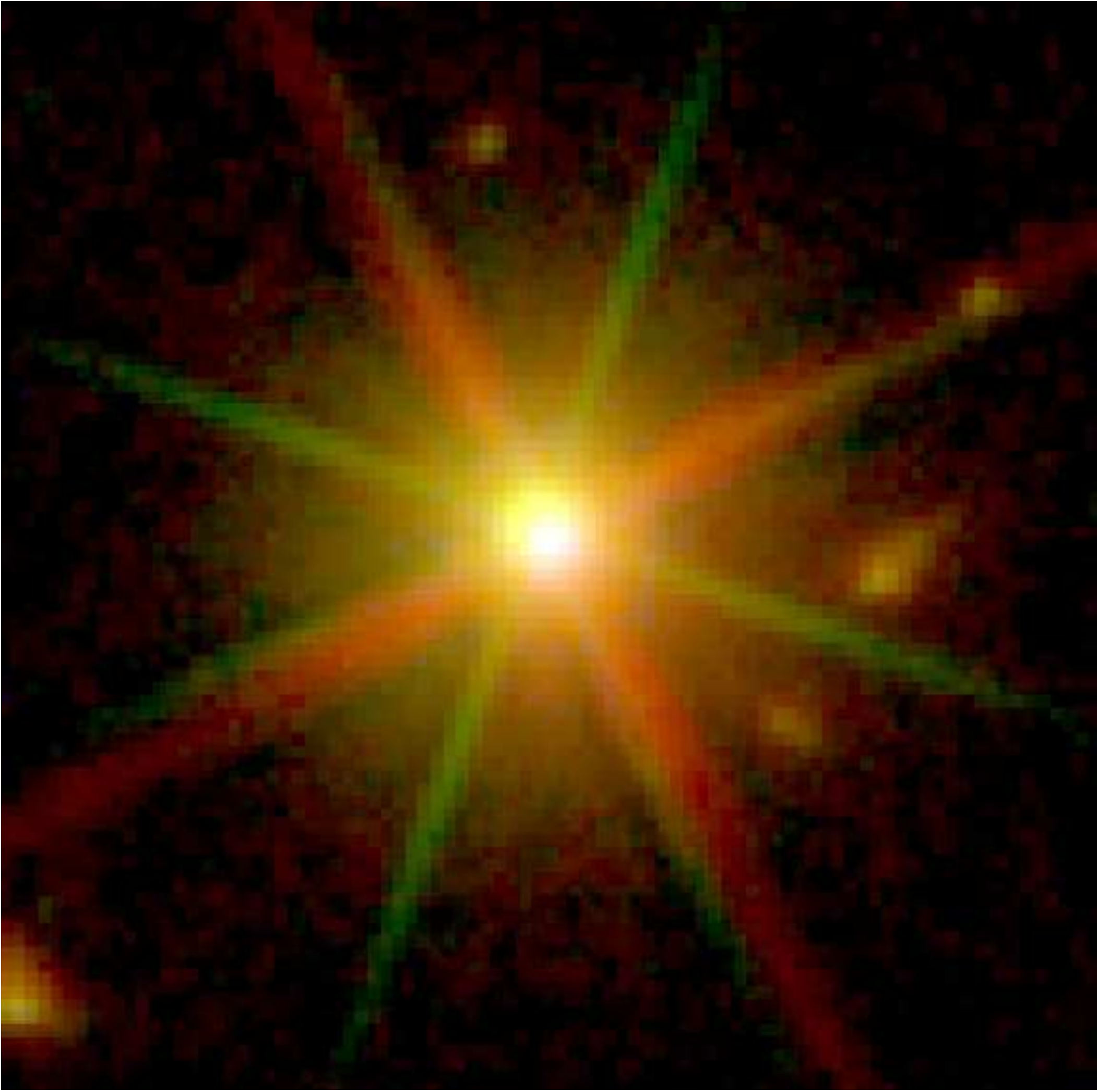}
\includegraphics[width=0.23\linewidth,angle=-90]{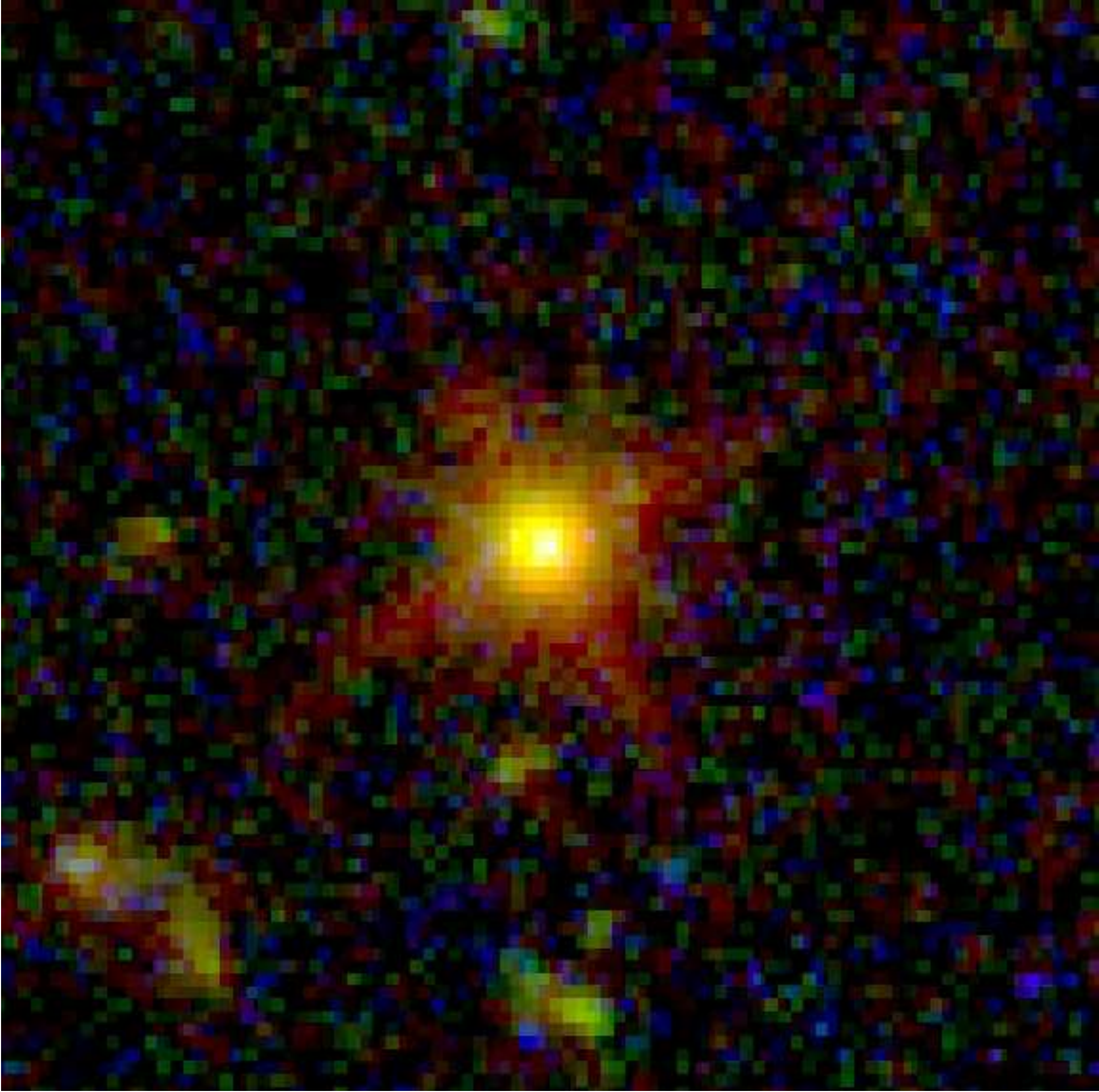}
\includegraphics[width=0.23\linewidth,angle=-90]{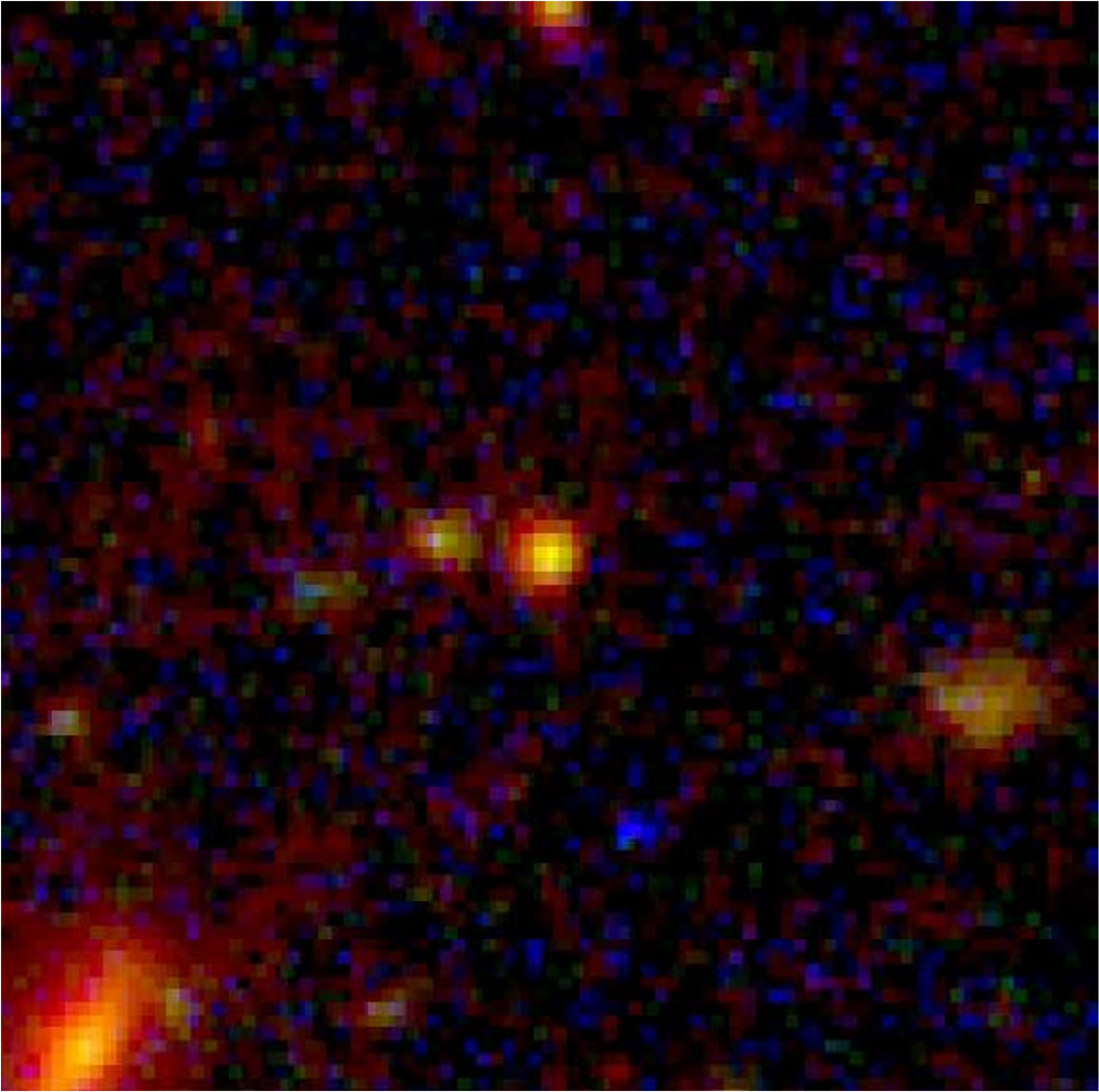}
}

\vspace*{+0.2cm}
\n{\textbf Fig. 6c\ Log(log) color reproduction of the 20 ERS stars with the
highest (\cge 3-$\sigma$) proper motion, as discussed in \S 4.3.2 and Fig. 4c.
Each image is 85$\times$85 pixels or 7\arcspt 65$\times$7\arcspt 65 on the side.
The images used a similar color balance as in Fig. 5a--5b, except that only the
2009 WFC3 UVIS filters F225W, F275W, and F336 are now in the Blue gun, all the
2003 ACS BViz filters F438W, F606W, F775W, and F850LP are in the Green gun, and
the 2009 WFC3 IR filters F098M, F125W, and F160W are in the Red gun. The RGB
colors were further adjusted such that the proper motion between the Green 2003
ACS colors and the Blue+Red (or violet) 2009 WFC3 colors were maximally
contrasted. In many cases, the proper motion is visible as an offset between the
centroids of the Green and the Red+Blue images. For details, see text. }

\ve 

\n\cl{
\includegraphics[width=0.40\linewidth,angle=0]{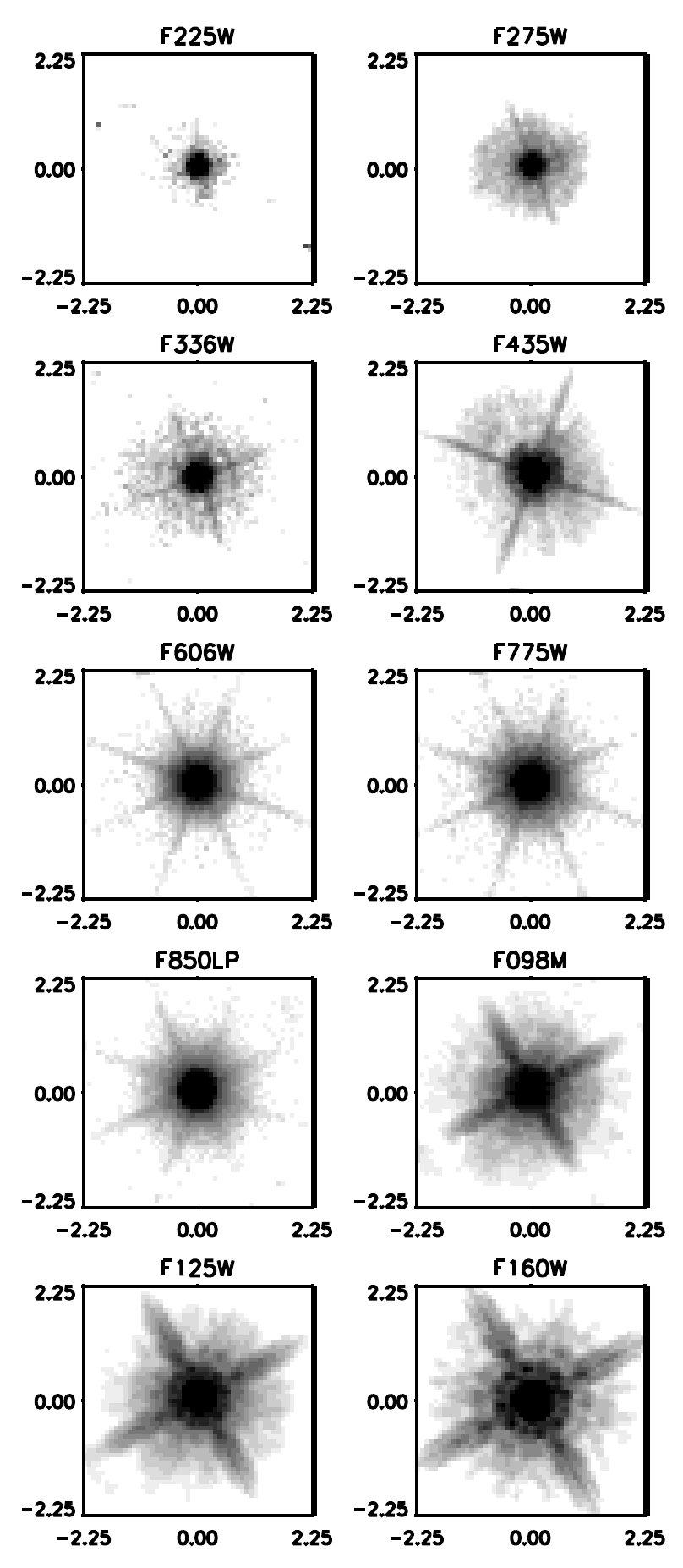}
\includegraphics[width=0.60\linewidth]{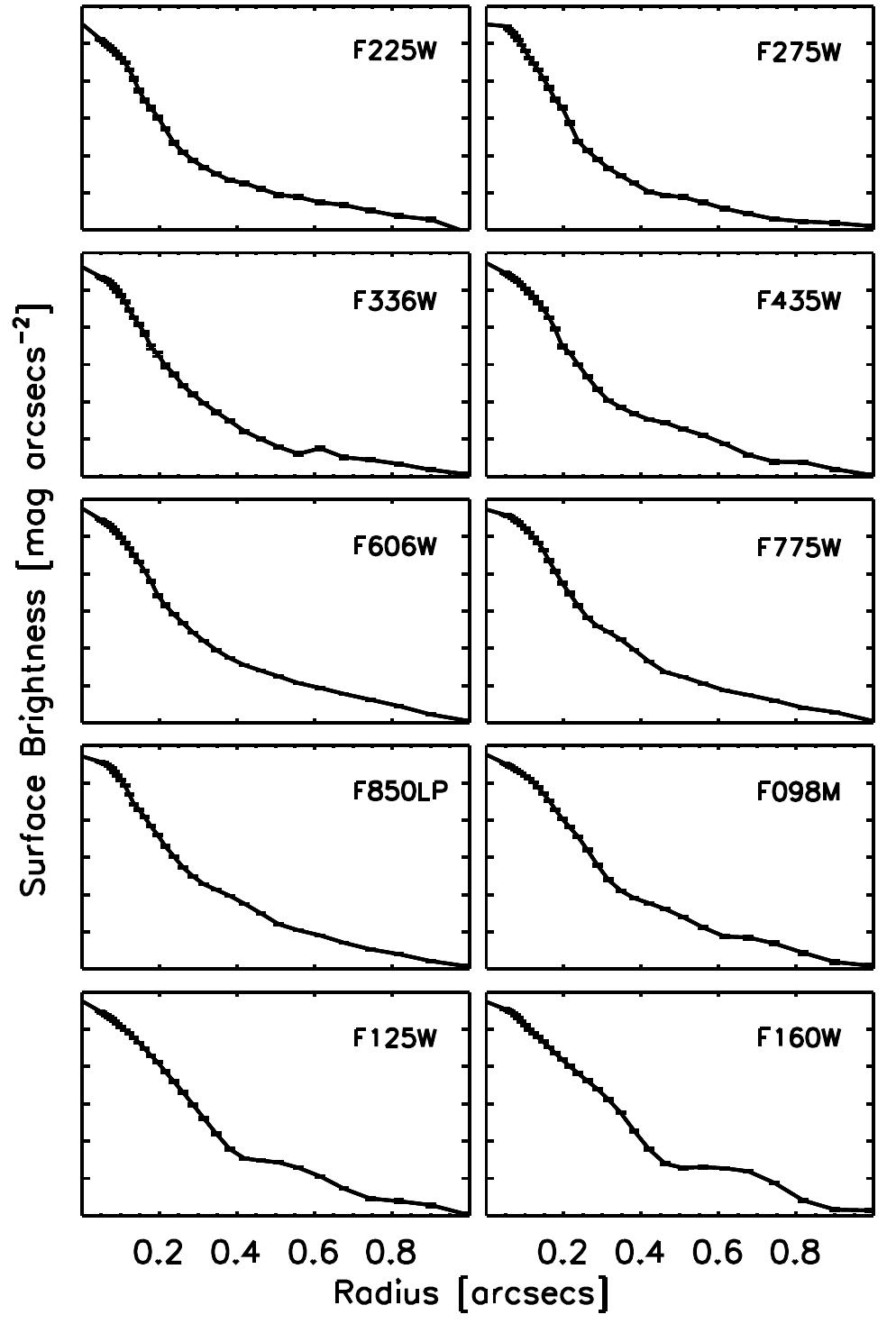}
}

\vspace*{-0.0cm}
\bn{\textbf Fig. 7a (left panels)\ Stellar images, and Fig. 7b (right panels)\
Stellar light-profiles in the individual 10-band images. Note the progression of
the PSF size with wavelength, as discussed in \S 4.3.2. }

\ve 

\vspace*{-0.0cm}\hspace*{-0.5cm}
\n\cl{
\includegraphics[width=0.50\linewidth]{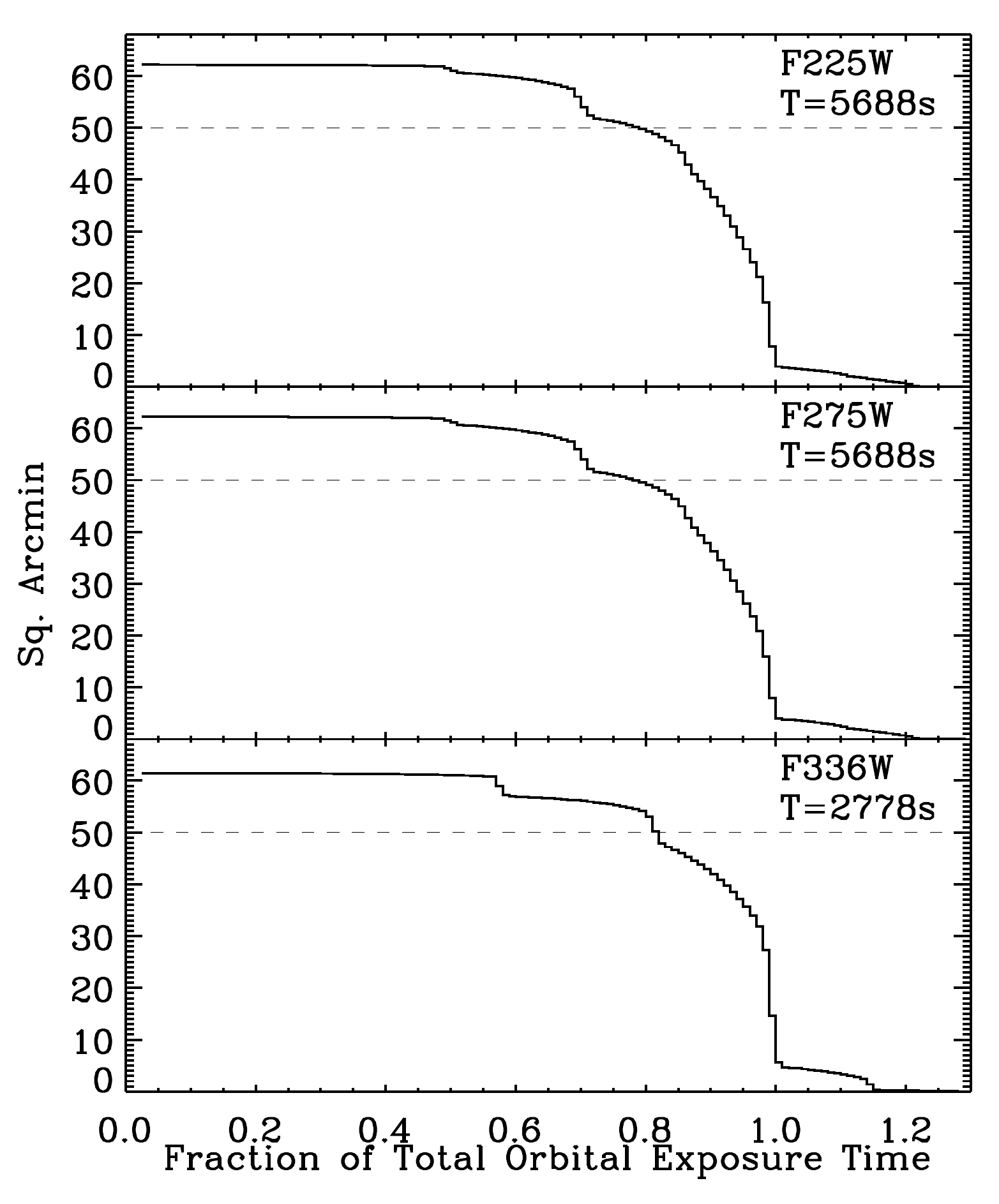}
\includegraphics[width=0.50\linewidth]{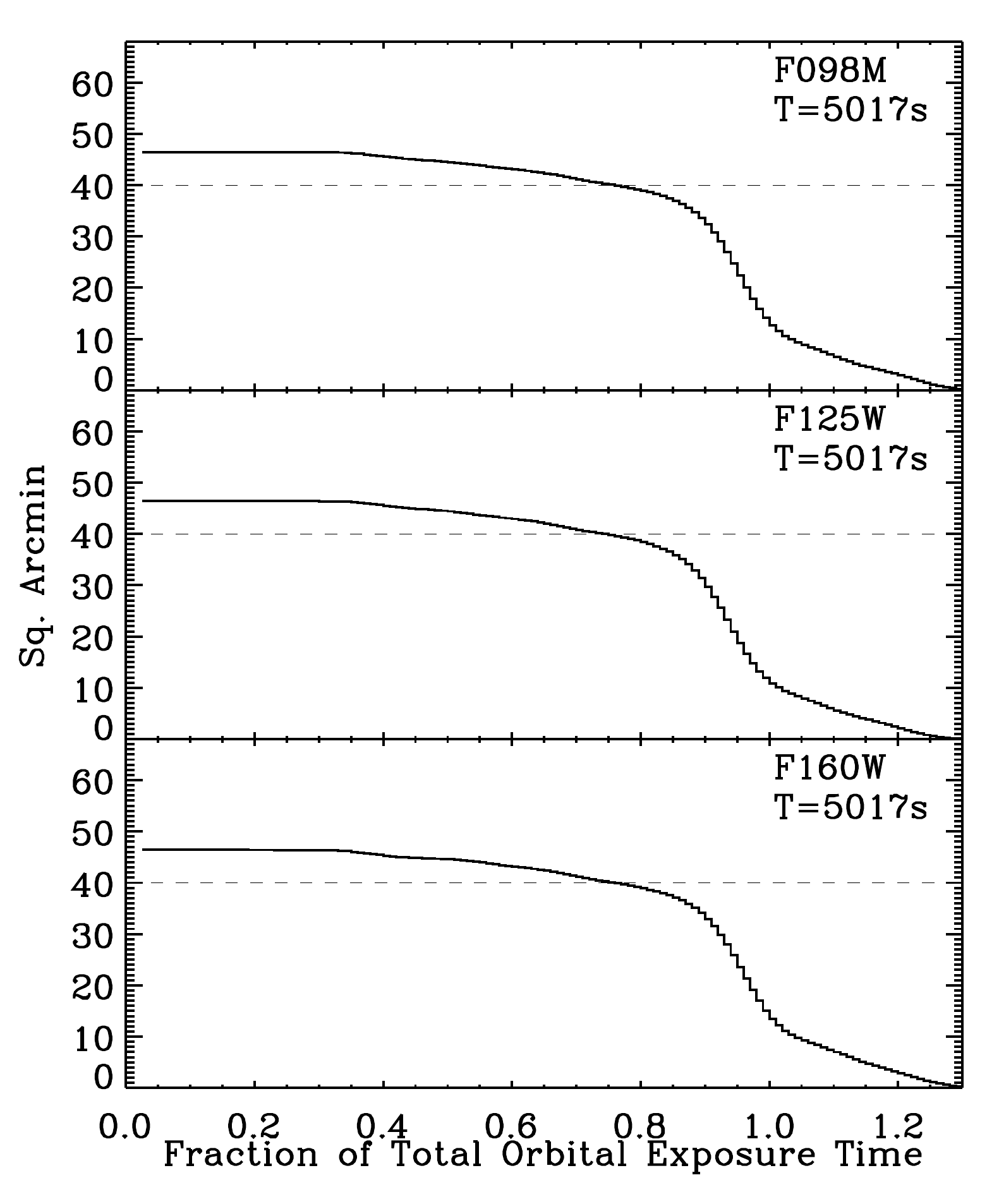}
}

\n\cl{
\includegraphics[width=0.50\linewidth]{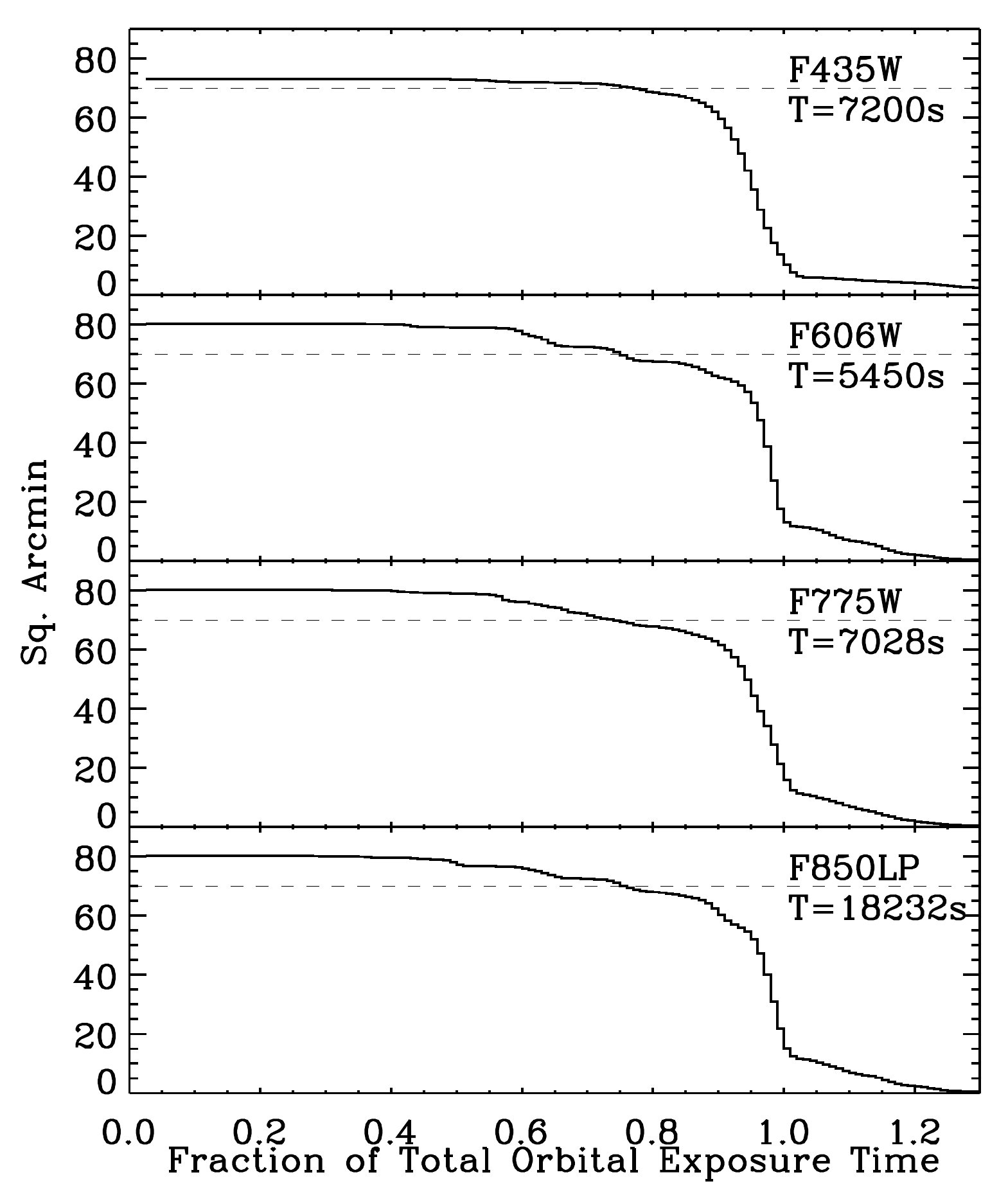}
}

\vspace*{-0.0cm}
\bn{\textbf Fig. 8a (Top left panels).\ Cumulative distribution of the maximum
pixel area that has the specified fraction of total orbital exposure time in each
ERS UVIS mosaic. These effective areas must be quantified in order to properly do
the object counts in Fig. 11a. Fig. 8b (Bottom middle panels)\ Same as Fig. 8a,
but for the GOODS v2.0 mosaics in BViz. Fig. 8c (Top right panels)\ Same as Fig.
8a, but for the ERS mosaics in the WFC3 IR filters. }

\ve 

\vspace*{-0.0cm}
\n\cl{
\includegraphics[width=1.00\linewidth]{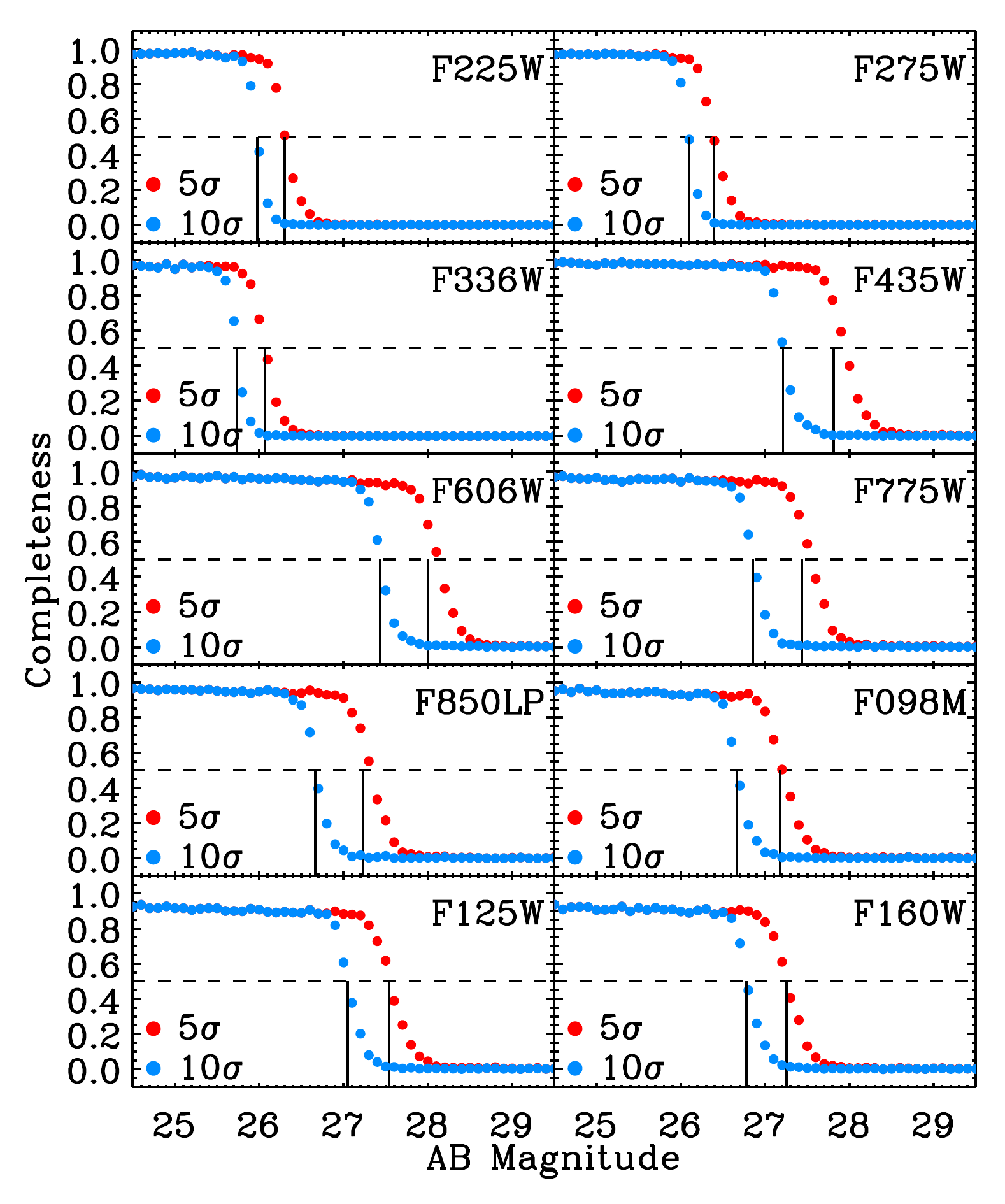}
}

\vspace*{-0.0cm}
\bn{\textbf Fig. 9.\ Panchromatic completeness functions of the number counts in
the panchromatic ERS images at the 5-$\sigma$ (red) and 10-$\sigma$ (blue)
detection levels of total object magnitudes. These were derived by Monte Carlo
insertion of faint point-like objects into the ERS images, and plotting the
object recovery ratio as a function of total magnitude. At the \cge 5-$\sigma$
level, the ERS is more than 50\% complete at levels fainter than AB$\sim$26 mag
for the WFC3 UVIS images, fainter than AB$\sim$27 mag for the WFC3 IR images, and
substantially fainter than than AB$\sim$27 mag for the deeper multi-year GOODS
ACS v2.0 images (see Table 1). }

\ve 

\vspace*{-0.6cm}
\hspace*{+0.0cm} 
\n\cl{
\includegraphics[width=1.00\linewidth]{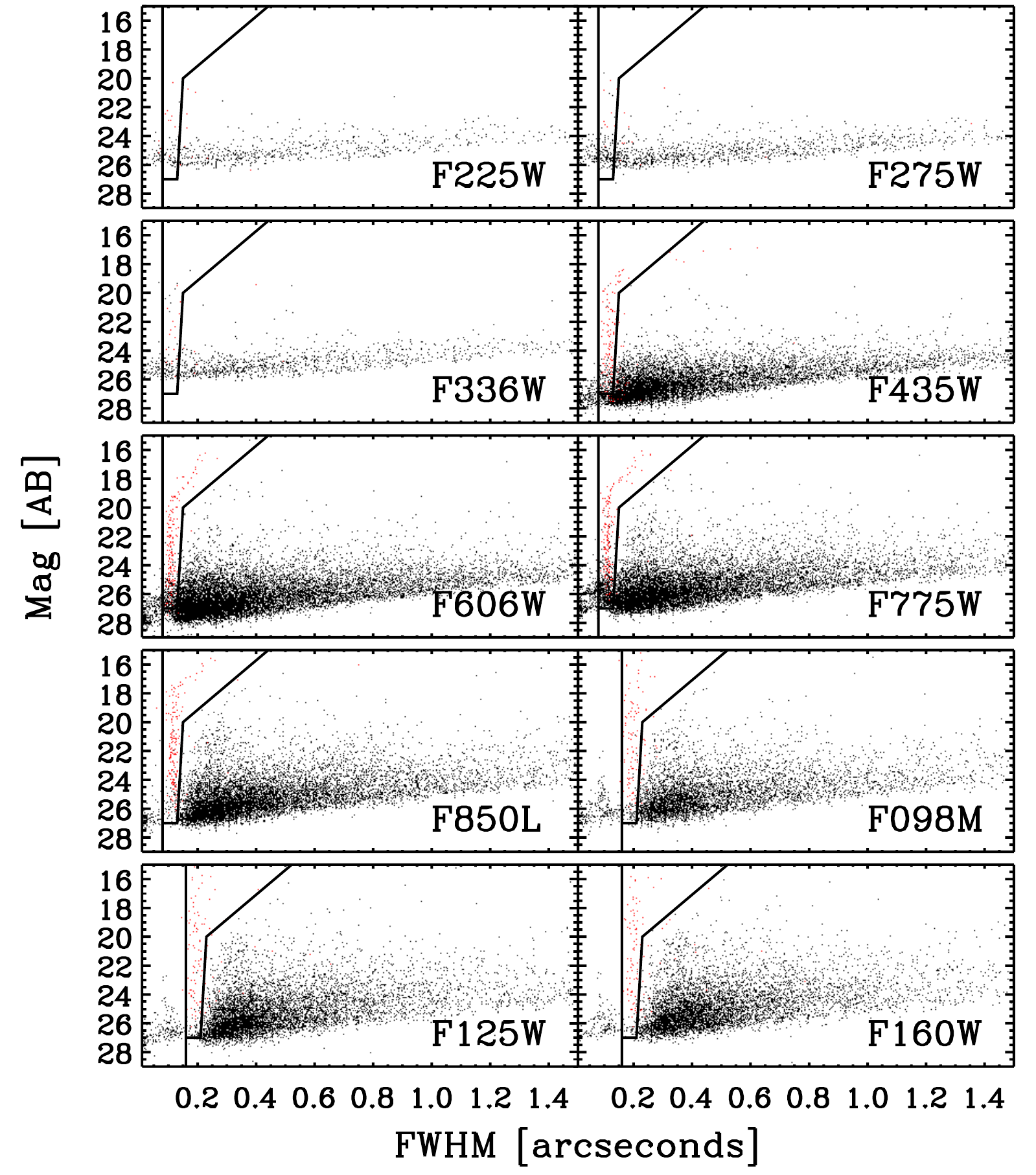}
}

\vspace*{-0.0cm}
\n{\textbf Fig. 10a.\ Panchromatic star-galaxy separation in the 10-filter ERS
images. Plotted are object total AB-magnitude vs. SExtractor image diameter
$FWHM$ ($FWHM$$\simeq$twice its half-light radius). Objects with image diameter
\cle 0.9$\times$PSF-FWHM --- where the PSF-FWHM ($\simeq$0\arcspt 07--0\arcspt
15) is from Table 2 --- are image defects (black dots). Objects in the thin
vertical filaments (red dots) immediately larger than this are classified as
stars. ERS stars were defined to be those objects that resided in this red area
in at least 3 out of 10 ERS filters. Objects to the right of the full-drawn
slanted lines are galaxies (black dots). The star-galaxy separation becomes less
reliable for fluxes fainter than AB$\simeq$26--27 mag in UV\BVizYsJH,
respectively. Details are discussed in the text. [For best display, please zoom
in on the full-resolution version of this image, which is available on this
$URL$\footnote{http://www.asu.edu/clas/hst/www/wfc3ers/ERS2\_gxysv4ln.tif} ]. }

\ve 

\vspace*{-0.0cm}
\hspace*{+0.0cm} 
\n\cl{
\includegraphics[width=1.00\linewidth]{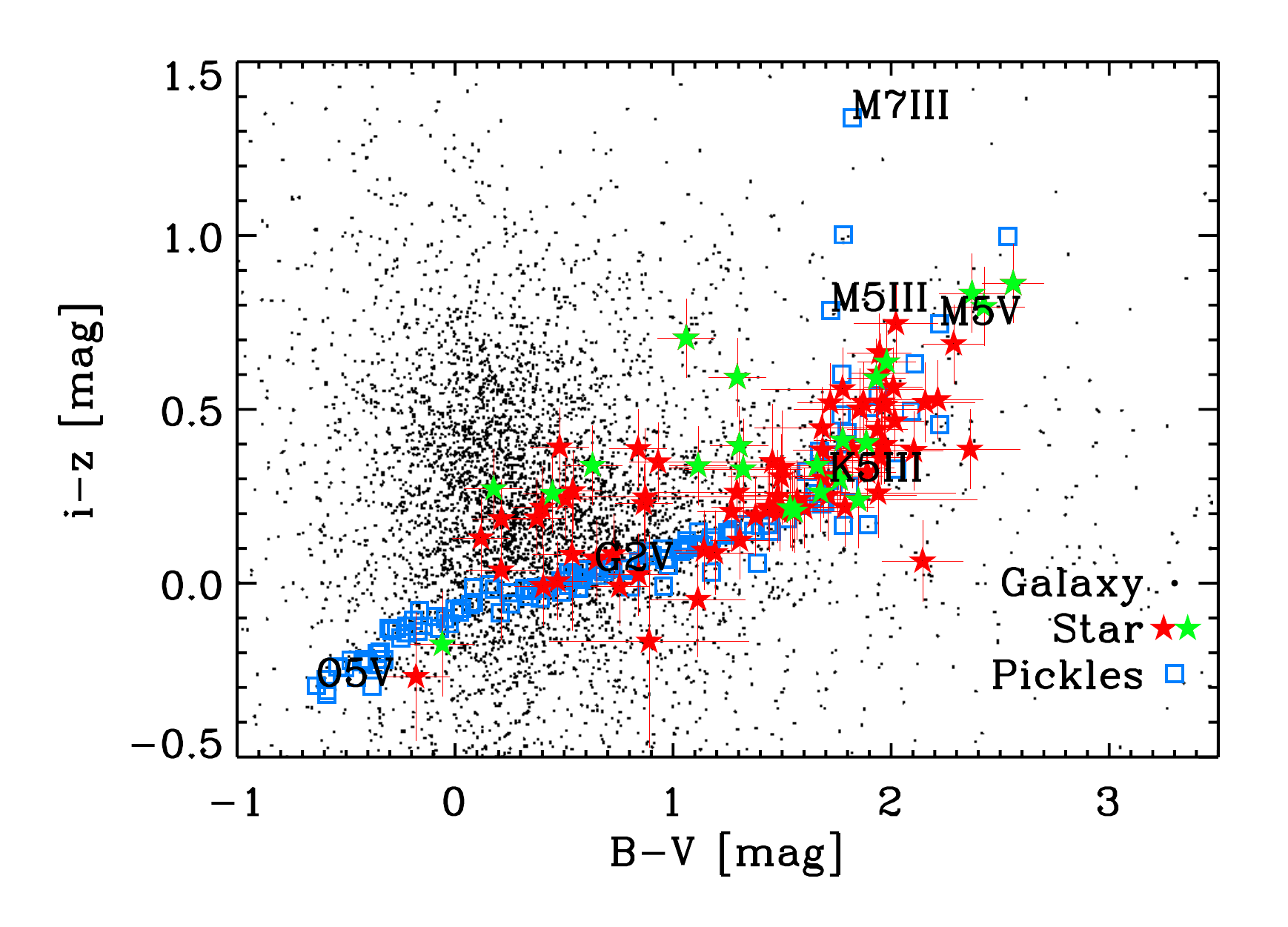}
}

\vspace*{-0.0cm}
\bn{\textbf Fig. 10b.\ The (i--z) vs. (B--V) color-color diagram in the GOODS
BViz filters for all ERS galaxies (black dots) and stars (red and green
asterisks). These are compared to the Pickles (1998) model stellar SED library
(blue squares), which range in spectral type from O5V--M7V, and also show the
giant branch models up to type M7III. The observed ERS stellar colors are shown
with their 1-$\sigma$ error bars. Only stars with combined color errors along
each axis \cle 0.5 mag are plotted. Green asterisks mark ERS stars with proper
motion established at the $\ge$2-$\sigma$ level. }

\ve 

\vspace*{-0.0cm}
\hspace*{-0.2cm}
\n\cl{
\includegraphics[width=0.95\linewidth]{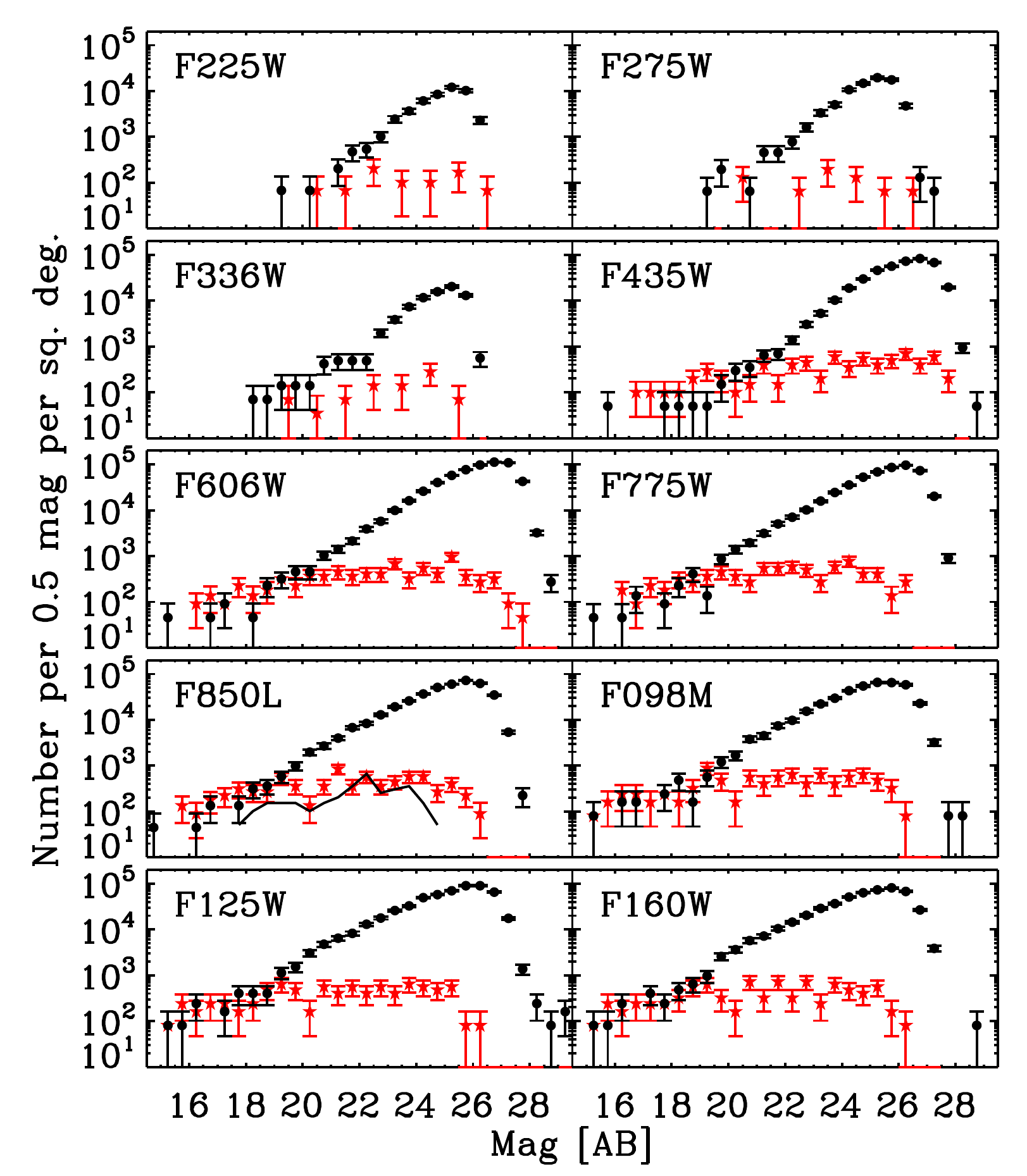}
}

\vspace*{+0.7cm}
\bn{\textbf Fig. 11a.\ Differential panchromatic star counts (red asterisks) and
differential panchromatic galaxy number counts in the ERS images (black dots),
with the optimized star-galaxy separation from Fig. 10a. All 10 ERS filters are
shown in units of object numbers per 0.5 mag per \degsq, but in the three bluest
filters (F225W, F275W, and F336W) some of the brightest bins were doubled to 1.0
mag in width to improve statistics. The solid black line in the F850LP panel are
the spectroscopic star counts from the HST PEARS ACS grism surveys of Pirzkal
\etal\ (2009), which are in good agreement with our F850LP star counts.

\ve 

\n\cl{
\includegraphics[width=0.55\linewidth]{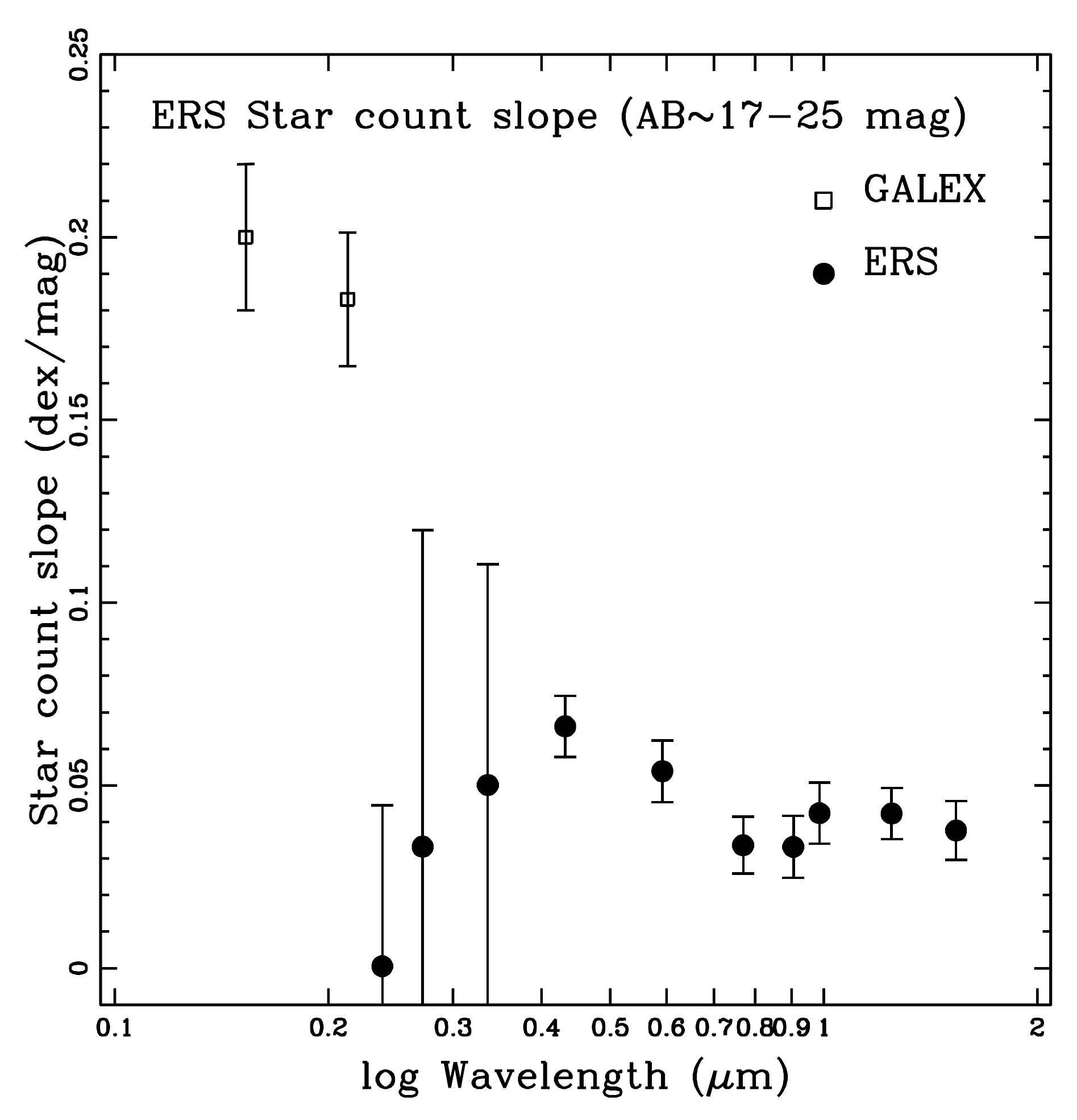}\ 
\includegraphics[width=0.55\linewidth]{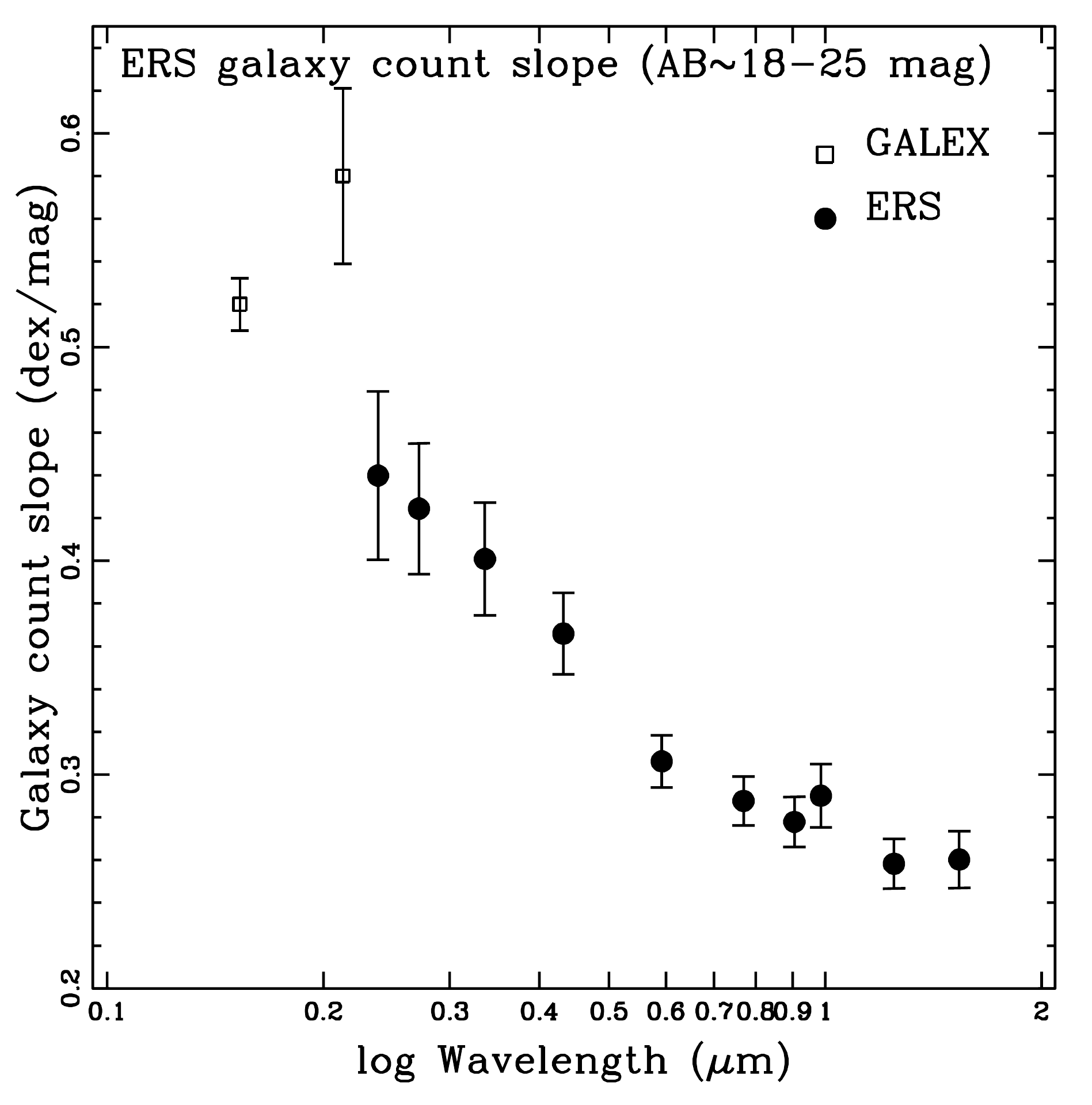}
}

\vspace*{-0.0cm}
\bn{\textbf Fig. 11b.\ (Left panel) ERS Star count slope (filled circles) versus
observed wavelength in the flux ranges AB$\simeq$19--25.5 mag for the 3 UV WFC3
filters, AB$\simeq$16--26 mag for the GOODS/ACS BViz filters, and
AB$\simeq$15--25 mag for the 3 WFC3 IR filters, respectively. The faint-end of
the Galactic star count slope is remarkably flat at all wavelengths from the
mid--UV to the near-IR, with best fit power-law slopes in general of order
0.03--0.20 dex/mag. The two bluest points at 153 and 231 nm are from the GALEX
star counts of Xu \etal\ (2005; open squares), which cover AB=17--23 mag. The ERS
counts at the shortest wavelength suffer from small number statistics and so are
less reliable (see Fig. 10a and 11a). For further details, see the text. }

\sn{\textbf Fig. 11c.\ (Right panel) ERS Galaxy count slope (filled circles)
versus observed wavelength in the flux ranges AB$\simeq$19--25 mag for the 3 UV
WFC3 filters, AB$\simeq$18--26 mag for the GOODS/ACS BViz filters, and
AB$\simeq$17--25 mag for the 3 WFC3 IR filters, respectively. The two bluest
points at 153 and 231 nm are from the GALEX galaxy counts of Xu \etal\ (2005;
open squares), which cover AB=17--23 mag. The galaxy counts show the well known
trend of a steepening of the best-fit power-law slope at the bluer wavelengths,
which is caused by a combination of the more significant K-correction and the
shape of the galaxy redshift distribution at the selection wavelength. }

\ve 

\vspace*{-0.9cm}
\n\cl{
\includegraphics[width=0.55\linewidth]{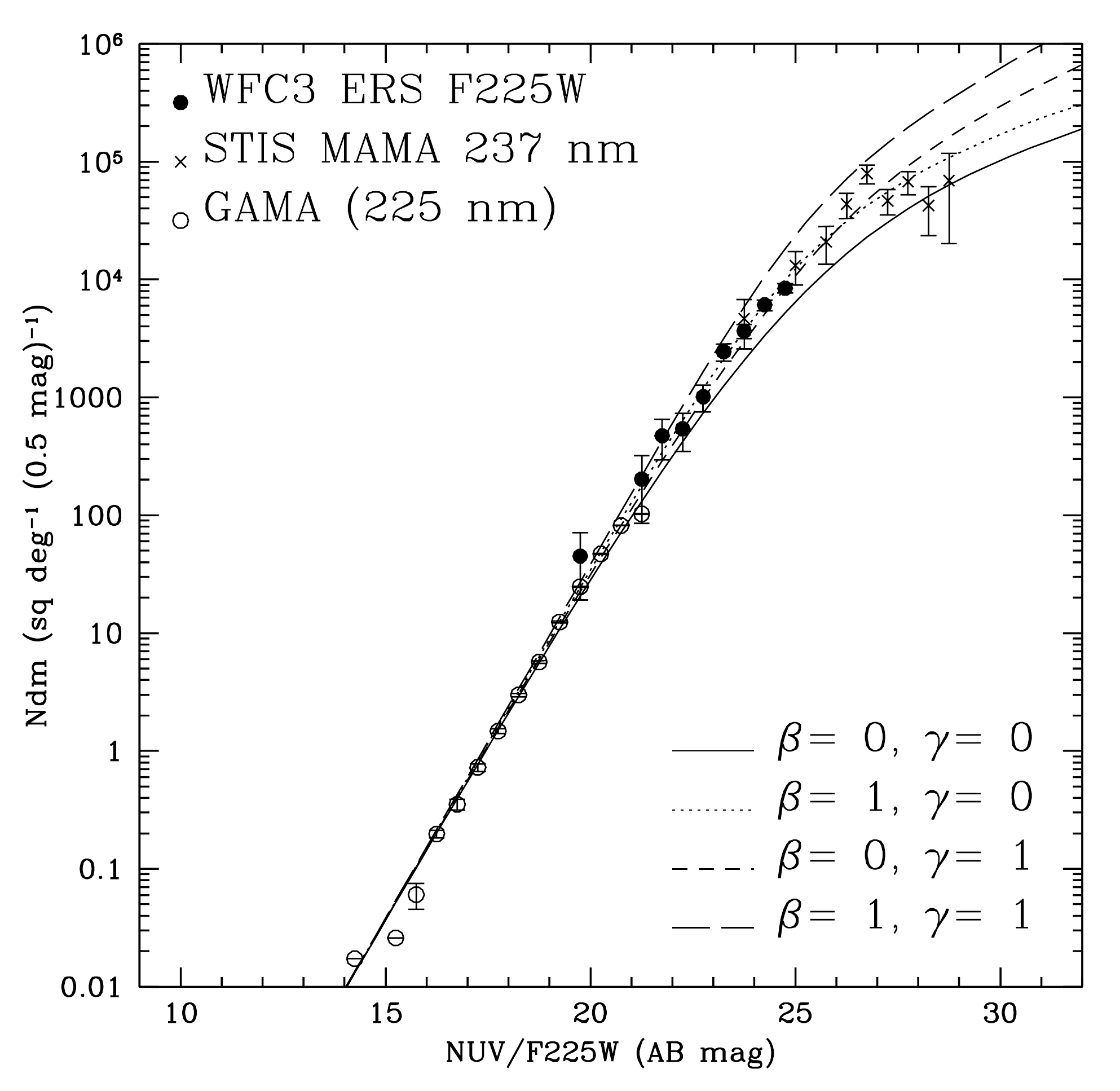}\ 
\includegraphics[width=0.55\linewidth]{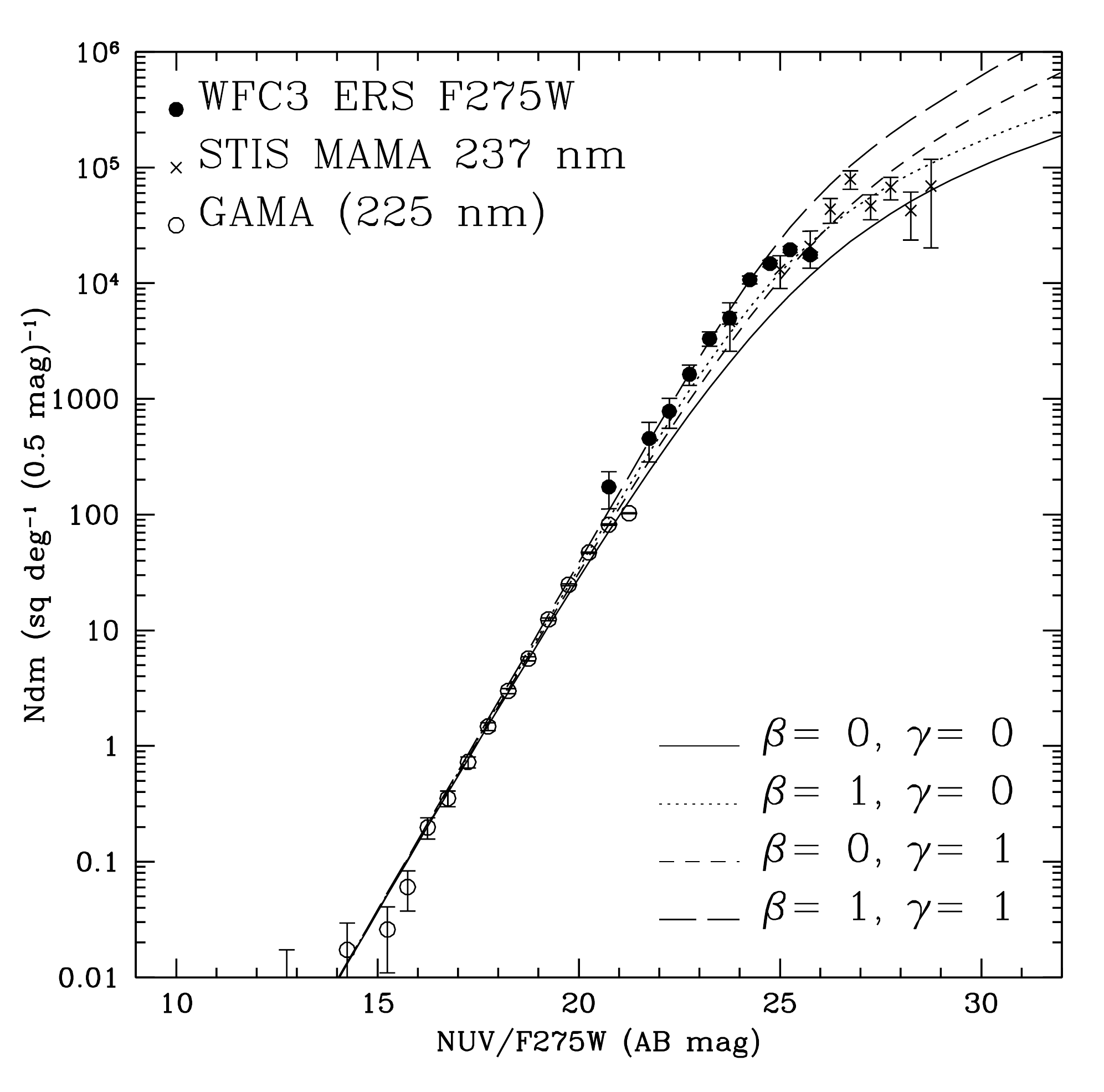}
}

\n\cl{
\includegraphics[width=0.55\linewidth]{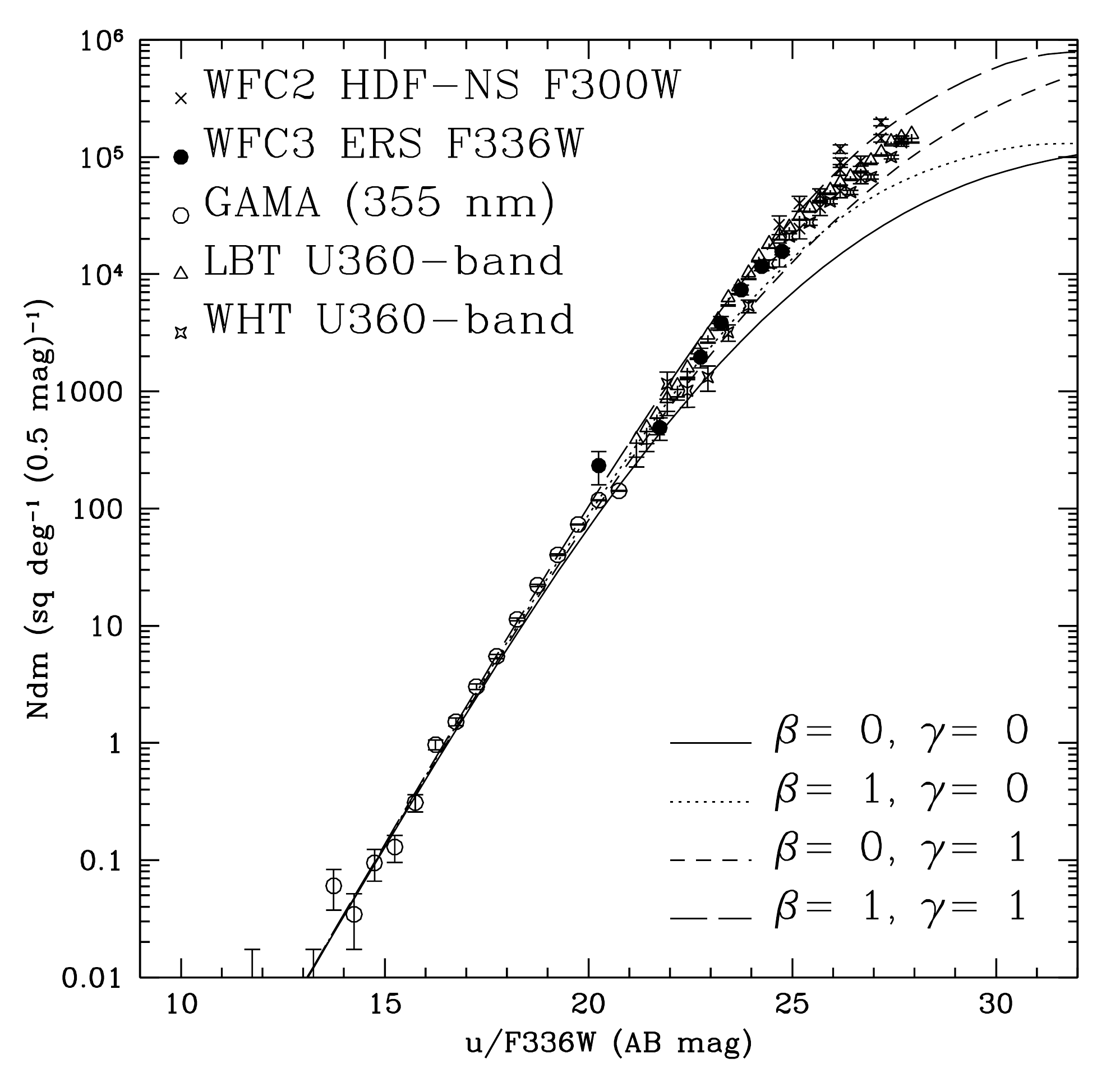}
}

\vspace*{-0.0cm}
\bn{\textbf Fig. 12a--12c.\ Differential galaxy number counts for the entire flux
range AB=10--30 mag in the F225W (12a), F275 (12b), and F336W (12c) filters. For
comparison at the bright end, we added to the ERS number counts from Fig. 11a the
panchromatic GAMA survey (Driver \etal\ 2009) counts in NUV+ugrizYJH, which cover
AB=10--21 mag (Xu \etal\ 2005; Hill \etal\ 2010a). At the faint end, we added the
WHT U-band and HDF-North and South F300W counts of Metcalfe \etal\ (2001), the
deep LBT U-band counts of Grazian \etal\ (2009), as well as the deep HST STIS
counts of Gardner \etal\ (2000). Best fit luminosity and number density evolution
models cover AB$\simeq$10--30 mag (see Driver \etal\ 1998; Hill \etal\ 2010a).

\ve 

\vspace*{-0.9cm}
\n\cl{
\includegraphics[width=0.55\linewidth]{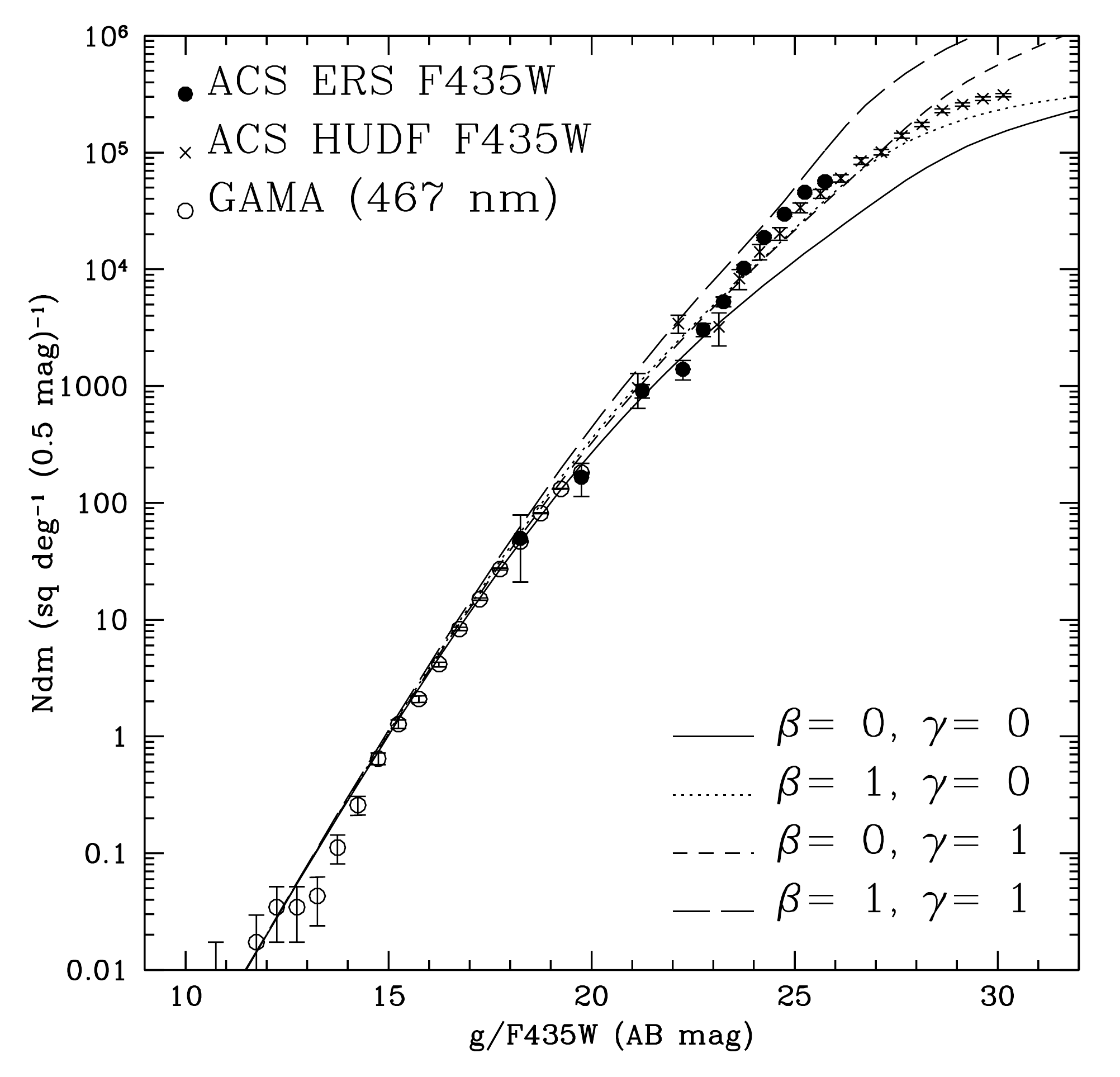}\ 
\includegraphics[width=0.55\linewidth]{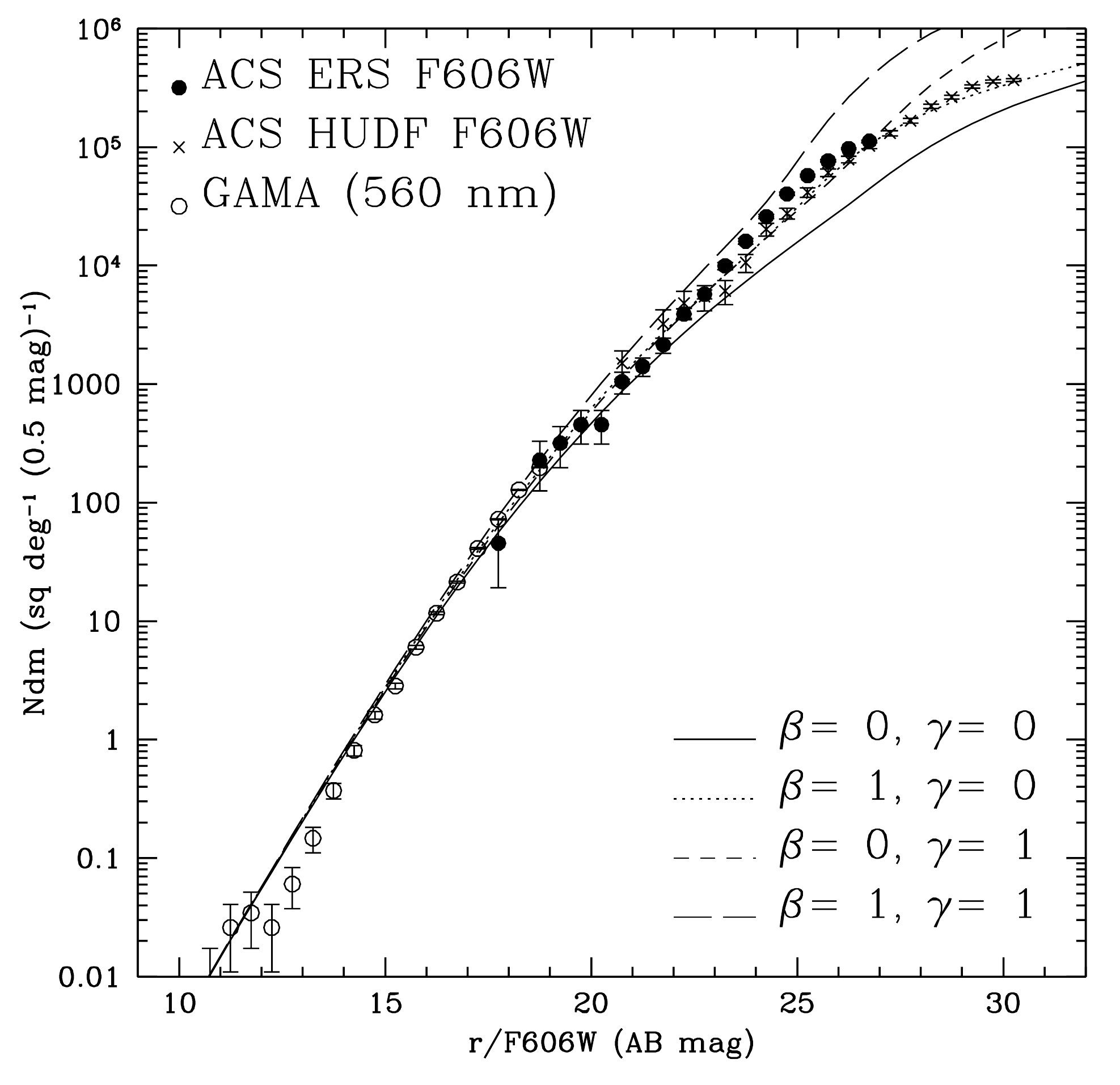}
}

\n\cl{
\includegraphics[width=0.55\linewidth]{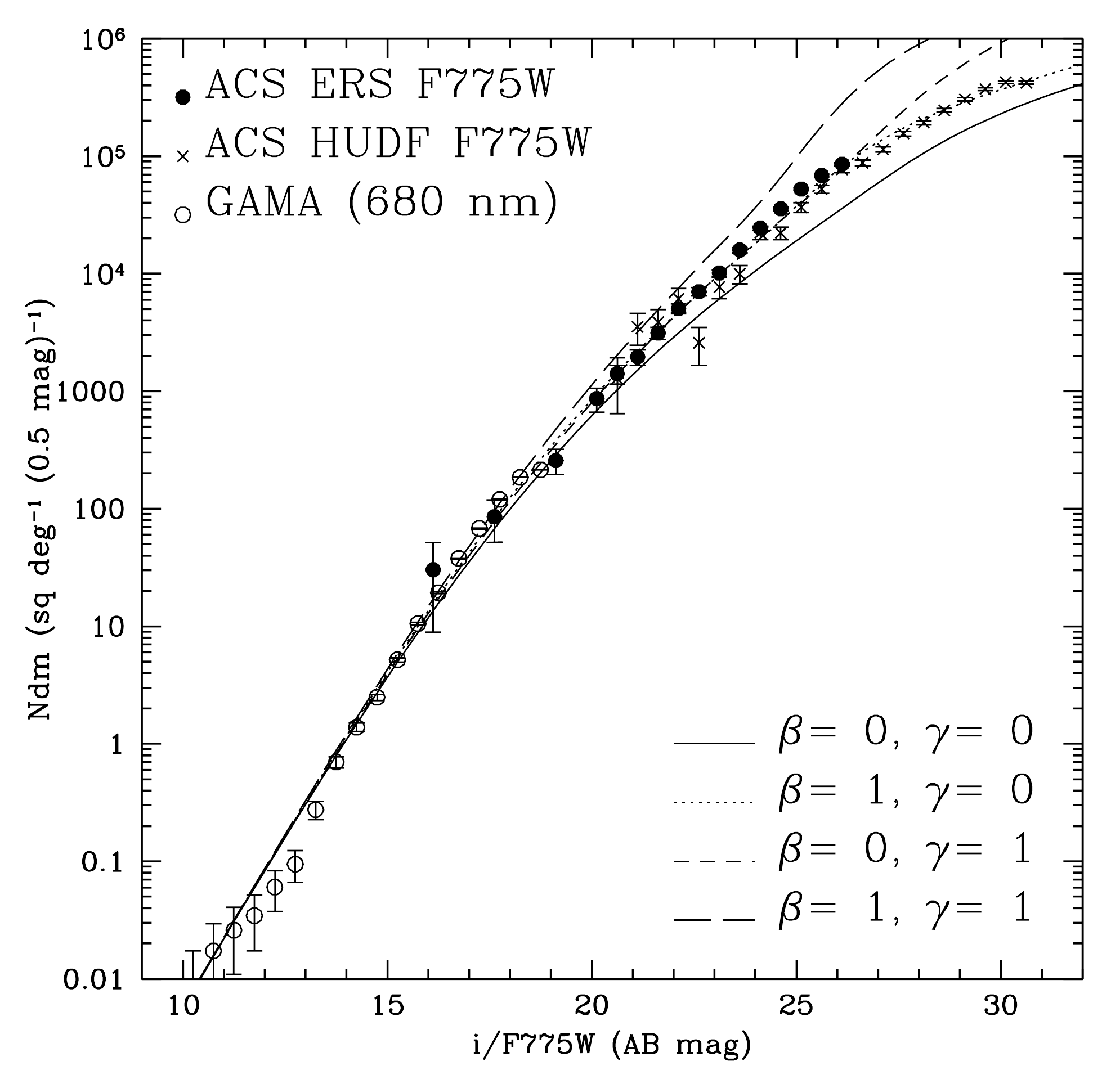}\
\includegraphics[width=0.55\linewidth]{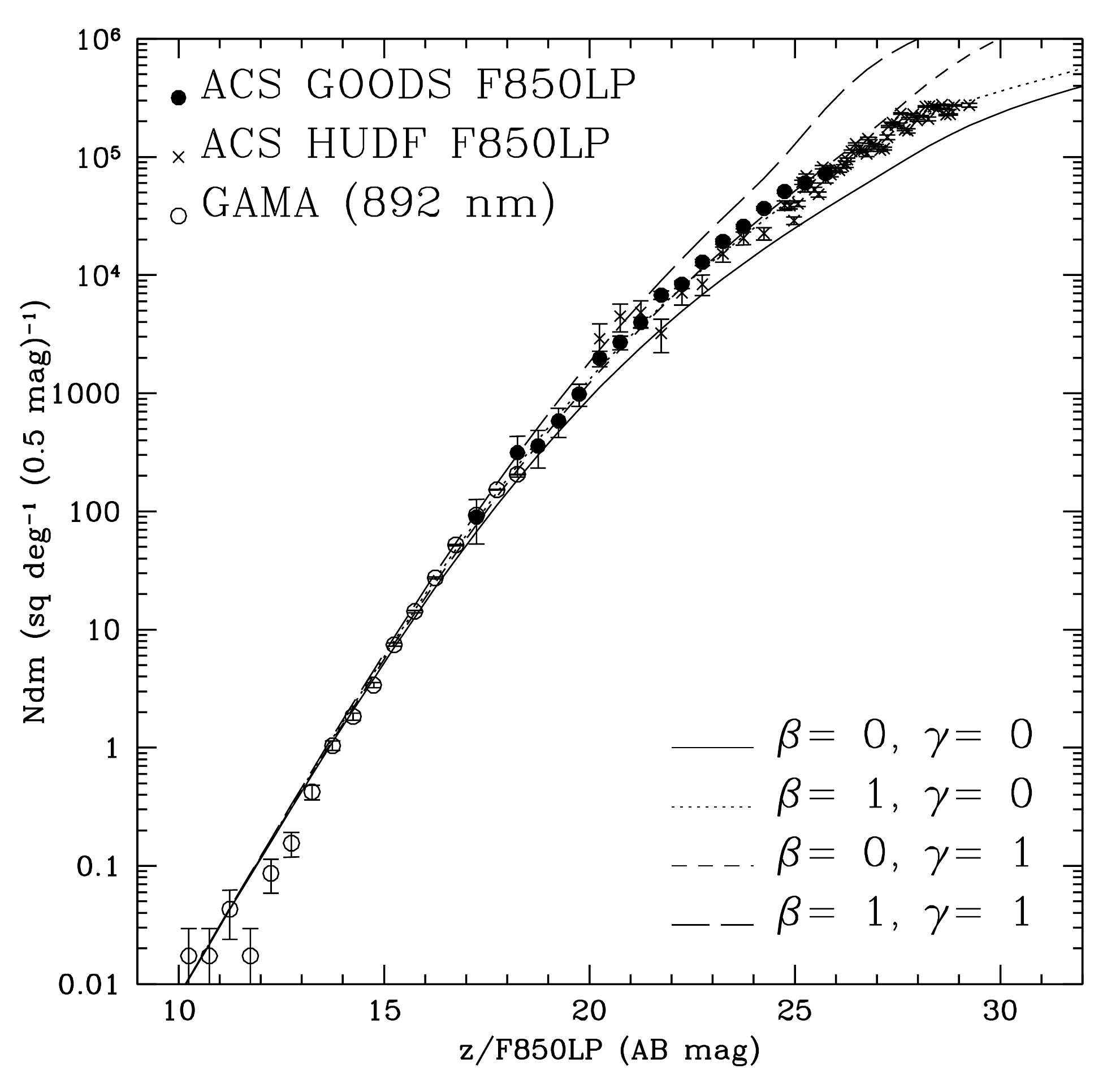}
}

\vspace*{-0.0cm}
\bn{\textbf Fig. 12d--12g.\ Differential galaxy number counts for the entire flux
range AB=10--30 mag in the F435W (12d), F606W (12e), F775W (12f), and F850LP
(12g) filters. Further details are given in the caption of Fig. 12a--12c. At the
faint end, we also added the HUDF counts in BViz from Beckwith \etal\ (2006),
which cover AB=23--30 mag.

\ve 

\vspace*{-0.9cm}
\n\cl{
\includegraphics[width=0.55\linewidth]{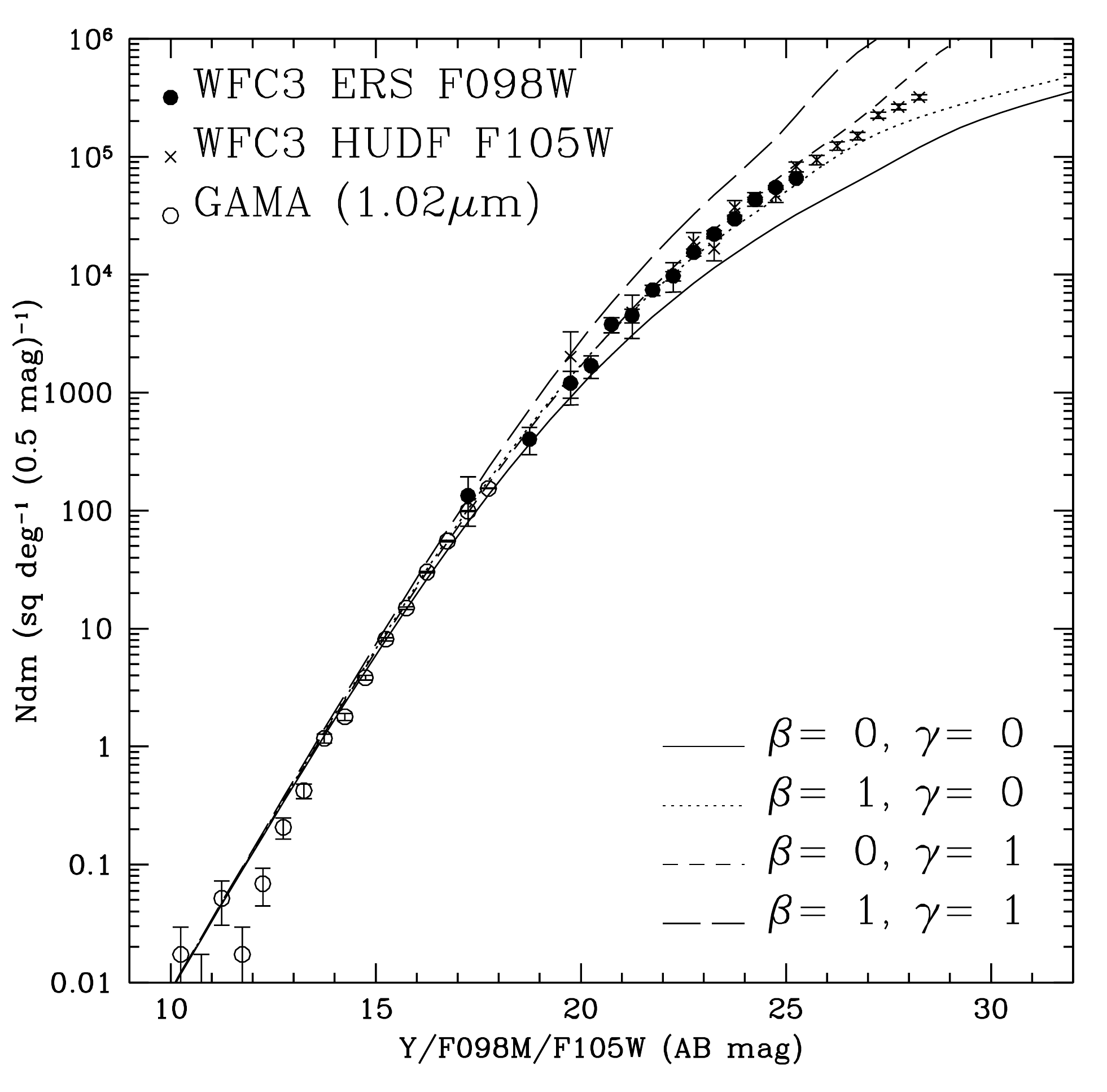}
}

\n\cl{
\includegraphics[width=0.55\linewidth]{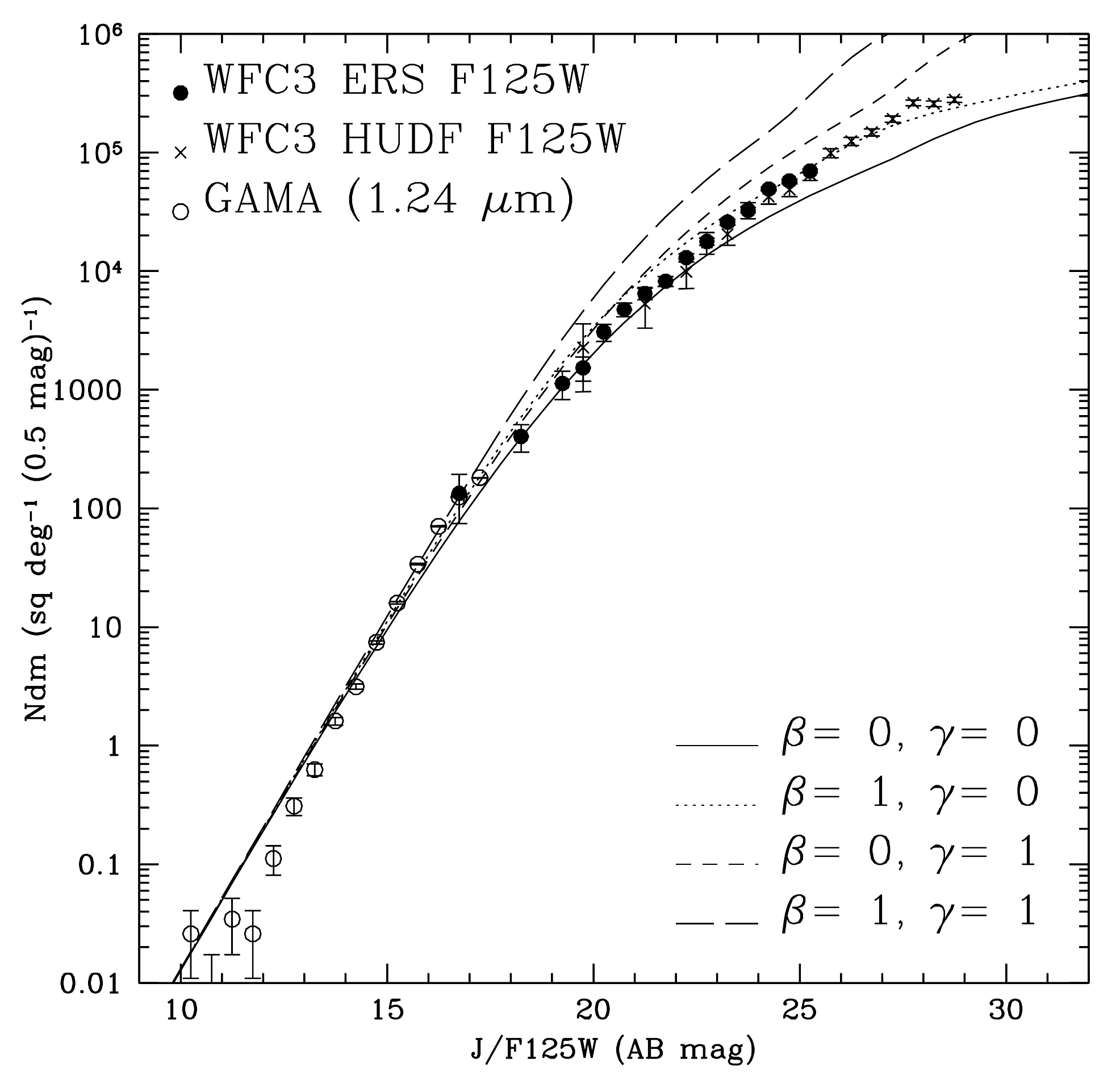}\ 
\includegraphics[width=0.55\linewidth]{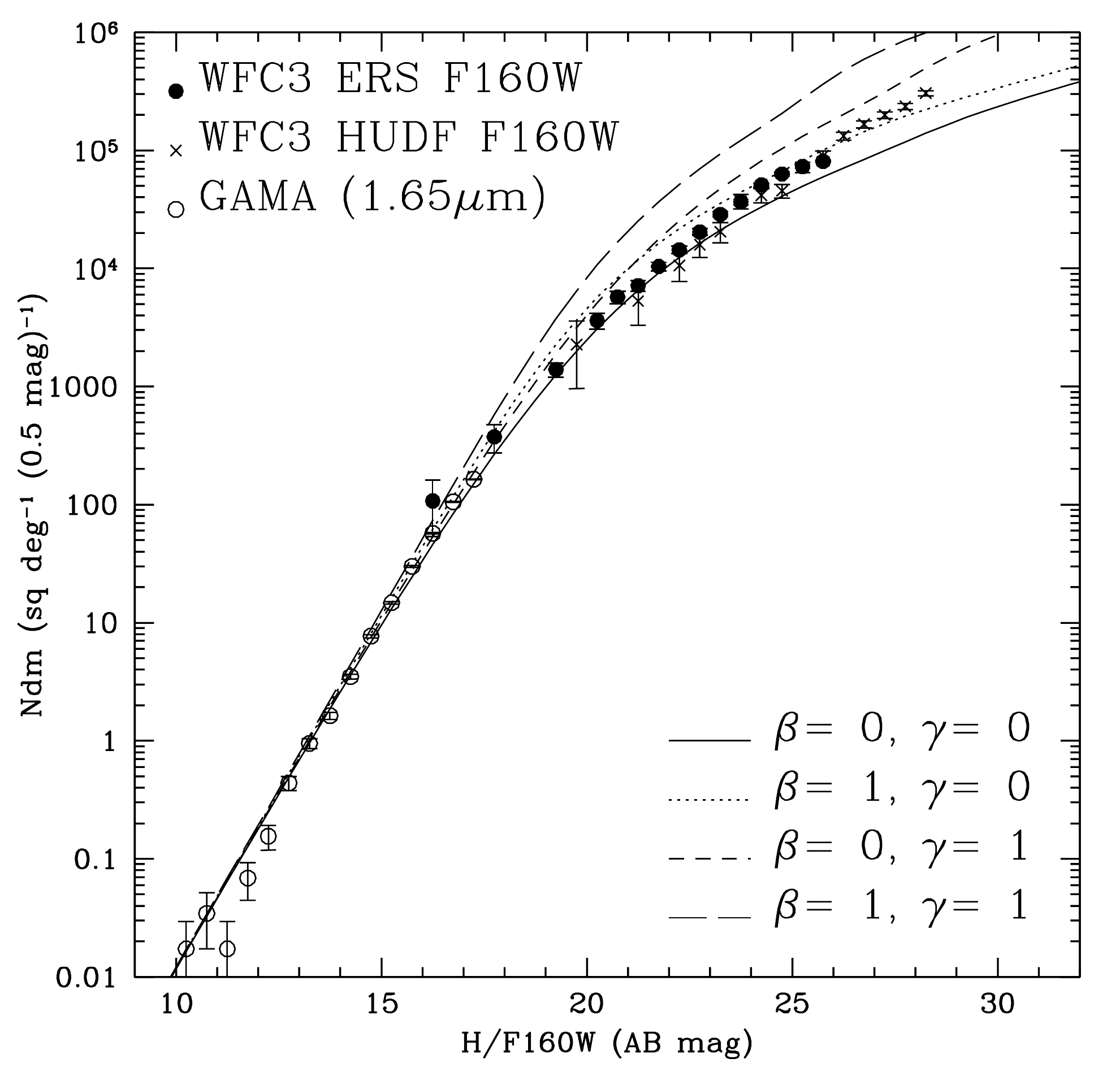}
}

\vspace*{-0.0cm}
\bn{\textbf Fig. 12h--12j.\ Differential galaxy number counts for the entire flux
range AB=10--30 mag in the F098M/F105W (12h), F125W (12i) and F160W (12j)
filters. Further details are given in the caption of Fig. 12a--12c. At the faint
end, we also added the HUDF counts from the Bouwens \etal\ (2009) YJH data, as
compiled by Yan \etal\ (2010), which cover AB=24--30 mag.

\ve 

\vspace*{-1.0cm}
\n\cl{
\includegraphics[width=1.25\linewidth,angle=90]{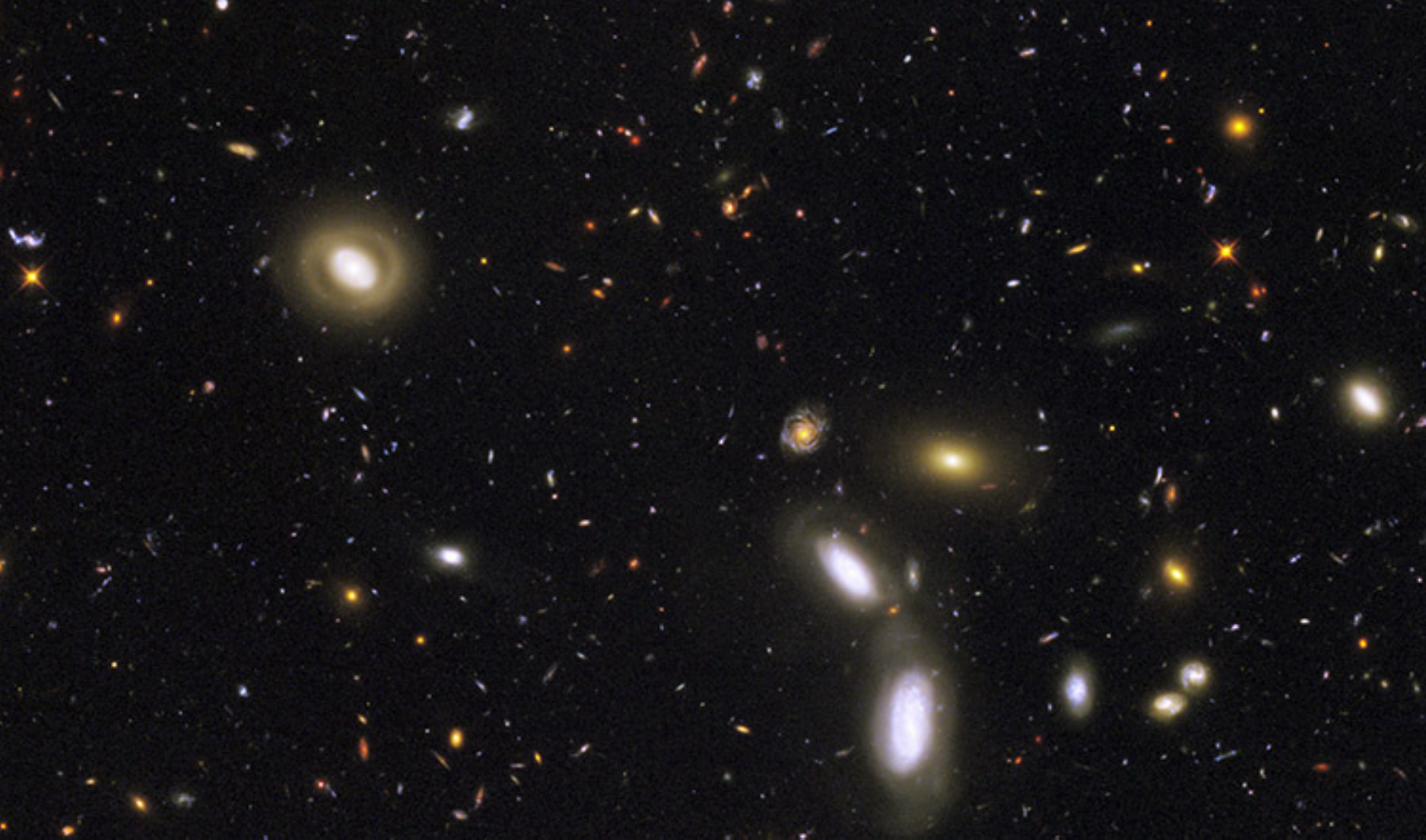}
}

\vspace*{+0.2cm}
\n{\textbf Fig. 5b.\ Enlargement of the panchromatic 10-band ERS color image in
the GOODS-South field (see Fig. 5a). [For best display, please zoom in on the
full-resolution version of this image, which is available on this
$URL$\footnote{http://www.asu.edu/clas/hst/www/wfc3ers/ERS2\_gxysv4ln.tif} ]. }

\ve 

\vspace*{-0.0cm}
\n\cl{
\includegraphics[width=0.33\linewidth,angle=0]{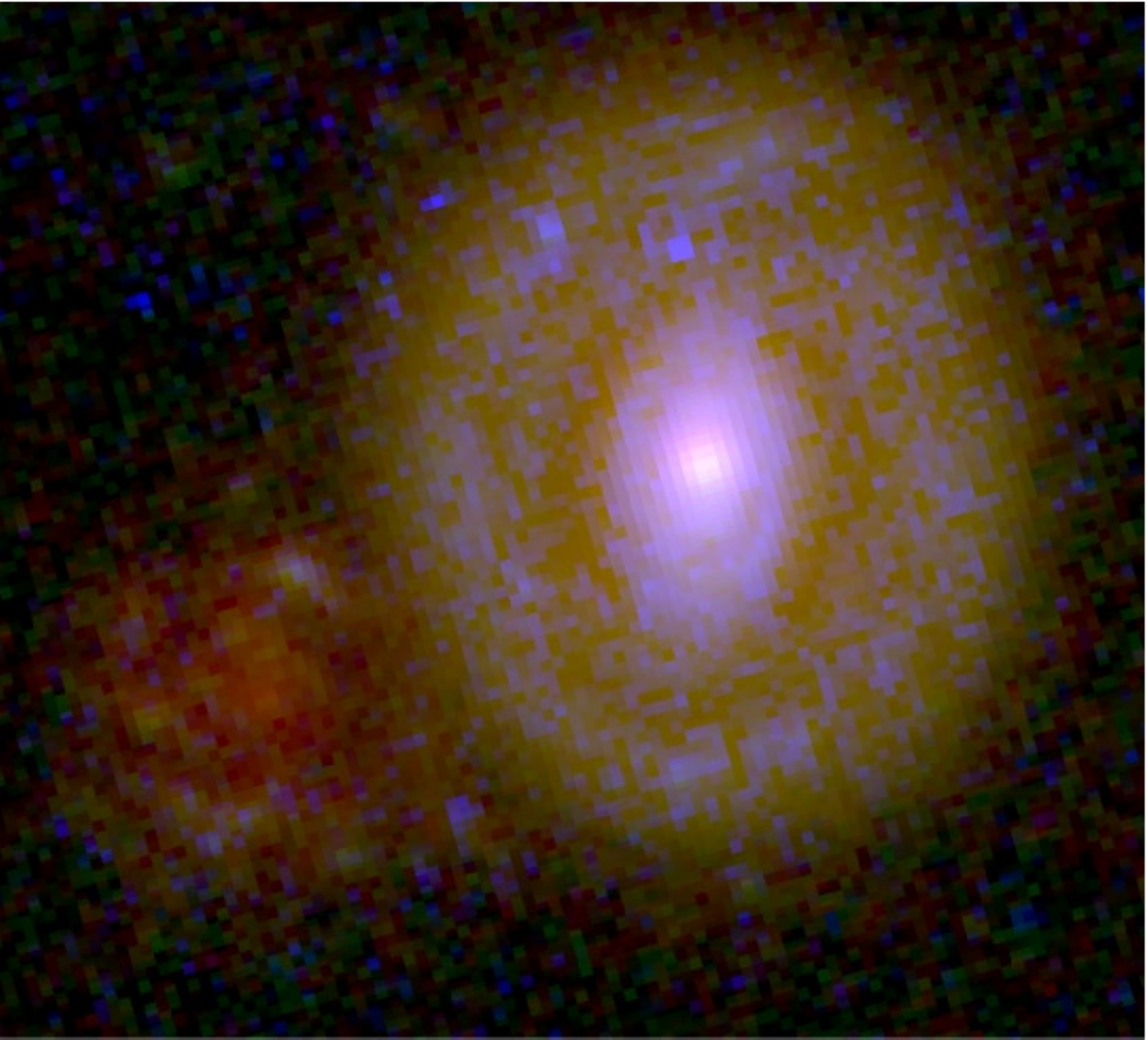}
\includegraphics[width=0.33\linewidth,angle=0]{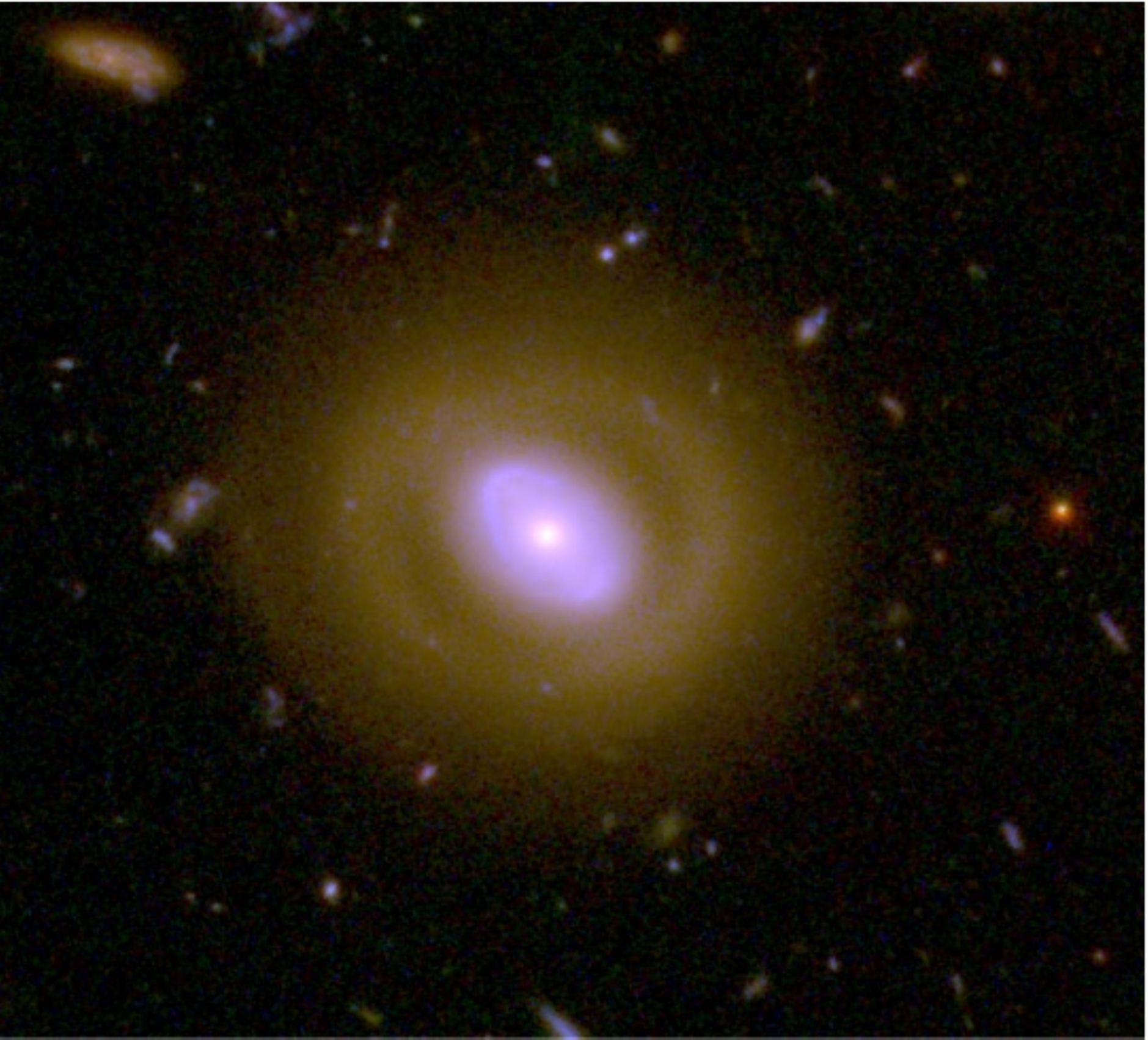}
\includegraphics[width=0.33\linewidth,angle=0]{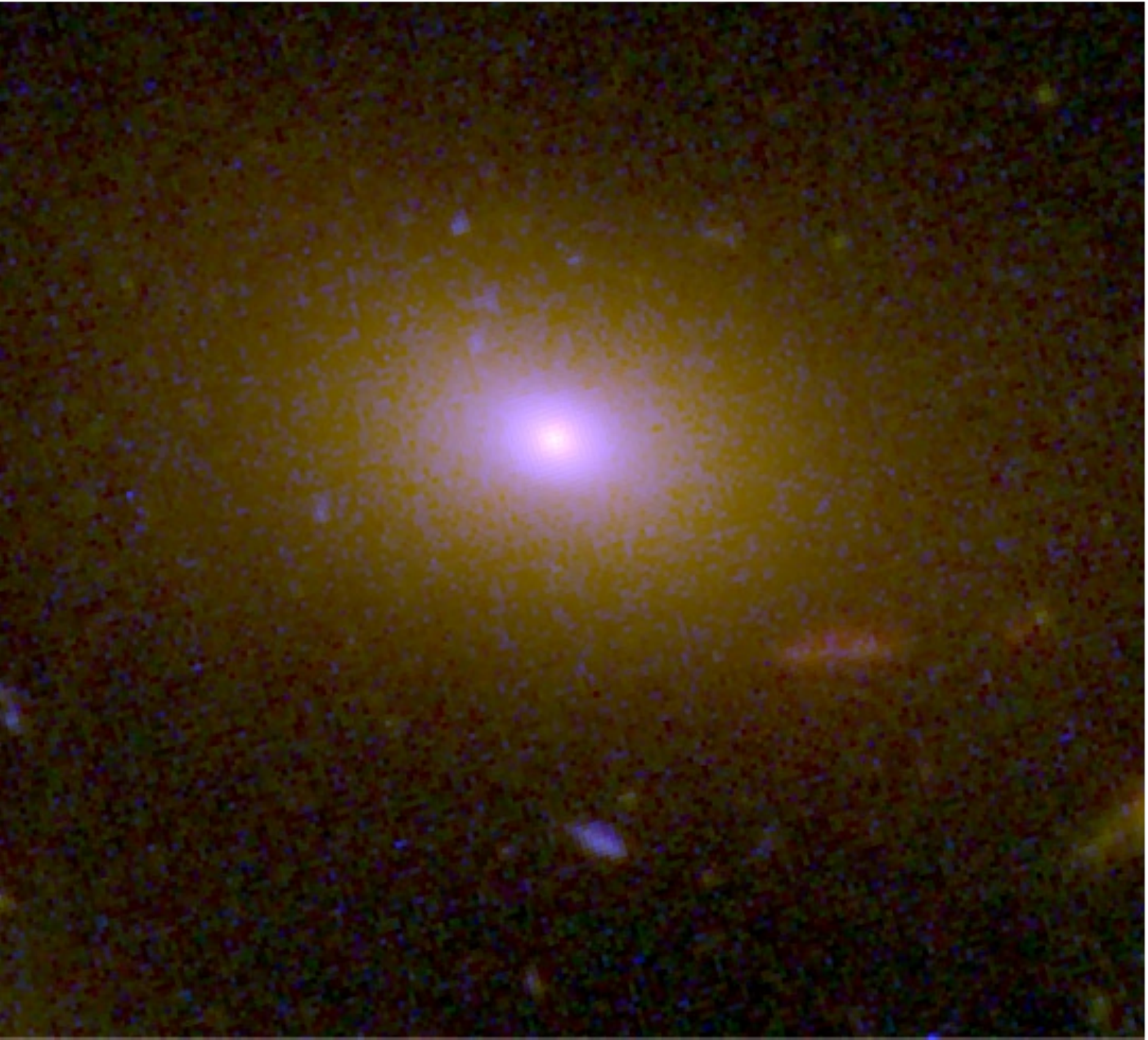}
}

\n\cl{
\includegraphics[width=0.33\linewidth,angle=0]{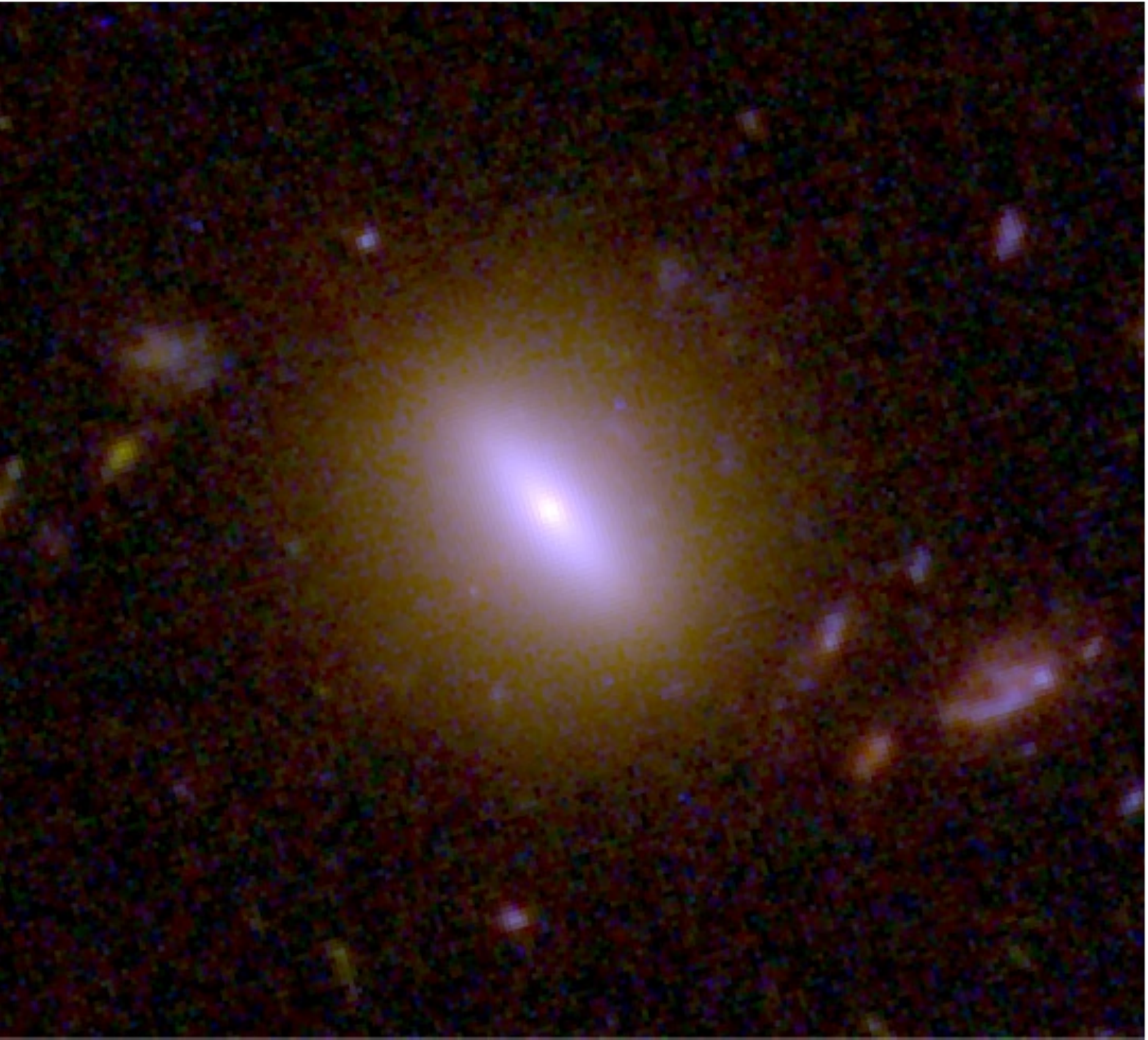}
\includegraphics[width=0.33\linewidth,angle=0]{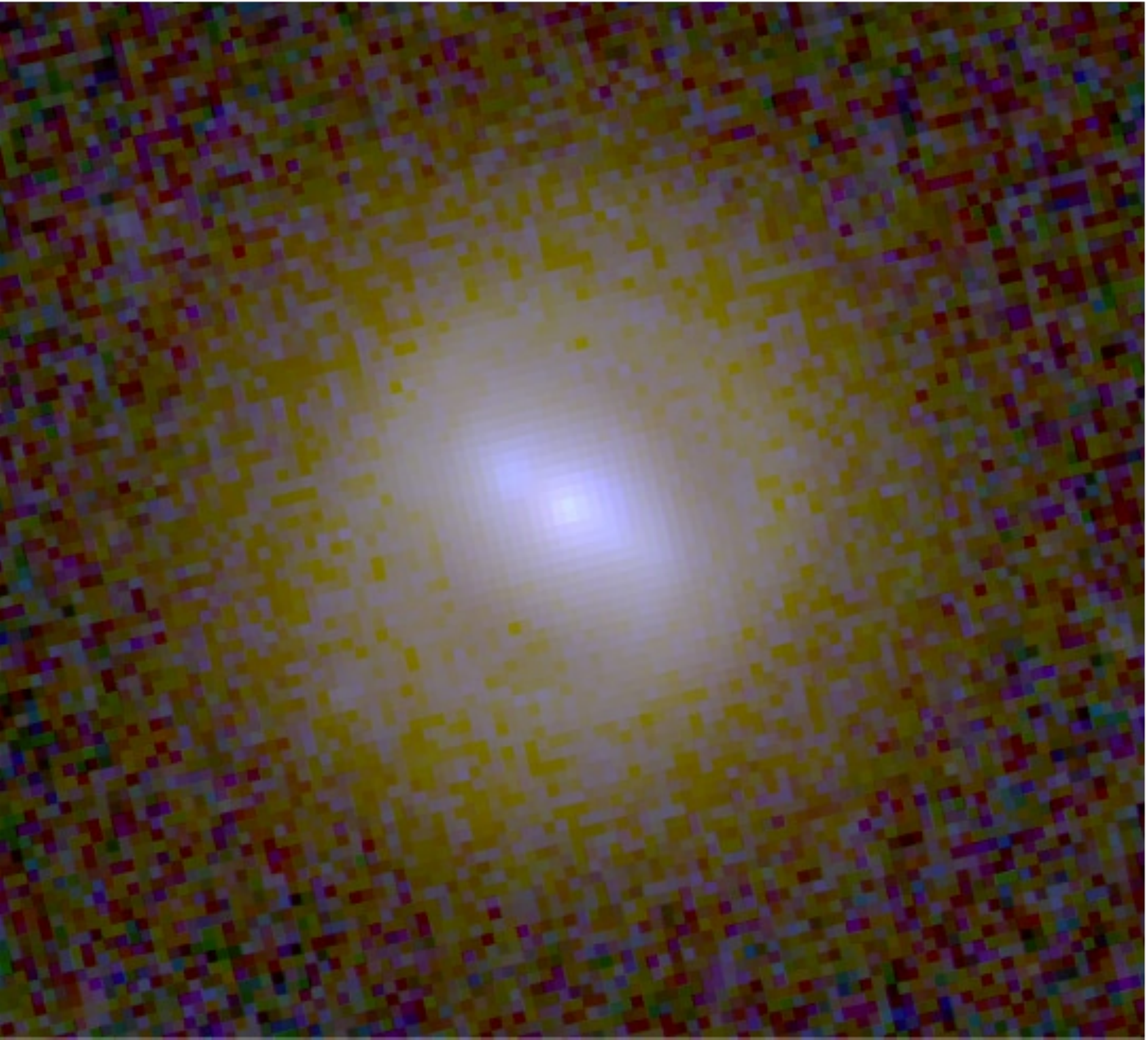}
\includegraphics[width=0.33\linewidth,angle=0]{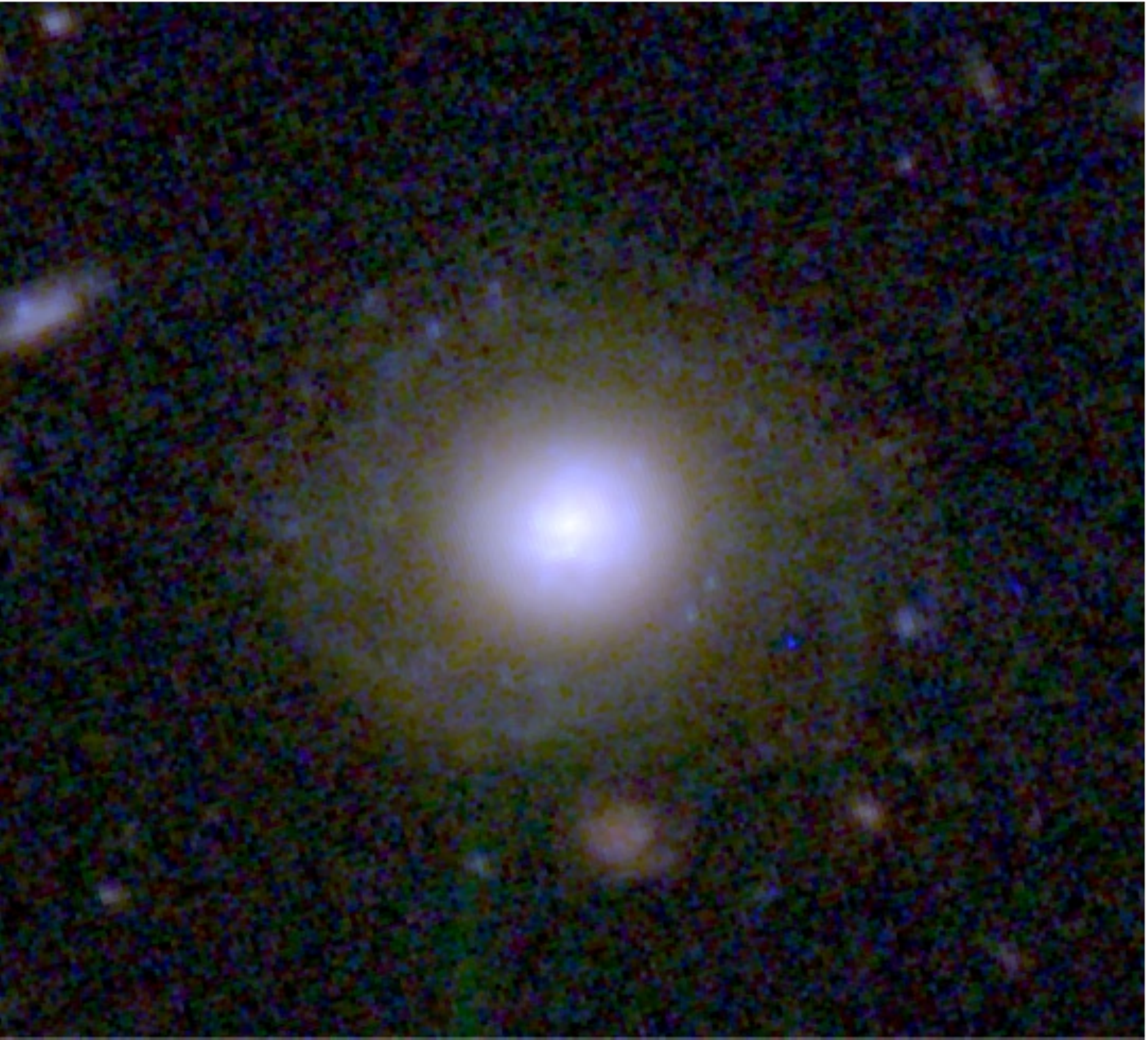}
}

\n\cl{
\includegraphics[width=0.33\linewidth,angle=0]{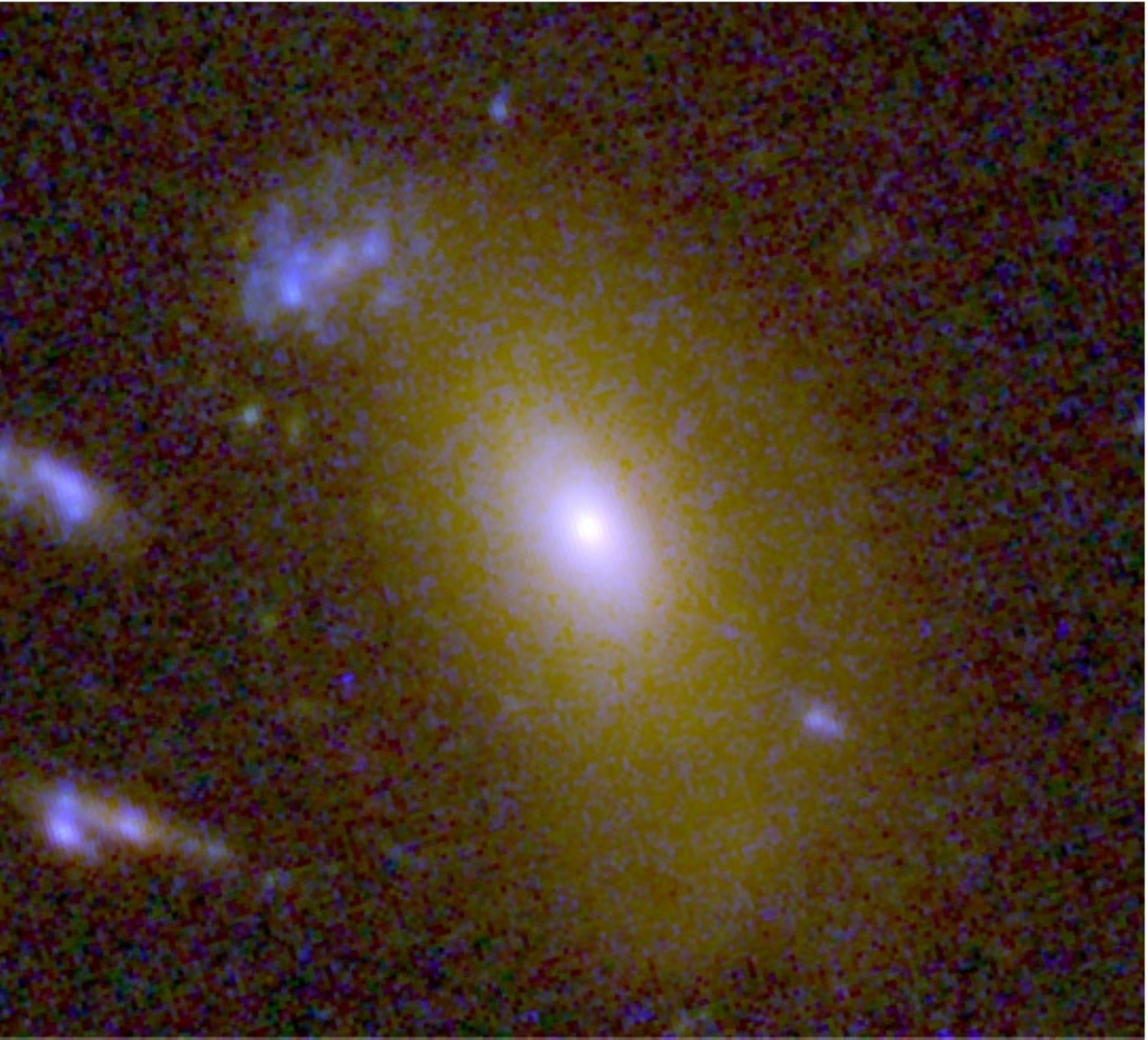}
\includegraphics[width=0.33\linewidth,angle=0]{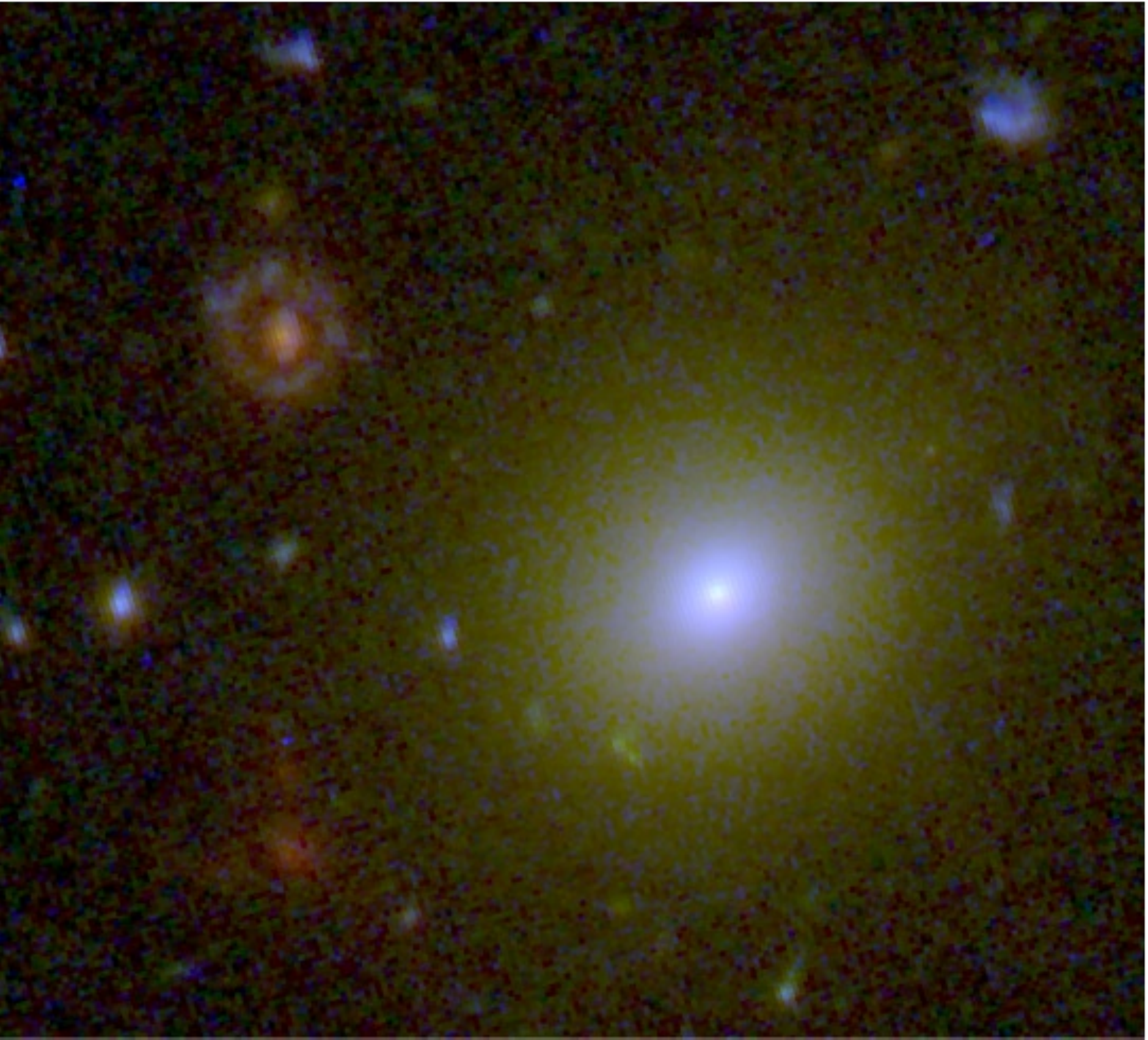}
\includegraphics[width=0.33\linewidth,angle=0]{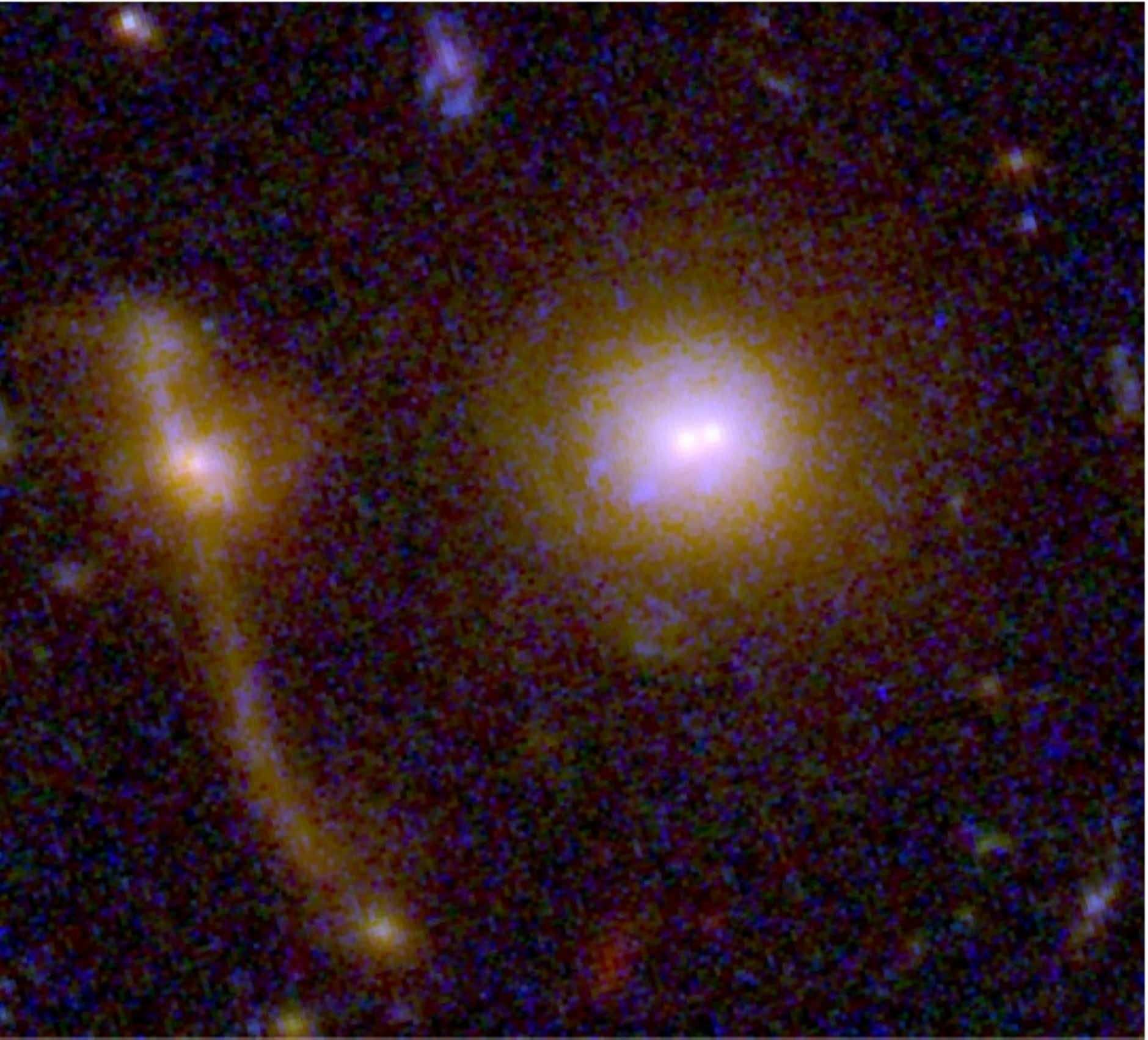}
}

\n{
\includegraphics[width=0.33\linewidth,angle=0]{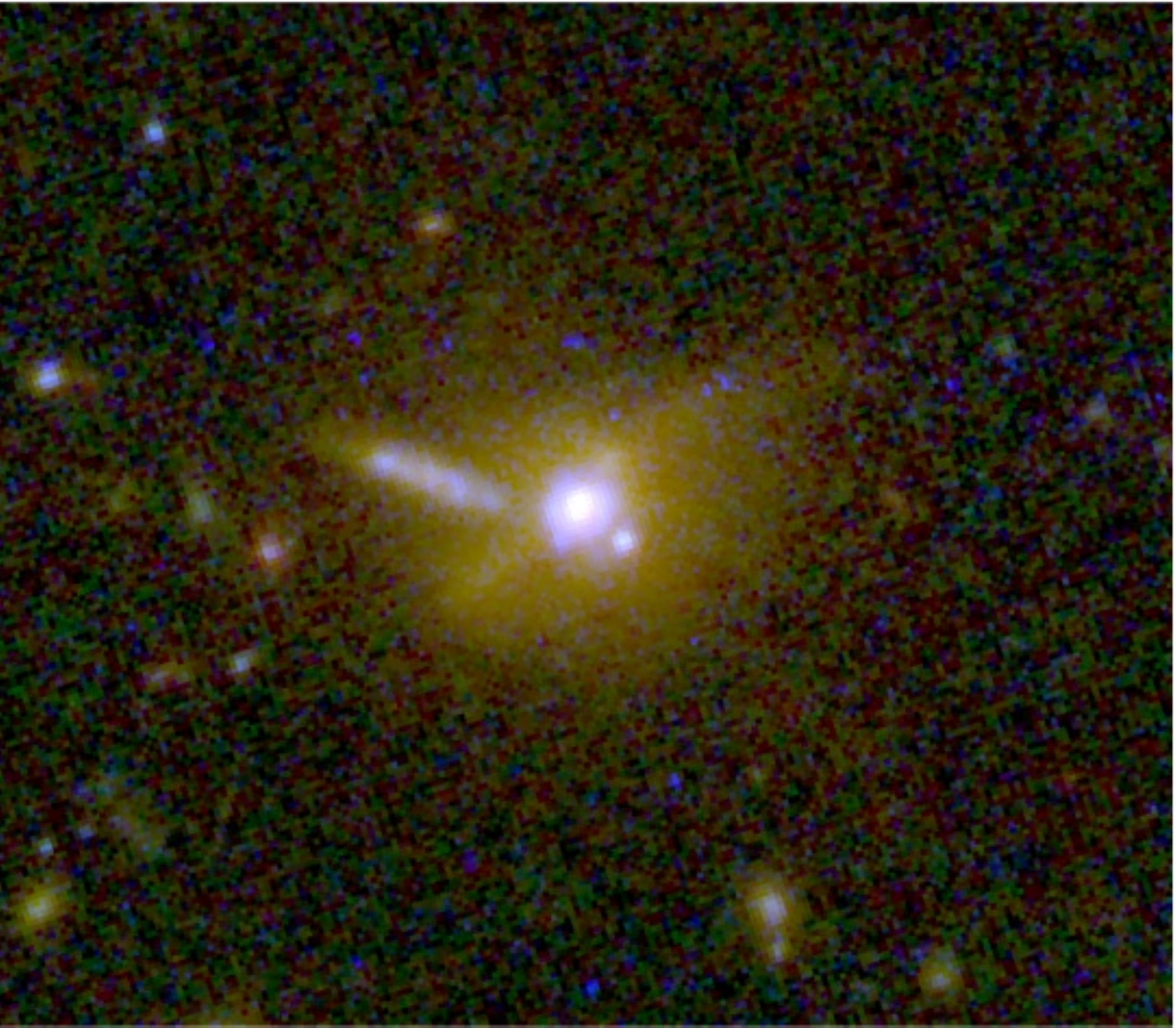}
}

\vspace*{-0.0cm}
\bn{\textbf Fig. 13.\ Panchromatic postage stamps of early-type galaxies in the
ERS with nuclear star-forming rings, bars, or other interesting nuclear
structure. Each postage stamp is displayed at a slightly different color stretch
that best brings out the UV nuclear structure. For further details, see Rutkowski
\etal\ (2011). }

\ve 

\cl{
\includegraphics[width=0.21\linewidth,angle=0]{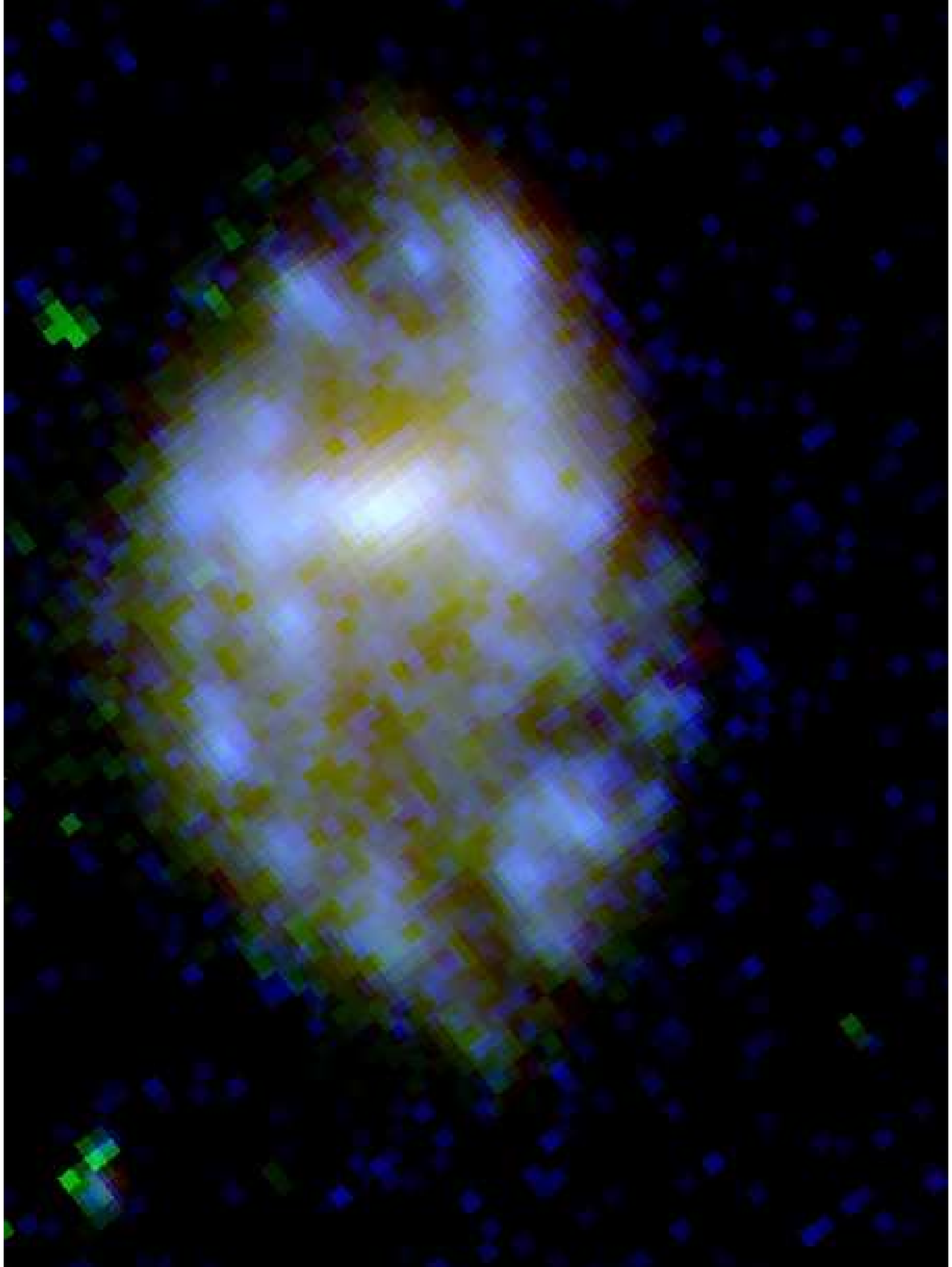}
\includegraphics[width=0.20\linewidth,angle=0]{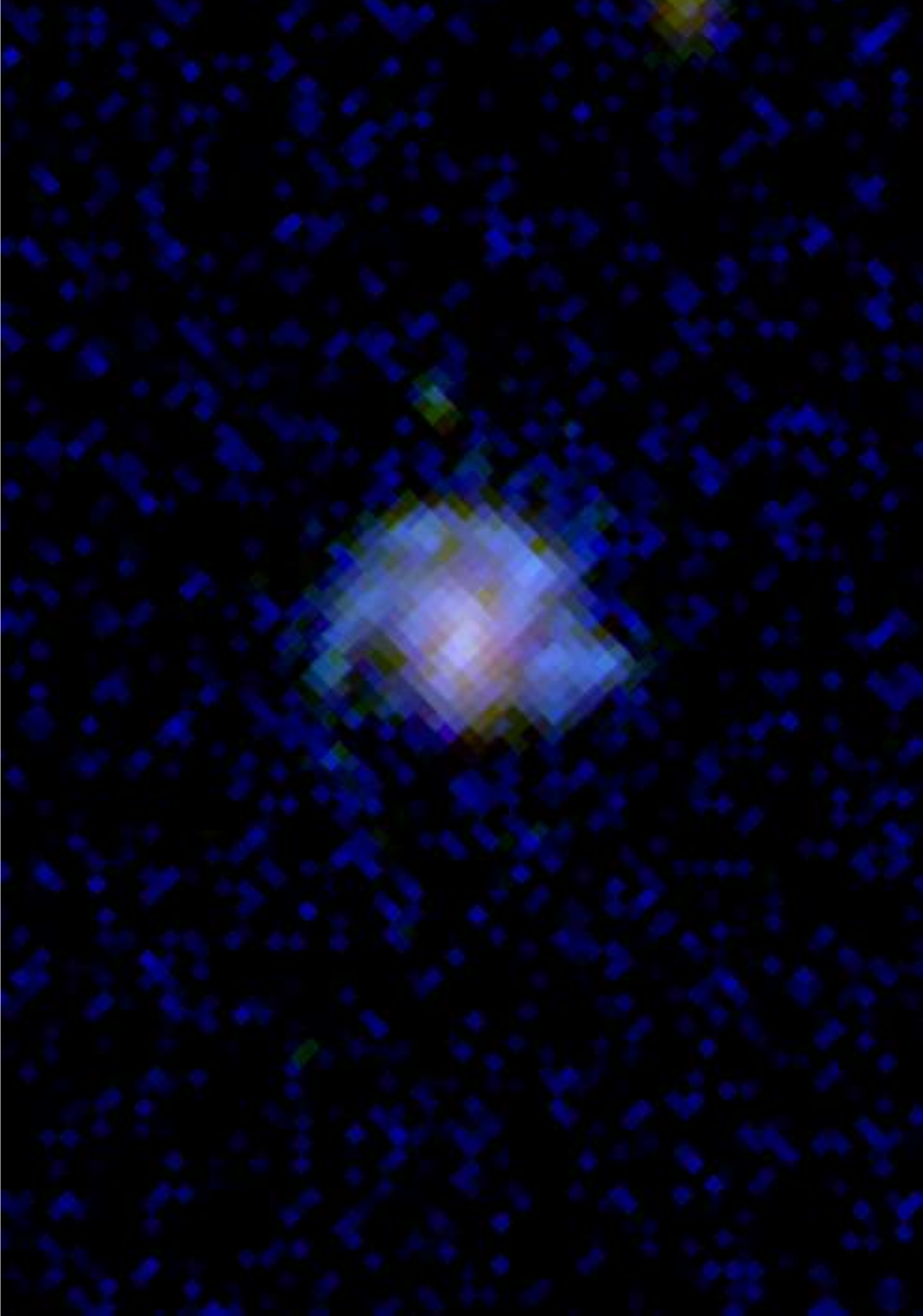}
\includegraphics[width=0.20\linewidth,angle=0]{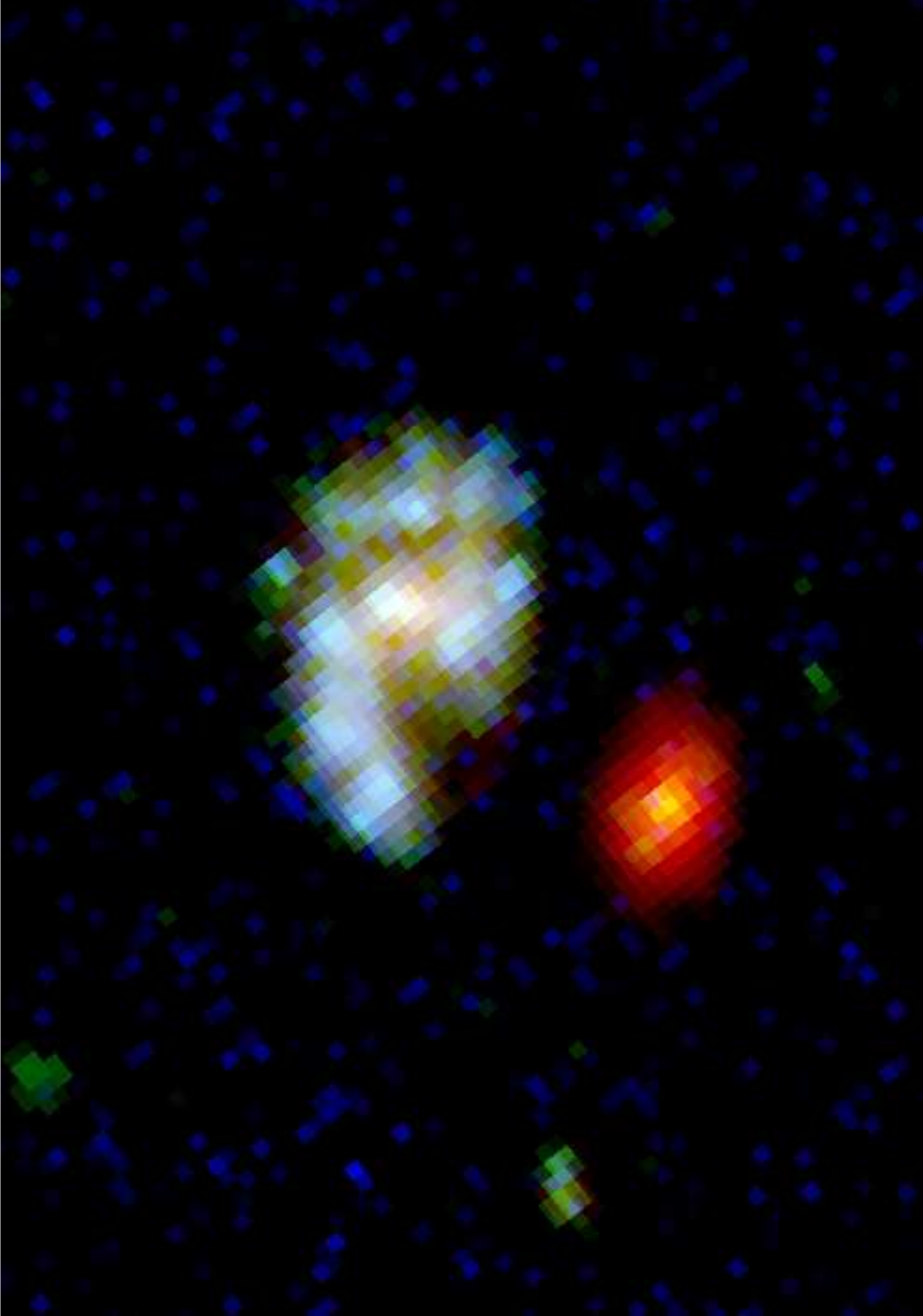}
\includegraphics[width=0.20\linewidth,angle=0]{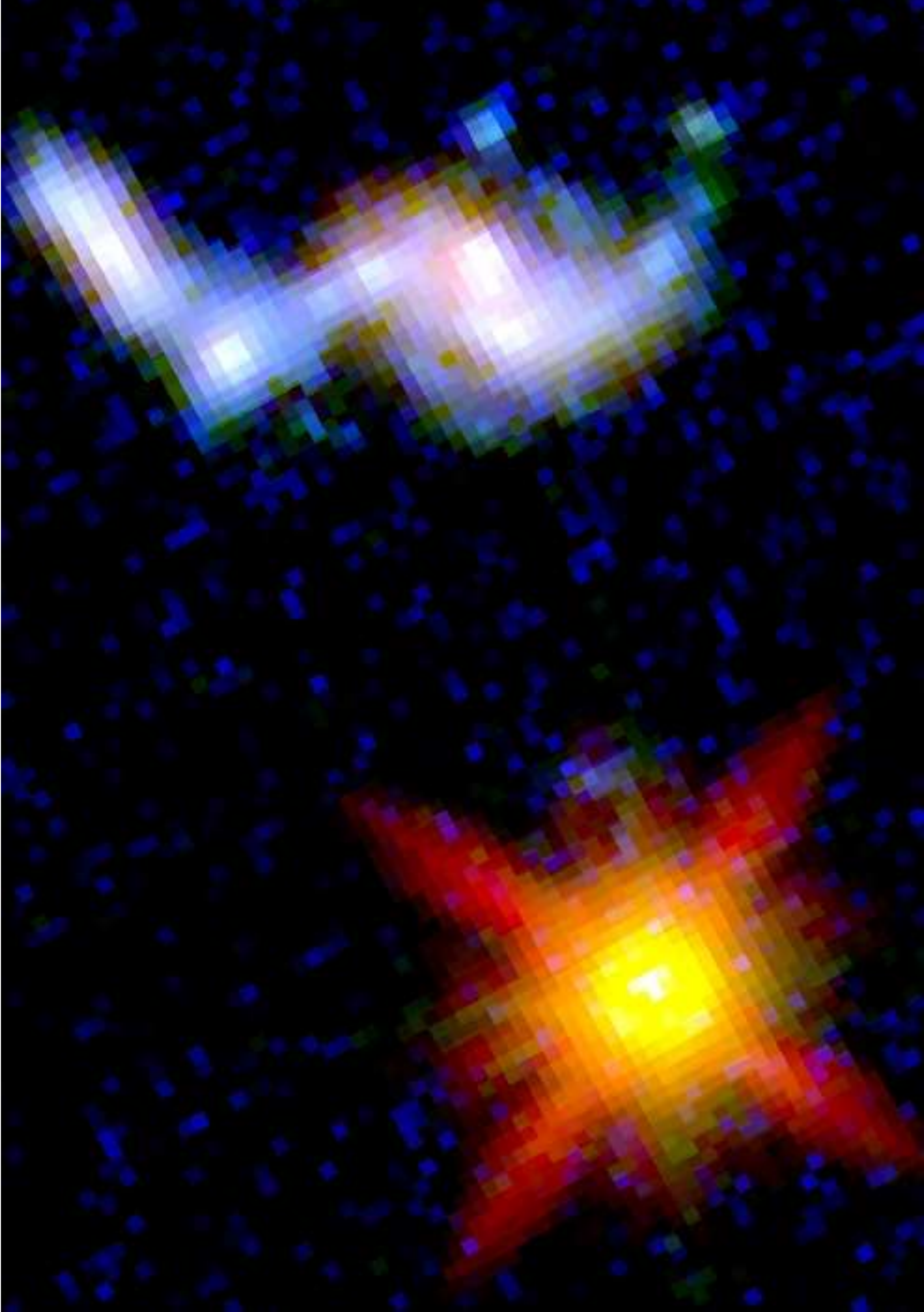}
\includegraphics[width=0.20\linewidth,angle=0]{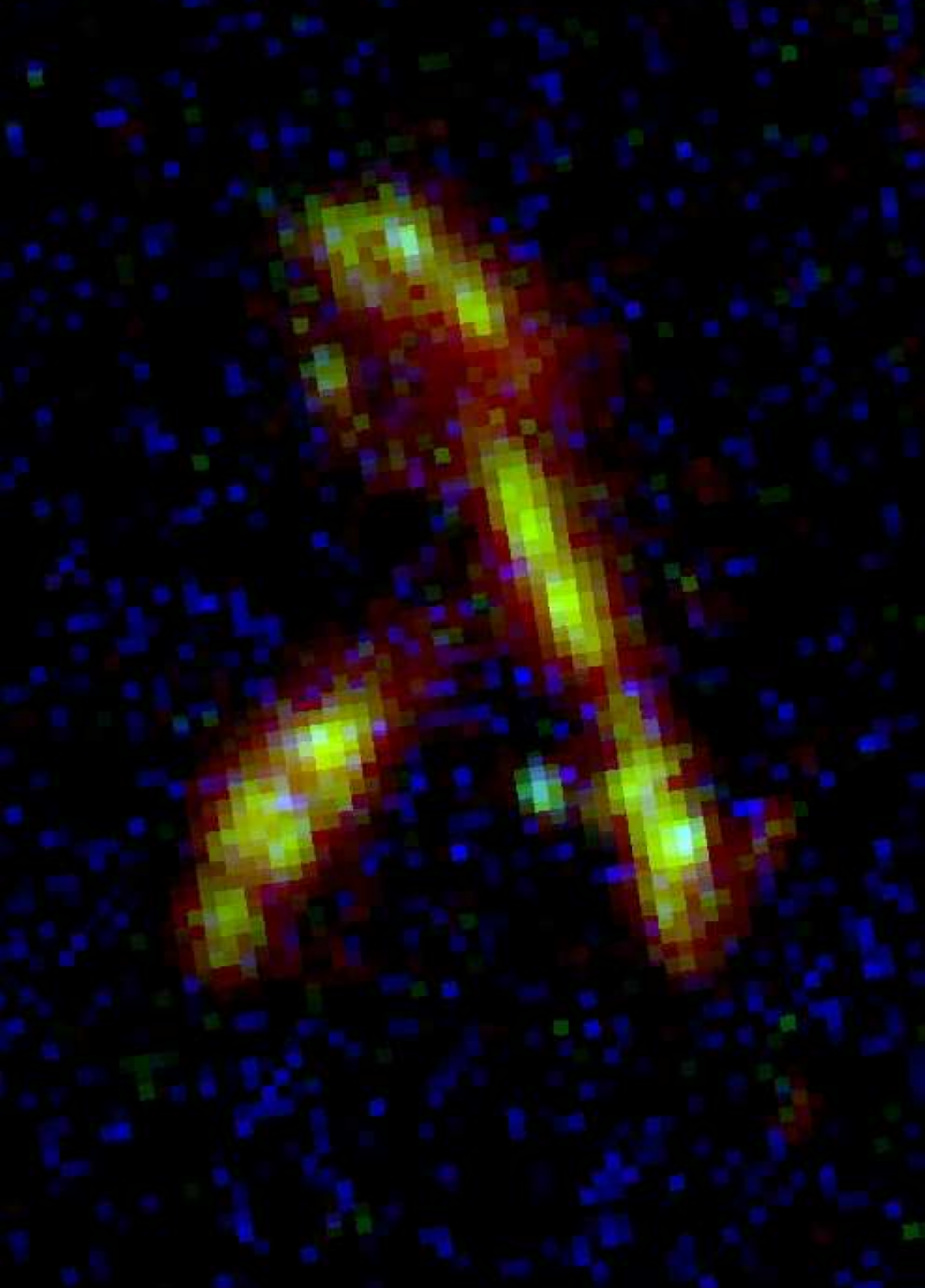}
}

\vspace*{-0.0cm}
\bn{\textbf Fig. 14.\ Panchromatic postage stamps of objects with interesting
morphological structure in the 10-band ERS color images of the GOODS-South field:
from left to right, high signal-to-noise detections of ERS galaxies resembling
the main cosmological parameters \Ho, $\Omega$, $\rho_{o}$, $w$, and $\Lambda$,
respectively. These images illustrate the rich and unique morphological
information available in the 10-band panchromatic ERS data set. }

\ve 

\vspace*{-0.0cm}
\n\cl{
\includegraphics[width=1.00\linewidth,angle=-0]{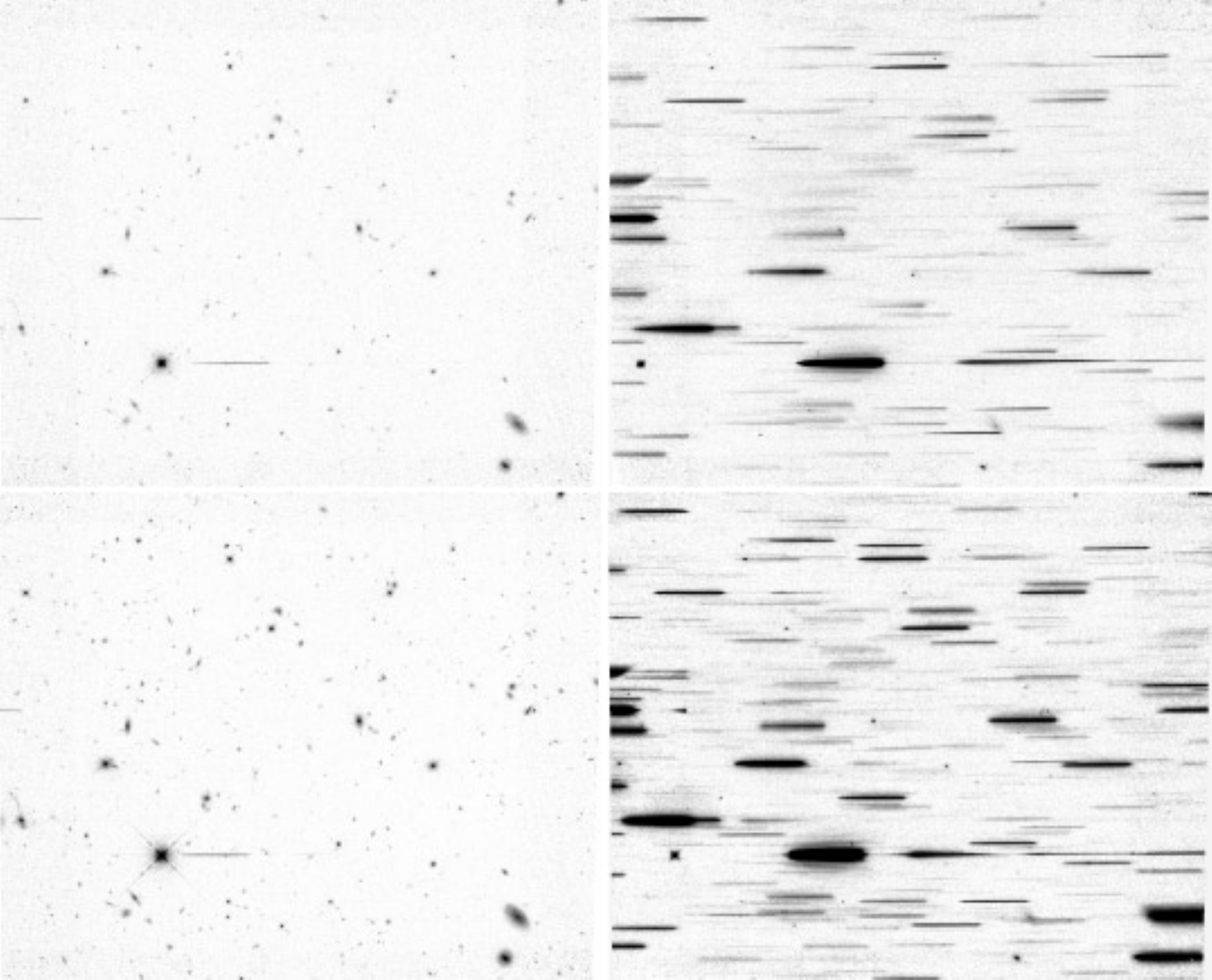}
}

\vspace*{-0.0cm}
\bn{\textbf Fig. 15a--15d.\ The 4212 sec WFC3 G102 (top right) and G141 (bottom
right) grism exposures of a single WFC3 pointing in the GOODS-South field (green
box in Fig. 3), together with their 1612 sec finder images in F098M (top left)
and G141 (bottom left), respectively. Each image is a 4-point dithered mosaic.
[For best display, please zoom in on the full-resolution PDF version of this
image]. All brighter object grism spectra show a 0$^{th}$ order image to their
left, displaced by about twice the spectral image length, which should not be
confused with real emission lines. Many faint object spectra are visible to a
continuum flux of AB\cle 25--25.5 mag, including many faint emission line
galaxies. For details, see Straughn \etal\ (2009, 2011). }

\ve 

\vspace*{-0.0cm}
\n\cl{
\includegraphics[width=0.55\linewidth,angle=-0]{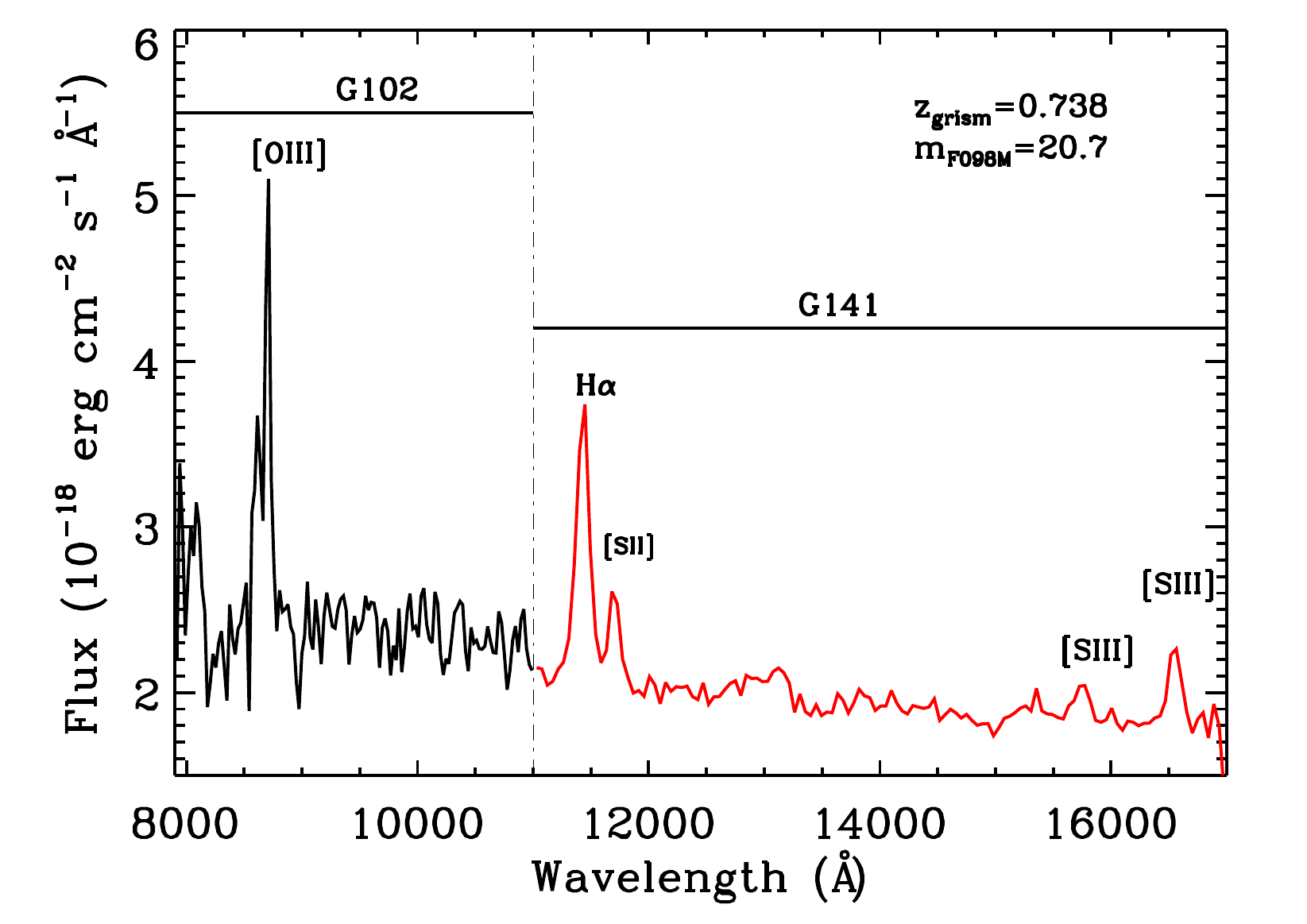}
\includegraphics[width=0.55\linewidth,angle=-0]{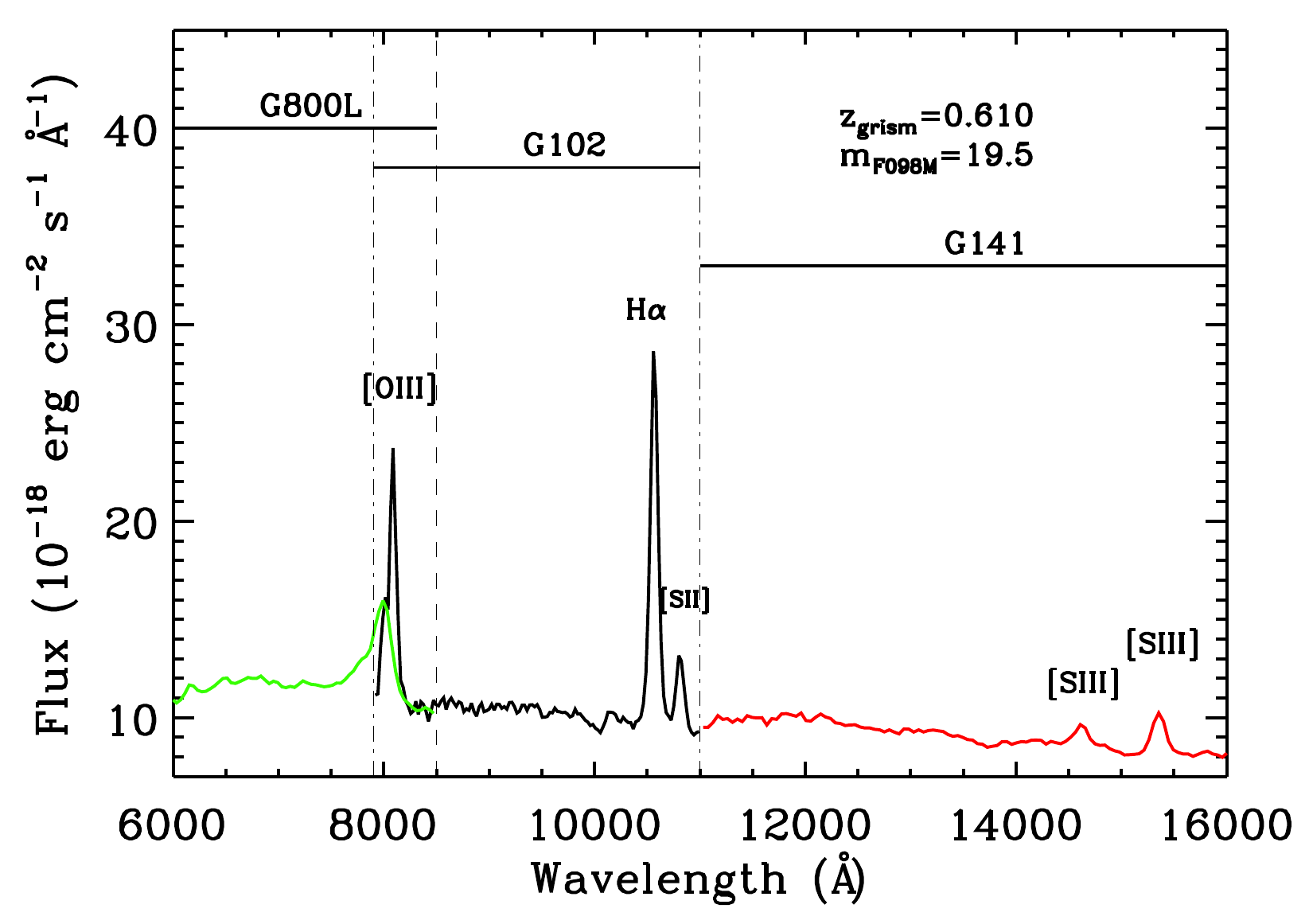}
}

\vspace*{-0.0cm}
\n{\textbf Fig. 15e--15f.\ Examples of WFC3 G102 and G141 spectra of emission
line galaxies extracted from the ERS grism images (Fig. 15a--15d). The left panel
shows an ELG at z=0.738 and the right panel at z=0.610. The latter also shows the
available lower-resolution ACS G800L grism spectrum (green). The WFC3 G102 and
G141 spectra allow for low-resolution faint object spectroscopy over the entire
0.80--1.70\mum-range unimpeded by the ground-based OH-forest. For further
details, see Straughn \etal\ (2009, 2011). }

\ve 

\end{document}